\documentclass[11pt]{article}
\pdfoutput=1
\usepackage{jheppub}
\usepackage{tabu}
\usepackage[vcentermath]{youngtab}
\usepackage[usenames,dvipsnames,table]{xcolor}
\usepackage{graphicx,amsmath,amssymb,amsthm,multirow,array,bm,bbm,esint}
\usepackage[mathscr]{eucal}
\usepackage[bbgreekl]{mathbbol}
\usepackage{epsf,amsfonts}
\usepackage{slashed}
\usepackage[numbers,sort&compress]{natbib}
\usepackage{minibox}
\usepackage{rotating}
\usepackage{pdflscape}
\usepackage{array,tikz-cd}
\usepackage{subcaption}
\usepackage{framed}

\usepackage{slashed}

\newcommand{\R}{r}
\newcommand{\G}{g}
\newcommand{\B}{b}

\newcommand{\dipole}{\bold{D}}        

\newcommand{\col}{{\mathfrak{c}}}     

\newcommand{\effcoupling}{\mathcal{J}}      

\newcommand{\Jonsite}{{J_{\text{\tiny on-site}}}} 

\def \sgn{\mbox{sgn\,}}
\def \Tr{\mbox{Tr\,}}   
\def \tr{\mbox{tr\,}}   
\def \bitr{{\mathbb{tr}\,}}   
\def \TR{\mbox{TR\,}}   
\def \ch{\mbox{ch}}   
\def \fluca{\eta}   
\def \flucb{\zeta}   
\def \mfluca{\boldsymbol{\eta}}     
\def \mflucb{\boldsymbol{\zeta}}    

\newcommand{\sym}{ {\text{\tiny S}} } 
\newcommand{\anti}{ {\text{\tiny A}} }

\newcommand{\hq}{ {\mathbb{q}} }

\newcommand{\maPsi}{ {\boldsymbol{\Psi}} }  
\newcommand{\mzero}{ {\boldsymbol{\rho}} }  
\newcommand{\zero}{ {\rho} }        
\newcommand{\mT}{ {\boldsymbol{T}} }  

\newcommand{\eigenz}{ {\Upsilon} }

\newcommand{\mstar}{ {\circledast} }
\newcommand{\mg}{ {\boldsymbol{g}} }

\newcommand{\idm}{{\boldsymbol{I}}}     

\newcommand{\coeff}{{ \Lambda }}        

\newcommand{\mcurrent}{{\boldsymbol{\mathsf{J}} }}    
\newcommand{\current}{{\mathsf{J} }}    

\newcommand{\diffeo}{{ \text{diff} }}       
\newcommand{\tdiffeo}{{ \text{\tiny diff} }}       
\newcommand{\tlie}{{ \text{\tiny G} }}       

\newcommand{\ind}{ { \mathtt{x} } }     

\newcommand{\dP}{{ \cal P }}
\newcommand{\dK}{{ \cal K }}
\newcommand{\dD}{{ \cal D }}

\newcommand{\ie}{\emph{i.e.}, }
\newcommand{\eg}{\emph{e.g.}, }

\title{SYK Models and SYK-like Tensor Models with\\ Global Symmetry}

\author{Junggi Yoon}
\affiliation{International Centre for Theoretical Sciences (ICTS-TIFR), \\
Shivakote, Hesaraghatta Hobli, Bengaluru 560089, India.}
\emailAdd{junggi.yoon@icts.res.in}

\abstract{
In this paper, we study an SYK model and an SYK-like tensor model with global symmetry. First, we study the large $N$ expansion of the bi-local collective action for the SYK model with manifest global symmetry. We show that the global symmetry is enhanced to a local symmetry at strong coupling limit, and the corresponding symmetry algebra is the Kac-Moody algebra. The emergent local symmetry together with the emergent reparametrization is spontaneously and explicit broken. This leads to a low energy effective action. We evaluate four point functions, and obtain spectrum of our model. We derive the low energy effective action and analyze the chaotic behavior of the four point functions. We also consider the recent 3D gravity conjecture for our model.

We also introduce an SYK-like tensor model with global symmetry. We first study chaotic behavior of four point functions in various channels for the rank-3 case, and generalize this into a rank-$(q-1)$ tensor model. 
}

\begin{document}
\maketitle


\section{Introduction}
\label{sec: introduction}

The Sachdev-Ye-Kitaev~(SYK) model proposed by~\cite{Sachdev:1992fk} describes the quantum mechanics of $N$~Majorana fermions with disordered interactions. The SYK model features the dominance of the ``Melonic'' diagrams in large $N$ and is solvable at strong coupling limit~\cite{KitaevTalks,Polchinski:2016xgd,Jevicki:2016bwu,Maldacena:2016hyu,Jevicki:2016ito}. One of the most interesting aspects of the SYK model is the emergent reparametrization symmetry at strong coupling limit~\cite{KitaevTalks,Polchinski:2016xgd,Jevicki:2016bwu,Maldacena:2016hyu,Jevicki:2016ito}. This reparametrization symmetry is spontaneously and explicitly broken, like the chiral Lagrangian for the pion. This leads to a Pseudo-Nambu-Goldstone boson, the dynamics of which is governed by a Schwarzian low energy effective action. These low energy modes are also responsible for the saturation of the chaos bound~\cite{KitaevTalks,Maldacena:2016hyu} of the SYK model, and this suggests that the SYK model might have a holographic dual, and has resulted in great interest. In particular, it has been shown that the low energy sector of the SYK model is captured by the dilaton gravity~\cite{Maldacena:2016upp} and the Liouville theory~\cite{Mandal:2017thl}. Furthermore, the infinite tower of states in the SYK model can be understood as the Kaluza-Klein modes of the 3D gravity~\cite{Das:2017pif} .

In this context, it is useful to generalize the SYK model in a way that the essential properties of the model are preserved.\footnote{Also, a variety aspects of the SYK model have been explored: the quenched and annealed average of the complex SYK model~\cite{Gurau:2017xhf}, $i\epsilon$ prescription for the two point function~\cite{Gurau:2017qna} and $1/N$ corrections~\cite{Bonzom:2017pqs,Dartois:2017xoe}.} Indeed, there have been various directions such as lattice and higher dimensional generalization~\cite{Gu:2016oyy,Berkooz:2016cvq,Banerjee:2016ncu,Turiaci:2017zwd,Berkooz:2017efq,Jian:2017unn,Murugan:2017eto} and supersymmetry~\cite{Fu:2016vas,Li:2017hdt,Kanazawa:2017dpd,Murugan:2017eto,Yoon:2017gut,Peng:2017spg}. In this paper, we are interested in SYK models with global symmetries. There has already been some works along this direction. For example, the $U(1)$ symmetry in the complex SYK model~\cite{Sachdev:2015efa,Davison:2016ngz} as well as in the 2d SUSY generalization~\cite{Murugan:2017eto} have been explored. Also, $U(1)_R$ symmetry was studied in the $\mathcal{N}=2$ SUSY SYK model~\cite{Fu:2016vas}. Furthermore, a model with flavor symmetry was introduced by~\cite{Gross:2016kjj}.

The SYK model and its generalizations have a common feature: the bi-locals. This bi-local structure originates from the fact that those models are basically large $N$ vector models. The systematic analysis of the large $N$ vector models can be given by the collective field theory~\cite{Jevicki:1979mb,deMelloKoch:1996mj,Das:2003vw,Koch:2010cy,Koch:2014aqa} where the bi-local structure naturally arises. We emphasize that the bi-local collective field theory enables us to analyze the SYK models with a powerful action formulation~\cite{Jevicki:2016bwu,Jevicki:2016ito,Yoon:2017gut}. Even if we add extra bi-local structures such as replica space~\cite{Jevicki:2016bwu}, superspace~\cite{Yoon:2017gut} and doubled space in Thermofield Dynamics~\cite{Jevicki:2015sla,Jevicki:2015pza} in addition to the bi-local spacetime, the bi-local collective field theory straightforwardly incorporates them into an enlarged bi-local space. Hence, it is also natural to add flavor structure to the SYK model to construct a bi-local space of flavor and time. For this, we need to find a generalized SYK model which is invariant under the global symmetry explicitly because the bi-local collective field theory leads to a matrix model in the enlarged bi-local space. Then, the bi-local field theory of such a generalized SYK model is nothing but the same matrix model as in the SYK model with enlarged bi-local space.

Also, it has been shown that the same features of SYK model such as the maximal chaos is shared by ``SYK-like'' tensor models\footnote{Note that various tensor models and their $1/N$ expanstions have been studied in early papers~\cite{Gurau:2009tw,Gurau:2011aq,Gurau:2011xp,Bonzom:2012hw,Gurau:2011xq,Carrozza:2015adg,GurauSchaeffer}.} which is a quantum mechanical model of fermi tensors without disorder~\cite{Witten:2016iux,Gurau:2016lzk,Klebanov:2016xxf}. In addition, the finite $N$ tensor models have been investigated in~\cite{Krishnan:2016bvg,Krishnan:2017ztz,Chaudhuri:2017vrv,Krishnan:2017txw}, and the generalizations of the SYK-like tensor model in large $N$ have been explored in~\cite{Ferrari:2017ryl,Itoyama:2017emp,Itoyama:2017xid,Narayan:2017qtw,Klebanov:2017nlk,Mironov:2017aqv,Gurau:2017qya}. Now, like the SYK model, it is also natural to consider the generalize the SYK-like tensor model by introducing a global symmetry. However, in contrast to the SYK model, there are too many invariant operators in tensor models, which make it difficult to construct a collective action for the invariants.\footnote{Recently, the classification of the invariants in tensor models has been investigated in~\cite{Diaz:2017kub,robert:2017}, and a collective field theory for tensor models just began to be studied~\cite{robert:2017}.} Nevertheless, \cite{Narayan:2017qtw} developed a general technique which enables us to analyze a broad class of tensor models on the lattice. In particular, this analysis is not restrict to lattice space, but can be applied to abstract spaces such as the flavor index. Like the bi-local field theory, the analysis in~\cite{Narayan:2017qtw} immediately gives similar results for a tensor model with global symmetry.

We will introduce a new SYK model which exhibits global symmetry\footnote{Although we will work on a specific $G=SO(M)$ in the main text, it could be generalized to any Lie group. We analyze $U(M)$ case in Appendix~\ref{app: su n}.} $G$ already at the level of Hamiltonian before disorder averaging. After disorder average, this global symmetry is enhanced to the local symmetry at strong coupling limit of which symmetry algebra is the Kac-Moody algebra $\widehat{g}$ of the group $G$. Together with the reparametrization symmetry, the emergent 
$\diffeo \ltimes \widehat{g}$ symmetry is also broken spontaneously and explicitly, which is responsible for the structure of the low energy effective action: the Schwarzian and quantum mechanics of a particle on the $G$ group manifold. We study the large $N$ expansion of the corresponding bi-local collective action. Furthermore, we evaluate four point functions and study the long time behavior of the out-of-time-ordered correlators. We also introduce a tensor model with a global symmetry, and investigate four point functions. Moreover, we explore the conjecture on the dual 3D gravity of the SYK model proposed by~\cite{Das:2017pif}.

The outline of this paper is as follows. \textbf{In Section~\ref{sec: syk model with global symmetry}}, we introduce a real $O(N)$ SYK model with global $SO(M)$ symmetry, and construct the bi-local collective action of our model. In large $N$, we derive the saddle-point equation corresponding to the Schwinger-Dyson equation of the two point function of the fermions. We take $SO(M)$ invariant ansatz for the large $N$ classical solution, and the saddle-point equation is reduced to that of the original SYK model. We discuss the emergent reparametrization and local $SO(M)$ symmetry corresponding Kac-Moody algebra at strong coupling limit. We show that the full symmetry algebra is $\diffeo \ltimes \widehat{so}(M)$.

\textbf{In Section~\ref{sec: four point functions}}, we discuss four point functions. First, \textbf{in Section~\ref{sec: quadratic action}}, we study the systematic large $N$ expansion of the bi-local collective action. Expanding around the classical solution, we derive the quadratic action of the bi-local fluctuation. Since the fermion transforms fundamental representation of the $SO(M)$, the bi-local fluctuation is decomposed into singlet, anti-symmetric and symmetric irreducible representations of $SO(M)$. We discuss the properties of the singlet, anti-symmetric and symmetric irrep eigenfunction of the quadratic action with (Bessel function) basis. Using this basis, we diagonalize the quadratic action, and obtain the propagator of the singlet, anti-symmetric and symmetric bi-local fluctuations. These two point functions of the bi-local fluctuations correspond to four point functions of the fermions. Also, in this diagonalization, we obtain functions which determine the spectrum of our model.

\textbf{In Section~\ref{sec: kernel}}, following~\cite{Maldacena:2016hyu}, we analyze the four point functions. We derive the kernel of the four point function simply from the quadratic action of the bi-local fluctuations without diagrammatics. Then, we evaluate the four point functions with the conformal eigenfunctions~\cite{Maldacena:2016hyu,Murugan:2017eto,Peng:2017spg,Bulycheva:2017uqj}. The singlet channel is the same as the original SYK model, and the anti-symmetric and symmetric channels lead to new four point functions\footnote{ For the $U(1)$ case, the analogous results to the anti-symmetric channel has been found in~\cite{Bulycheva:2017uqj}.}. We work on the OPE limit of the four point functions.

\textbf{In Section~\ref{sec: diagrammatic}}, we confirm the diagrammatics of our model. In the previous sections, we have not used diagrammatics thanks to the collective action, and have analyzed our model without diagrammatics. In this section, we double check that the large $N$ diagrammatics also leads to the same Schwinger-Dyson equations for two and four point functions.

\textbf{In Section~\ref{sec: chaos and effective action}}, following~\cite{Jevicki:2016bwu,Jevicki:2016ito} we derive the effective action for the Pseudo-Nambu-Goldstone bosons related to broken reparametrization and $\widehat{so}(M)$. We obtain the Schwarzian action and the action for a particle on the group manifold. From the effective action, we evaluate the contribution of the low energy mode to the out-of-time-ordered correlators by using the method in~\cite{Maldacena:2016hyu}. The low energy mode of the Schwarzian makes the singlet channel maximally chaotic. On the other hand, the anti-symmetric channel gets a contribution only from the low energy mode related to $\widehat{so}(M)$ which does not grow exponentially. Furthermore, there is no contribution of the low energy effective actions to the symmetric channel. We also analyze the contribution from the non-zero modes. The calculation for the singlet channel, which is the same as the original SYK model, gives the ${1\over \beta\mathcal{J}}$ correction to the Lyapunov exponent in~\cite{Maldacena:2017axo}, and the anti-symmetric/symmetric channels do not have exponential growth. Furthermore, we also obtain the consistent result by using retarded kernel~\cite{kitaevfirsttalk,Maldacena:2016hyu}.

\textbf{In Section~\ref{sec: tensor model}}, we propose a $O(N)$ tensor model with global symmetry (in particular, we work on $SO(M)$ for simplicity) which is analogous to our SYK model. Following~\cite{Narayan:2017qtw}, we first analyze our tensor model with rank-3 tensor. \cite{Narayan:2017qtw} investigated tensor models on the lattice by using general techniques that can be applied to broad class of tensor models. We consider the $SO(M)$ index as the lattice index in~\cite{Narayan:2017qtw}, and the interactions of our model can be interpreted as the non-local hopping interactions on the lattice. Then, we can immediately generate all the results such as two and four point functions by using \cite{Narayan:2017qtw}. We consider two types of $O(N)$ invariant four point functions: ``Cooper'' channel and ``Pillow'' channels. The Cooper channel is reminiscent to the four point function of our SYK model with $SO(M)$ global symmetry. Hence, the singlet Cooper channel saturates~\cite{KitaevTalks,Maldacena:2016hyu} the chaos bound~\cite{Maldacena:2015waa}. The Pillow channels do not grow exponentially up to the leading order in $N$. We generalize our model into rank-$(q-1)$ tensor model with the global $SO(M)$ symmetry by ``uncoloring'' process, and obtain the same results. We show that for $q>6$, the subleading ladder diagram (in $N$) of the Pillow channels grow exponentially with maximal growth rate ${2\pi \over \beta}$, and we claim that this is consistent with the contribution from the Schwarzian effective action if it exists.

\textbf{In Section~\ref{sec: 3d gravity}}, following~\cite{Das:2017pif}, we extend the conjecture on the 3D gravity for the case of $q=4$ in order to understand phenomenologically the infinite tower of the spectrum in our SYK model. Considering the three 3D scalar fields on $AdS_2\times [-L,L]$, we repeat the analysis in~\cite{Das:2017pif} with the spectrum of the singlet, anti-symmetric and symmetric irrep sectors of $SO(M)$. As in~\cite{Das:2017pif}, we demand Dirichlet boundary condition for two scalar fields on both ends of the interval which correspond to the singlet and symmetric irrep sectors. On the other hand, we impose Neumann boundary condition for the other scalar field. As in~\cite{Das:2017pif}, the 3D propagators evaluated at a specific point on the extra space agree with the propagators in our model. We also consider a similar action with three scalar fields conformally coupled to gravity on $AdS_2\times[0,L]$ where we impose Dirichlet/Robin boundary condition for two scalar fields, and Neumann/Robin boundary condition for the other. And, we also have the same result.

\textbf{In Section~\ref{sec: conclusion}}, we present our conclusion and future directions. 

\textbf{In Appendix~\ref{app: schur collective action}}, we consider a general form of our SYK model with the global symmetry $SO(M)$ by using the permutation group~$S_{q\over 2}$. The corresponding collective action is a Schur polynomial of the product of two bi-local fields with a representation $R$ of the permutation group $S_{q\over 2}$. Up to quadratic level, all these actions gives the same results such as two and four point functions. \textbf{In Appendix~\ref{app: su n}}, we work on the (complex) SYK model with $U(M)$ global symmetry. For simplicity, we take the anti-symmetric classical solution (the particle-hole symmetric case). We have four bi-local fluctuations corresponding to symmetric/antisymmetric (or, equivalently real/imaginary) part of the singlet and adjoint fluctuations. \textbf{In Appendix~\ref{app: eigenfunctions}}, we give the details of the diagonalization of the quadratic action by using Bessel function basis. \textbf{In Appendix~\ref{app: zero mode}}, we explore the properties of zero mode eigenfunctions and their generalization.

\section{SYK Models with Global Symmetry}
\label{sec: syk model with global symmetry}

\subsection{Bi-local Collective Actions}
\label{sec: bi-local collective action}

We begin by introducing an $O(N)$ SYK model with $SO(M)$ global symmetry. The model consists of $N M $ Majorana fermions $\chi^{i \alpha}$ where $i\; (=1,\cdots, N)$ and $\alpha\;(=1,\cdots, M)$ are $O(N)$ and $SO(M)$ indices, respectively. Note that we will take large $N$ limit, but $M$ should be finite, or at most $M\ll N$. We define the following $SO(M)$ invariant action with a random coupling constant:
\begin{equation}
     S = \int d\tau \;\left[{1\over 2}\sum_{i=1}^N\sum_{\alpha=1}^M \chi^{i \alpha}  \partial_\tau \chi^{i \alpha} -  \sum_{i,j, k,l=1}^N\sum_{\alpha_1,\alpha_2=1}^M J_{ijkl}      \chi^{i \alpha_1} \chi^{j \alpha_1} \chi^{k \alpha_2} \chi^{l \alpha_2}    \right]\ .\label{def: syk action}
\end{equation}
One can consider more general action with the $SO(M)$ global symmetry, which we discussed in Appendix~\ref{app: schur collective action}. $J_{i_1 i_2 i_3 i_4}$ denotes the random coupling constant from Gaussian distribution:
\begin{equation}
\prod_{i,j,k,l=1}^N\exp\left[- {2M  N^3 \over  J^2 }  J_{ijkl}J_{ijkl}\right]\label{eq: gaussian ensemble} \ .
\end{equation}
Here, the parameter $J$ has dimension of energy. Note that we do not assume any symmetry property for $J_{ijkl}$ (\ie $J_{ijkl}$ need not to be anti-symmetric) so that $J_{ijkl}$'s are all independent random coupling constant.\footnote{I thank Prithvi Narayan for pointing out this.}

For the quenched disorder average, one has to follow the replica trick~\cite{Jevicki:2016bwu}. In large $N$, one can take replica symmetry ansatz assuming that there is no spin glass phase. This is equivalent to treat the random coupling constant $J_{ijkl}$ as additional non-dynamical field in large $N$ SYK model, and to integrate it out in the partition function.

For the bi-local collective action, we define the bi-local field $\Psi^{\alpha_1 \alpha_2}(\tau_1,\tau_2)$ by
\begin{equation}
\Psi^{\alpha_1  \alpha_2 }(\tau_1,\tau_2) \equiv {1 \over N} \sum_{i=1}^N \chi^{i \alpha_1}(\tau_1) \chi^{i \alpha_2}(\tau_2)\ .
\end{equation}
Note that the bi-local field $\Psi^{\alpha_1  \alpha_2 }(\tau_1,\tau_2)$ can be considered as an anti-symmetric matrix in the bi-local space $(\tau_1,\alpha_1;\tau_2,\alpha_2)$. \ie
\begin{equation}
    \Psi^{\alpha_1 \alpha_2}(\tau_1,\tau_2)=-\Psi^{\alpha_2 \alpha_1}(\tau_2,\tau_1)\ .\label{eq:anti-symmetry1} 
\end{equation}
Also, in this paper, it is convenient to represent it as the $M\times M$ matrix-valued bi-local field of the bi-local space $(\tau_1,\tau_2)$:
\begin{equation}
    \maPsi(\tau_1,\tau_2)\hspace{5mm}:\hspace{5mm} M\times M\;\;\mbox{matrix}\ .
\end{equation}
In this matrix notation, the anti-symmetry of bi-local field in \eqref{eq:anti-symmetry1} can be written as
\begin{equation}
    \maPsi(\tau_1,\tau_2)=-\maPsi^t(\tau_2,\tau_1)\label{eq:Symmetry_Prop}
\end{equation}
where $A^t$ is the transpose of the $M\times M$ matrix $A$. We will use the following notations through out this work: an $M \times M$ matrix will be denoted by bold letters $\maPsi, \boldsymbol{\eta}$ etc., and the matrix elements will be denoted by $\Psi^{\alpha_1\alpha_2},\eta^{\alpha_1\alpha_2}$ etc. 

Now, we will perform the Gaussian integral over $J_{ijkl}$ with ~\eqref{eq: gaussian ensemble} in the action \eqref{def: syk action}. Since the action~\eqref{def: syk action} and the Gaussian distribution~\eqref{eq: gaussian ensemble} are invariant under (global) $SO(M)$ and $SO(N)$, we expect the result to be invariant, too. 
%
%
In terms of the bi-local collective field, the path integral is given by
\begin{equation}
    Z=\int \mathcal{D}\maPsi \; \mu[\maPsi] \; e^{-S_{col}[\maPsi]}\label{eq: path integral for collective field}
\end{equation}
where $\mu[\maPsi]$ is a measure of order $\mathcal{O}(N^0)$ which is not important in this paper, and the collective action $S_{col}[\maPsi]$ is~\cite{Jevicki:2016bwu,Jevicki:2016ito,Yoon:2017gut}
\begin{align}
   & S_{col}[\maPsi]= {N\over 2} \Tr\left[-D\mstar \maPsi +\log \maPsi \right]  - {NJ^2\over 8M} \int d\tau_1d\tau_2 \; \left[\tr\left(- \maPsi(\tau_1,\tau_2)\maPsi(\tau_2,\tau_1) \right)\right]^2\ .\label{eq: collective action for q=4}  
\end{align}
We have introduced some compact notations here which are standard in the bi-local theory literature that requires some explanations. First, note that $\tr$ denotes the trace of the $M\times M$ matrix, and runs over the $SO(M)$ index $\alpha=1,2,\cdots,M$. \ie $\tr A= \sum_{\alpha=1}^M A^{\alpha\alpha}$. Also, note that $\Tr$ denotes the trace of a matrix in the bi-local space $(\tau_1,\alpha_1;\tau_2,\alpha_2)$ (\ie $\Tr A= \int d\tau \sum_{\alpha=1}^M A(\tau,\alpha;\tau,\alpha)$). Also, the bi-local derivative denoted by $D$ is defined as
\begin{equation}
    D^{\alpha_1 \alpha_2}(\tau_1,\tau_2)\equiv \delta^{\alpha_1 \alpha_2}\partial_{\tau_1}\delta(\tau_1-\tau_2)\ ,
\end{equation}
and, $\mstar$ denotes the star product in the bi-local space $(\tau_1,\alpha_1;\tau_2,\alpha_2)$. \ie
\begin{equation}
(A\mstar B)^{\alpha_1 \alpha_2}(\tau_1;\tau_2)\equiv \int d\tau_3 \sum_{\alpha_3=1}^M A^{\alpha_1 \alpha_3}(\tau_1;\tau_3)B^{\alpha_3 \alpha_2}(\tau_3;\tau_2)\ .
\end{equation}
In addition, the second term ${N\over 2}\Tr\log \Psi$ in~\eqref{eq: collective action for q=4} comes from the non-trivial Jacobian in Hubbard-Stratonovich type transformation to the bi-local field.\footnote{Strictly speaking, this should be ${N-1\over 2}\Tr\log \Psi$, and the shift in large $N$ by $-1$ plays an important role in exact calculations~\cite{Jevicki:2014mfa,Yoon:2017gut}. However, we absorbed the shift into the measure $\mu[\Psi]$ in~\eqref{eq: path integral for collective field} because this correction does not have any effect on our calculations.}

\paragraph{Generalization to $q$ point interaction:}

One can generalize the model into one with a $q$-point interaction:
 \begin{equation}
     S = \int d\tau \;\left[ {1\over 2} \chi^{i \alpha}\partial_\tau \chi^{i \alpha} +  i^{q\over 2}   J_{i_1 \dots i_{q}} \chi^{i_1 \alpha_1} \chi^{i_2 \alpha_1}\chi^{i_3 \alpha_2} \chi^{i_4 \alpha_2}  \dots  \chi^{i_{q-1} \alpha_\hq}  \chi^{i_{q} \alpha_\hq}    \right]\label{def: syk model action general}
\end{equation}
where we defined
\begin{equation}
    q\equiv 2\hq\ ,
\end{equation}
and, we have omitted the summation over the $O(N)$ indices $i_1,\cdots, i_q$ and $SO(M)$ indices $\alpha_1,\cdots, \alpha_\hq$. The distribution of Gaussian random coupling constant $J_{i_1 \cdots i_q}$ is given by
\begin{equation}
    \exp\left[-{\hq M^{{q\over 2}-1}N^{q-1} \over J^2 }\sum_{i_1,\cdots, i_q=1}^N J_{i_1\cdots i_q} J_{i_1\cdots i_q}\right]\ .\label{eq: gaussian distribution 2}
\end{equation}
After disorder averaging, we have the following bi-local collective action.
\begin{equation}
   S_{col}={N\over 2}\Tr\left[-D\mstar \maPsi +\log \maPsi \right]  - {NJ^2\over 4\hq M^{\hq-1}} \int d\tau_1d\tau_2 \; \left[\tr\left(- \maPsi(\tau_1,\tau_2)\maPsi(\tau_2,\tau_1) \right)\right]^{\hq } \ . \label{eq: collective action general q}  
\end{equation}

We have chosen the simplest action~\eqref{def: syk action} instead of the general form of $SO(M)$ invariant action. In Appendix~\ref{app: schur collective action}, we take more general $SO(M)$ invariant actions and derive the corresponding collective actions. We show that the basic building block of the collective action is Schur polynomial of the $M\times M$ matrix $- \maPsi(\tau_1,\tau_2)\maPsi(\tau_2,\tau_1)$ with Young tableau $R$ ($|R|=\hq$), and the (potential term of) collective action is the linear combination of them with positive coefficients. For example, \eqref{eq: collective action general q} corresponds to the sum of all Schur polynomials of degree $\hq$. We also find in Appendix~\ref{app: schur collective action} that each Schur polynomial as a collective action gives the identical result up to quadratic level.

Furthermore, although we obtained the collective action for $G = SO(M)$ group here, the reader must keep in mind that most of the steps could go through, and one would be able to obtain a similar collective action for any semi-simple Lie group $G$. This is worked out for the case of $G = U(M)$ in Appendix~\ref{app: su n}.

\subsection{Large $N$ Classical Solution}
\label{sec: classical solution}

In this section, we will evaluate the large $N$ classical solution $\maPsi_{cl}$, which is the two point function of the fundamental fermions as well as the one point function of the bi-local field:
\begin{equation}
    \Psi^{\alpha_1 \alpha_2}_{cl}(\tau_1,\tau_2)=\left\langle \;\Psi^{\alpha_1 \alpha_2}_{cl}(\tau_1,\tau_2)\;\right\rangle ={1\over N}\sum_{i=1}^N \left\langle\; \chi^{i \alpha_1}(\tau_1)\chi^{i \alpha_2}(\tau_2)\;\right\rangle \ .
\end{equation}
To evaluate the large $N$ classical solution, we vary the collective action~\eqref{eq: collective action general q} with respect to the bi-local field, and (matrix-)multiply $\maPsi_{cl}$ to the result in order to obtain the large $N$ saddle-point equation:
\begin{align}
    &-(D\mstar \maPsi_{cl})(\tau_1,\tau_2) + \idm \delta(\tau_1-\tau_2) \cr
    & +  {J^2\over M^{\hq-1}}\int d\tau_3 \; \left[-\tr (  \maPsi_{cl}(\tau_1,\tau_3)\maPsi_{cl}(\tau_3,\tau_1))\right]^{\hq -1} \maPsi_{cl}(\tau_1,\tau_3) \maPsi_{cl}(\tau_3,\tau_2)=0\label{eq: saddle point equation}
\end{align}
where $\idm $ is the $M\times M$ identity matrix so that $\idm \delta(\tau_1-\tau_2)$ is the identity matrix in the bi-local space. 

At the strong coupling limit where $|J \tau|\longrightarrow \infty $, one can drop the first term in \eqref{eq: saddle point equation}, and can solve it exactly. Recall that the collective action has global $SO(M)$ symmetry. Hence, we will take the following $SO(M)$ invariant ansatz for the classical solution:
\begin{equation}
    \maPsi_{cl}(\tau_1,\tau_2)=\idm\; \psi_{cl}(\tau_1,\tau_2)\ .
\end{equation}
Note that $\psi_{cl}(\tau_1,\tau_2)$ is not a $M\times M$ matrix but a anti-symmetric bi-local field in the bi-local time space $(\tau_1,\tau_2)$, \ie $\psi_{cl}(\tau_1,\tau_2)=-\psi_{cl}(\tau_2,\tau_1)$ because of \eqref{eq:anti-symmetry1}. Then, the saddle-point equation in~\eqref{eq: saddle point equation} is reduced to the large $N$ saddle-point equation of the original SYK models at the strong coupling limit:
\begin{equation}
    J^2 [\psi_{cl}]^{q-1}(\tau_1,\tau_3) \star \psi_{cl}(\tau_3,\tau_2)=-\delta(\tau_1-\tau_2)
\end{equation}
where $\star$ denotes the star product in bi-local time space $(\tau_1,\tau_2)$. \ie 
\begin{equation}
(A\star B)(\tau_1,\tau_2)\equiv \int d\tau_3\; A(\tau_1,\tau_3)B(\tau_3,\tau_2)\ .
\end{equation}
The solution is~\cite{Sachdev:2015efa,KitaevTalks,Polchinski:2016xgd,Jevicki:2016bwu,Maldacena:2016hyu,Jevicki:2016ito}
\begin{equation}
   \psi_{cl}(\tau_1,\tau_2)= \coeff{\mbox{sgn}(\tau_1-\tau_2) \over |\tau_1 - \tau_2|^{2\Delta}  }\hspace{10mm} (\Delta\equiv{1\over q})\label{eq: classical solution}
\end{equation}
where the constant $\coeff$ is given by
\begin{equation}
    J^2 \coeff^q \pi =\left({1\over 2}-{1\over q}\right) \tan {\pi \over q}\ .\label{def: coefficient}
\end{equation}
%

\subsection{Reparametrization and Kac-Moody Algebra}
\label{sec: algebra}

In this subsection, we will explore the symmetries in the low energy limit. Recall that the SYK model features the emergent reparametrization symmetry at strong coupling limit. Apart from the reparametrization symmetry, our model has a global $SO(M)$ symmetry, which can already be seen at the level of action in~\eqref{def: syk action}. We will show that this $SO(M)$ global symmetry gets enhanced to a local $SO(M)$ symmetry at the strong coupling limit. Also, we will show that the symmetry algebra is the (chiral half) of an affine Kac-Moody algebra with symmetry group $G$ ($SO(M)$ for this case).

First, let us recall how the  reparametrization symmetry emerges in our case. To see this, we drop the kinetic term in the collective action~\eqref{eq: collective action general q} at the strong coupling limit, and define a critical collective action $S_{col,c}$:
\begin{equation}
    S_{col,c}={N\over 2}\Tr\left[\log \maPsi \right]  - {NJ^2\over 4\hq M^{\hq-1}} \int d\tau_1d\tau_2 \; \left[\tr\left(- \maPsi(\tau_1,\tau_2)\maPsi(\tau_2,\tau_1) \right)\right]^{\hq } \ . \label{eq: critical collective action general q}  
\end{equation}
The critical collective action is invariant under reparametrization given by
\begin{equation}\label{eq:Reparamwetrization Invariance}
    \maPsi(\tau_1,\tau_2)\quad\longrightarrow \quad \left(f'(\tau_1)\right)^{\Delta}\maPsi(f(\tau_1),f(\tau_2))\left(f'(\tau_2)\right)^{\Delta}\hspace{10mm}(\Delta={1\over q})\ .
\end{equation}

It is now easy to see that non-abelian local symmetry emerges at strong coupling in addition to the reparametrization symmetry. Recall that the collective action~\eqref{eq: collective action general q} as well as the original action in~\eqref{def: syk model action general} has non-abelian global symmetry $SO(M)$. At the strong coupling limit, this global symmetry is enhanced to local symmetry:
\begin{equation}
    \maPsi(\tau_1,\tau_2)\quad\longrightarrow \quad \mg(\tau_1) \maPsi(\tau_1,\tau_2) \mg^{-1}(\tau_2)
\end{equation}
where $\mg(\tau)$ is a $SO(M)$ matrix corresponding to the local $SO(M)$ transformation. Note that the star product $\mstar$ in the bi-local space $(\tau_1,a_1;\tau_2,a_2)$ is manifestly invariant under the local $SO(M)$ transformation, and therefore, the first term of the critical action~\eqref{eq: critical collective action general q} is invariant. In addition, since the matrix $\maPsi(\tau_1,\tau_2)$ is always multiplied to $\maPsi(\tau_2,\tau_1)$ in the second term of the critical action, the second term is also invariant.

Furthermore, we find that the emergent symmetry of our model is the semi-direct product of the reparametrization and the affine Kac-Moody algebra:
\begin{equation}
\diffeo \ltimes \widehat{so}(M)   \ .
\end{equation}

We denote the group element of $\diffeo \ltimes \widehat{so}(M)$ by $[ f(\tau),\mg(\tau) ]$ where $f(\tau)$ and $\mg(\tau)$ corresponds to the reparametrization and local $\widehat{so}(M)$ transformation, respectively. \ie
\begin{equation}
    \maPsi(\tau_1,\tau_2)\quad\xrightarrow{\;\;[f,\mg]\;\;} \quad \left(f'(\tau_1)\right)^{1\over q} \mg(\tau_1) \maPsi(f(\tau_1),f(\tau_2)) \mg^{-1}(\tau_2)\left(f'(\tau_2)\right)^{1\over q}\ .\label{eq: local transformation}
\end{equation}
Then, the group composition law (the left group action) is found to be
\begin{equation}
    [ f_2(\tau), \mg_2(\tau) ] \boldsymbol{\cdot} [ f_1(\tau), \mg_1(\tau) ] = [ (f_1 \circ f_2)(\tau), \mg_2(\tau)\mg_1(f_2(\tau)) ]\ .\label{eq:group composition law}
\end{equation}
Here, $\boldsymbol{\cdot}$ denotes the group multiplication of $\diffeo \ltimes \widehat{so}(M)$, and $(f_1 \circ f_2)(\tau)=f_1(f_2(\tau))$ corresponds to two successive reparametrizations. We now work with infinitesimal symmetry transformations:
\begin{equation}
    f(\tau) = \tau + \epsilon(\tau) \hspace{5mm},\hspace{5mm} \mg(\tau) = \idm + i\mzero(\tau) \ .
\end{equation}
From \eqref{eq:group composition law}, the successive two infinitesimal transformation\footnote{I thank Prithvi Narayan for extensive discussion and collaboration on the algebra in this subsection.} of $(\tau,\idm)$ is given by
\begin{align}
&[ \tau,\idm ]\quad \xrightarrow{[f_2,\mg_2]\cdot [f_1,\mg_1]}  \cr
&[ \tau + \epsilon_1(\tau)+ \epsilon_2(\tau) + \epsilon_1'(\tau) \epsilon_2(\tau) , \idm + i\mzero_1(\tau)+ i\mzero_2(\tau) + i\epsilon_2(\tau) \mzero_1'(\tau) - \mzero_2(\tau) \mzero_1 (\tau) ]
\end{align}
Hence, denoting these infinitesimal transformations by $\delta_{\epsilon,\mzero}$, we have
\begin{equation}
    \delta_{\epsilon_2,\mzero_2} \delta_{\epsilon_1,\mzero_1}  - \delta_{\epsilon_1,\mzero_1} \delta_{\epsilon_2,\mzero_2}   = \delta_{ \epsilon_1'(\tau) \epsilon_2(\tau) -  \epsilon_2'(\tau) \epsilon_1(\tau) , \epsilon_2(\tau) \mzero_1'(\tau) - \epsilon_1(\tau) \mzero_2'(\tau) - [\mzero_2(\tau) , \mzero_1(\tau)]  }\ .
\end{equation}
Let us define the modes of the infinitesimal transformation by  
\begin{equation}
    \epsilon(\tau) \equiv \sum_{n\in \mathbb{Z}} \epsilon_n \tau^{n+1}\quad,\quad \mzero(\tau)  = \sum_{n\in \mathbb{Z}}\sum_a\zero_n^a  \mT^{a} \tau^n\ .
\end{equation}
where $a$ indices run over the adjoint(anti-symmetric irrep) of $\widehat{so}(M)$, and $\mT^a_\anti$ is the generators of $\widehat{so}(M)$ satisfying the commutation relation $[\mT^a_\anti, \mT^b_\anti ] = i f^{abc} \mT^c_\anti$. Defining the generators $\delta_{\epsilon,\mzero} = -\epsilon_n L_n + -i\zero_n^a J^a_n$, we finally get 
\begin{equation}
    \begin{split}
        [ L_n , L_m ] & = (n-m) L_{m+n} \ ,\\
        [ J^a_n , J^b_m ] &= i f^{abc} J^c_{m+n} \ , \\
        [ L_n , J_m ] &= -m J_{m+n}\ .
    \end{split}\label{eq: diffeo and akm}
\end{equation}
As an aside, one can easily realize the algebra by ~\eqref{eq: diffeo and akm} by the following representation of $L_n$ and $J_n^a$:
\begin{equation}
    L_n = -\tau^{n+1} \partial_\tau \hspace{10mm} J^a_n = \mT^a_\anti \tau^n \ .
\end{equation}
Note that the algebra~\eqref{eq: diffeo and akm} is nothing but the semi-direct product of Virasoro and Kac-Moody algebra without the central extensions. \ie $Vir \ltimes \widehat{so}(M)$.
 Since the central extensions are anomaly terms, we can obtain it only after a careful analysis of the theory at finite temperature and with non-trivial background charge configurations.

We mention here that the global properties of the group might depend on the asymptotic fall-off that are allowed. In particular one must distinguish finite temperature $\diffeo(S^1)$ from that at zero temperature $\diffeo(\mathbb{R})$ although the algebra \eqref{eq: diffeo and akm} is identical. Recall that the classical solution is given by
\begin{equation}
    \Psi_{cl}(\tau_1,\tau_2) = \Lambda{\sgn(\tau_1-\tau_2) \over |\tau_1 - \tau_2|^{\Delta} } \mathbb{1}\ .
\end{equation}
Using the reparametrization symmetry with $f(\tau)=\tan{\pi \tau\over \beta}$, one can also obtain the classical solution at the finite temperature:
\begin{equation}
    \Psi_{cl}(\tau) = \Lambda\left({ \pi \over \beta \sin{\pi \tau\over \beta} }\right)^{2\Delta}\sgn(\tau)\idm=\psi_{cl}(\tau)\idm\ .\label{eq: finite temperature classical solution}
\end{equation}
One can easily see that $\diffeo \ltimes \widehat{so}(M)$ is spontaneously broken by the classical solution. In particular, the classical solution at finite temperature breaks $\diffeo(S^1)\ltimes \widehat{so}(M)$. But, it is still invariant under the global subalgebra $SL(2,\mathbb{R})\times so(M)$ thereof. Furthermore, At finite coupling, $\diffeo(S^1) \ltimes \widehat{so}(M)$ is explicitly broken by the kinetic term in the collective action~\eqref{eq: collective action general q} which has been ignored at the strong coupling limit. This leads to Pseudo-Nambu-Goldstone bosons for the broken $\diffeo(S^1)\ltimes \widehat{so}(M)$, and their effective action is invariant under (global) $SL(2,\mathbb{R})\times so(M)$ symmetry. We will discuss the effective action in Section~\ref{sec: effective action}. In addition, this effective action for the Goldstone bosons of $\diffeo(S^1) \ltimes \widehat{so}(M)$ with $SL(2,\mathbb{R})\times so(M)$ symmetry can be also captured by coadjoint orbit of $Vir \ltimes \widehat{so}(M)$ which is central extension of $\diffeo(S^1) \ltimes \widehat{so}(M)$.

Note that our model is not restricted to $SO(M)$ but could be generalized any semi-simple Lie group $G$. In particular, we explicitly consider the case of $G = U(M)$ in Appendix~\ref{app: su n}, and it would be straightforward to generalize the above discussion for any Lie groups.

\section{Four point Functions}
\label{sec: four point functions}

In this section, we will diagonalize the quadratic collective action in large $N$ expansion. In the quadratic action, we will read off the kernel, and evaluate the four point function. From the OPE limit of the four point function, we will obtain the spectrum of our model.

\subsection{Diagonalization of Quadratic Action in Large $N$}
\label{sec: quadratic action}

In large $N$, we will expand the collective action in~\eqref{eq: collective action general q} around the classical solution $\maPsi_{cl}$ in~\eqref{eq: classical solution}. For this, we expand the bi-local collective field around the classical solution. 

Since the fundamental fermion $\chi^{i \alpha}$ transforms in the fundamental representation of $SO(M)$, the bi-local field $\maPsi$ is decomposed into singlet, anti-symmetric and symmetric irreducible representations of $SO(M)$:
\begin{equation}
     \yng(1)\; \otimes \; \yng(1) = (\text{singlet}) \; \oplus \; (\text{anti-sym}) \; \oplus \; (\text{sym})\ .\label{eq: decomposition of bi-fundamental}
\end{equation}
Hence, we have
\begin{equation}
 \maPsi(\tau_1,\tau_2) = \idm \psi_{cl}(\tau_1,\tau_2) +  \sqrt{{2\over N}} \mfluca(\tau_1,\tau_2) +\sqrt{{2\over N}}\mflucb_\anti (\tau_1,\tau_2)+\sqrt{{2\over N}}\mflucb_\sym(\tau_1,\tau_2)\label{eq: large n expansion of classical solution}
\end{equation}
where the singlet, the symmetric and the antisymmetric fluctuations can be written as
\begin{equation}
\begin{split}
\mfluca(\tau_1,\tau_2) = &{1\over \sqrt{M} }\fluca(\tau_1,\tau_2) \; \idm\ ,\cr
\mflucb_\anti(\tau_1,\tau_2)=&{\flucb^{a}_\anti(\tau_1,\tau_2)\over\sqrt{2\ind_\anti}} \mT^a_\anti\hspace{5mm} (a=1,\cdots, {1\over 2}M(M-1))\ ,\cr
\mflucb_\sym(\tau_1,\tau_2)=&{\flucb^{a}_\sym(\tau_1,\tau_2)\over\sqrt{2\ind_\sym}} \mT^a_\sym\hspace{5mm} (a=1,\cdots,{1\over 2}M(M+1)-1) \ .
\end{split}
\end{equation}
Here, $\ind_{\text{\tiny R}}$ denotes the Dynkin index of the representation $R$ of the Lie algebra. \ie
\begin{equation}
	\ind_R={\dim (R) C_2(R)\over 2\dim(\text{adj})}
\end{equation}
where $C_2(R)$ is the Casimir of the representation $R$. We will use conventions that repeated indices are summed unless otherwise specified. Here, we choose the basis for $so(M)$ generators $\mT^a_\anti$ and $\mT^a_\sym$ satisfying
\begin{equation}
    \tr(T^a_\anti T^b_\anti)= 2\ind_\anti \delta^{ab}\hspace{5mm},\hspace{5mm} \tr(T^a_\sym T^b_\sym)= 2\ind_\sym \delta^{ab}\ .
\end{equation}
 Hence, the component bi-local fields $\fluca(\tau_1,\tau_2)$, $\flucb^a_\anti(\tau_1,\tau_2)$ and  $\flucb^a_\sym(\tau_1,\tau_2)$ can be expressed as
\begin{equation}
\fluca(\tau_1,\tau_2) ={1\over \sqrt{M}}\tr (\mfluca(\tau_1,\tau_2)  ) \hspace{5mm},\hspace{5mm}\flucb^a_{\anti/\sym}(\tau_1,\tau_2)  = {1\over \sqrt{2\ind_{\anti/\sym}}}\tr(\mflucb(\tau_1,\tau_2)  \mT^a_{\anti/\sym}) \ .
\end{equation}
One can easily see that the singlet fluctuation $\fluca(\tau_1,\tau_2)$ is anti-symmetric in the bi-local time space like the original SYK model. \ie
\begin{equation}
    \fluca(\tau_1,\tau_2)=-\fluca(\tau_2,\tau_1)\ .\label{eq: symmetry of fluctuation1}
\end{equation}
In addition, because of the anti-symmetry/symmetry of the generators in the antisymmetric/symmetric representation
\begin{equation}
    (\mT^a_\anti)^t=-\mT^a_\anti \quad, \quad (\mT^a_\sym)^t=\mT^a_\sym \ ,
\end{equation}
one can show that the anti-symmetric/symmetric irrep fluctuation $\flucb^a_\anti(\tau_1,\tau_2)$/$\flucb^a_\sym(\tau_1,\tau_2)$ is symmetric/anti-symmetric in the bi-local time space. respectively:
\begin{equation}
    \flucb^a_\anti(\tau_1,\tau_2)=
    \flucb^a_\anti(\tau_2,\tau_1)\quad,\quad \flucb^a_\sym(\tau_1,\tau_2)=-\flucb^a_\sym(\tau_2,\tau_1)\ .\label{eq: antisymmetry of fluctuation2}
\end{equation}
Now, we expand the collective action~\eqref{eq: collective action general q} around the classical solution in large $N$:
\begin{equation}
    S_{col}\left[\maPsi_{cl}+  \sqrt{{2/N}}( \mfluca + \mflucb_\anti + \mflucb_\sym) \right]= N S^{(0)}+ S^{(2)}[\mfluca,\mflucb_\anti,\mflucb_\sym]+ {1\over \sqrt{N}} S^{(3)}[\mfluca,\mflucb_\anti,\mflucb_\sym]+\cdots\ ,
\end{equation}
Recalling the anti-symmetry of $\fluca$ in~\eqref{eq: symmetry of fluctuation1} and $\flucb^a_\sym$ in~\eqref{eq: antisymmetry of fluctuation2} as well as the symmetry of $\flucb^a_\anti$ in~\eqref{eq: antisymmetry of fluctuation2}, one can read off the quadratic action $S_{col}^{(2)}$:
\begin{align}
    S_{col}^{(2)}
    =&-{1\over 2}\bitr (\psi_{cl}^{-1}\star\fluca \star\psi_{cl}^{-1} \star \fluca )-{q -1\over 2}J^2\int d\tau_1 d\tau_2 [\psi_{cl}(\tau_1,\tau_2)]^{q-2}\fluca(\tau_1,\tau_2)\fluca(\tau_1,\tau_2)\cr
    &-{1\over 2}\bitr (\psi_{cl}^{-1}\star\flucb^a_\anti \star\psi_{cl}^{-1} \star \flucb^a_\anti )+{1\over 2} J^2\int d\tau_1 d\tau_2 \; [\psi_{cl}(\tau_1,\tau_2)]^{q-2}\flucb^a_\anti(\tau_1,\tau_2)\flucb^a_\anti(\tau_1,\tau_2)\cr
    &-{1\over 2}\bitr (\psi_{cl}^{-1}\star\flucb^a_\sym \star\psi_{cl}^{-1} \star \flucb^a_\sym )-{1\over 2} J^2\int d\tau_1 d\tau_2 \; [\psi_{cl}(\tau_1,\tau_2)]^{q-2}\flucb^a_\sym(\tau_1,\tau_2)\flucb^a_\sym(\tau_1,\tau_2)\label{eq: quadratic action}
\end{align}
where $\bitr$ denotes the trace over the bi-local time space. \ie $\bitr (A)= \int d\tau A(\tau,\tau)$. Before we diagonalize the quadratic action, it is convenient to introduce new coordinates\footnote{The generalization of this transformation in the higher dimension, so-called ``bi-local map'', has been found in~\cite{Koch:2010cy,Koch:2014aqa,Koch:2014mxa,Jevicki:2015pza} in order to match the bi-local conformal generators with higher spin generators in the context of the higher spin AdS/CFT correspondence.} $(t,z)$:
\begin{equation}
    t\equiv {1\over 2}(\tau_1+\tau_2)\quad,\quad z\equiv {1\over 2}(\tau_1-\tau_2)\ .\label{eq: bi-local map}
\end{equation}
These coordinates play a role of Poincare coordinates of $AdS_2$. For example, in these coordinates, the propagators of the bi-local fluctuations, which correspond to four point functions of the fundamental fermions, can be expressed as a sum of correlation functions of scalar fields in AdS$_2$ Poincare coordinates~\cite{Jevicki:2016bwu,Das:2017pif}.

The quadratic action \eqref{eq: quadratic action} will be diagonalized by expanding the fluctuations $\fluca,\flucb_\anti$ and $\flucb_\sym$ in terms of suitable eigenfunctions~\cite{Polchinski:2016xgd,Jevicki:2016bwu,Jevicki:2016ito,Yoon:2017gut}. The eigenfunctions $u^-_{\nu w}$ (and $u^+_{\nu w}$), which correspond to singlet fluctuation $\fluca$ and symmetric irrep fluctuation $\flucb^a_\sym$ (and, anti-symmetric irrep fluctuation $\flucb^a_\anti$, respectively), are given by (for more details, refer to Appendix~\ref{app: eigenfunctions})
\begin{alignat}{2}
    &u^-_{\nu w}(t,z)= {1\over \sqrt{8\pi}} e^{-i w t} \sgn(z) |z|^{{1\over 2}-{2\over q}} Z^-_{\nu}(|wz|)\hspace{2mm} &&,\hspace{3mm} \nu=\begin{cases}
    \; 2n+{3\over 2} \quad &\; (n=0,1,\cdots)\\
    \;ir &\; (r\geqq 0)\\
    \end{cases}\\
    &u^{ +}_{\nu w}(t,z)= {1\over \sqrt{8\pi}} e^{-i w t} |z|^{{1\over 2}-{2\over q}}  Z^+_{\nu}(|wz|)\hspace{2mm} &&,\hspace{3mm} \nu=\begin{cases}
    \; 2n+{1\over 2} \quad &\; (n=0,1,\cdots)\\
    \;ir &\; (r\geqq 0)\\
    \end{cases}
\end{alignat}
where the function $Z_\nu^\mp(x)$ is defined by 
\begin{equation}
    Z_\nu^\mp(x)\equiv J_\nu(x)+\xi_{\pm\nu} J_{-\nu}(x)
\end{equation}
with
\begin{equation}
    \xi_\nu= {\tan{\pi \nu \over 2}+1\over \tan{\pi \nu \over 2}-1 } \ .
\end{equation}
We summarize some useful properties of the eigenfunctions below.
\begin{itemize}
    \item \textbf{Symmetry:} The symmetry properties of $u^-_{\nu w}(t,z)$ and $u^+_{\nu w}(t,z)$ under a transformation $z\rightarrow  -z$ is inherited from \eqref{eq: symmetry of fluctuation1} and \eqref{eq: antisymmetry of fluctuation2} via \eqref{eq: bi-local map}, respectively. \ie
%
\begin{equation}
    u^-_{\nu w}(t,-z)=-u^-_{\nu w}(t,z)\quad,\quad u^{ +}_{\nu w}(t,-z)=u^{ +}_{\nu w}(t,z)\ .
\end{equation}
\item \textbf{Asymptotics: }%
As $z\longrightarrow \infty$, the eigenfunctions asymptote to
\begin{equation}
    u^-_{\nu w}(t,z) \sim  e^{-i w t}  z^{-{2\over q}}  \cos z\quad,\quad u^{+}_{\nu w}(t,z) \sim  e^{-i w t} z^{-{2\over q}}\sin z \ .
\end{equation}
Furthermore, for real discrete value $\nu$'s, the eigenfunctions are regular as $z\longrightarrow 0$ because $\xi_\nu$ vanish for each discrete $\nu$'s. \ie
\begin{alignat}{2}
    &\xi_{2n+{3\over 2}}=0\qquad&&\Longrightarrow \qquad Z^-_{2n+{3\over 2}}(z)=J_{2n+{3\over 2}}(z)\ ,\\
    &\xi_{-\left(2n+{1\over 2}\right)}=0\qquad&&\Longrightarrow \qquad  Z^+_{2n+{1\over 2}}(z)=J_{2n+{1\over 2}}(z)\ .
\end{alignat}
\item \textbf{Complex Conjugate: }Note that the complex conjugate of $Z_\nu^\mp$ is
\begin{equation}
    \overline{Z_\nu^\mp }=\xi_{\mp \nu} Z_\nu^\mp (z)\ .\label{eq: conjugate of z}
\end{equation}
\item \textbf{Orthonormality: }The orthonormality of the eigenfunctions is given by 
\begin{equation}
   \int_0^\infty  { dz \over |z|}  \overline{Z^\alpha_\nu(|wz|)} Z^{\alpha}_{\nu'}(|w'z|)=\delta_{\alpha \alpha'} N_\nu \delta(\nu-\nu')\hspace{10mm} (\alpha=\mp)
\end{equation}
where
\begin{equation}
    N_\nu=\begin{cases}
    \quad {1\over 2\nu} & \quad \left(\nu={3\over 2}+2n\;\;\mbox{for $Z^-$, or}\;\;\nu={1\over 2}+2n \;\;\mbox{for $Z^+$}\quad(n=0,1,\cdots)\right)\\
    \quad {2 \sin \pi \nu\over \nu} & \quad \left(\mbox{for}\quad \nu=ir \quad(r\in \mathbb{R})\right)
    \end{cases}
\end{equation}
In particular, for the diagonalization of $\nu=ir$ modes, it is useful to use the following integration.
\begin{align}
    \int_0^\infty  { dz \over |z|}  Z^\mp_{ir}(|wz|) Z^\mp_{ir'}(|w'z|)=\widetilde{N}^\mp_{ir} \delta(r-r')\hspace{1cm} ( \widetilde{N}^\mp_{ir} \equiv \xi_{\pm ir } N_{ir})
\end{align}
\begin{equation}
    \widetilde{N}^\mp_\nu = \begin{cases}
    \; {1\over 2\nu } &\hspace{3mm} \left(\nu={3\over 2}+2n\;\;\mbox{for $Z^-$, or}\;\;\nu={1\over 2}+2n \;\;\mbox{for $Z^+$}\;(n=0,1,\cdots)\right) \\
    \; \xi_{\pm \nu }{2 \sin \pi \nu \over \nu } & \hspace{3mm} \left(\mbox{for}\quad \nu=ir \quad(r\in \mathbb{R})\right) \\
    \end{cases}
\end{equation}

\end{itemize}

Finally, as we will see later, we emphasize that the lowest discrete mode of each eigenfunction, $u^-_{{3\over 2}, w}$ and $u^{ +}_{{1\over 2}, w}$, corresponds to the zero mode of the quadratic action related to the broken reparametrization and local $SO(M)$ symmetry. Moreover, since the large $N$ classical solution is proportional to the $M\times M$ identity matrix, we will see that the lowest singlet fluctuation $u^-_{{3\over 2}, w}$ and the anti-symmetric fluctuations $u^{\anti +}_{{1\over 2}, w}$ correspond to the zero mode for reparametrization and local $so(M)$, respectively (See Section~\ref{sec: effective action}).

\paragraph{Diagonalization of the Quadratic Action}

Now, we are ready for diagonalizing the quadratic action~\eqref{eq: quadratic action}. First, note that the second terms in the each line of~\eqref{eq: quadratic action}, which come from the interaction of the collective action, are reduced to the inner product of the two eigenfunctions. Namely, since $[\Psi_{cl}]^{q-2}$ together with $|z|^{{1\over 2}-{2\over q}}$ factors in the two eigenfunctions form the measure of the Bessel function, we have
\begin{align}
    2J^2\int dt dz \left[\psi_{cl}(t+z,t-z)\right]^{q-2}u^-_{\nu w}(t,z)u^-_{\nu' w'}(t,z) =&J^2{\coeff^{q-2}\over 2^{2-{4\over q}}}  \widetilde{N}^-_{\nu} \delta_{w, -w'}\delta_{\nu, \nu'}\ ,\\
    2J^2\int dt dz \left[\psi_{cl}(t+z,t-z)\right]^{q-2}u^{ +}_{\nu w}(t,z)u^{+}_{\nu' w'}(t,z) =&J^2{\coeff^{q-2}\over 2^{2-{4\over q}}}  \widetilde{N}^+_{\nu} \delta_{w, -w'}\delta_{\nu, \nu'}\ .
\end{align}
where the factor 2 in front of the integral comes from the change of coordinates in~\eqref{eq: bi-local map}, and $\coeff$ is defined in~\eqref{def: coefficient}. 
%
%

In order to diagonalize the first terms in each line of~\eqref{eq: quadratic action} which arise from the Jacobian of the collective action, we will find a function $\tilde{u}^\mp_{\nu w}(t,z)$ for each $u^\mp_{\nu w}(t,z)$ such that
\begin{equation}\label{def: utilde}
\begin{split}
    \psi_{cl}\star \tilde{u}^-_{\nu w}  \star \psi_{cl}=& k_-(h) u^-_{\nu w}\ ,\\
    \psi_{cl}\star \tilde{u}^{ +}_{\nu w}  \star \psi_{cl}=& k_{\anti +}(h) u^{ +}_{\nu w}
\end{split}
\end{equation}
where $k(h)$'s are a function of $h\equiv \nu+{1\over 2}$. In Appendix~\ref{app: eigenfunctions}, we find that the function $\tilde{u}^\mp_{\nu w}(t,z)$ is proportional to the product of $[\Psi_{cl}]^{q-2}$ and $u^\mp$. \ie
%
%
\begin{align}
    \tilde{u}^-_{\nu w}(t,z)=&(q-1)J^2\left[\psi_{cl}(t+z,t-z)\right]^{q-2}u^-_{\nu w}(t,z)\label{eq: utilde m sol}\ ,\\
    \tilde{u}^{ +}_{\nu w}(t,z)=&J^2\left[\psi_{cl}(t+z,t-z)\right]^{q-2}u^{ +}_{\nu w}(t,z)\label{eq: utilde p sol}\ ,
\end{align}
and the $k_-(h)$ and  $k_{\anti +}(h)$ are found to be
\begin{align}
    k_-(h)=&-(q-1){\Gamma\left({3\over 2} -{1\over q}\right)\Gamma\left(1-{1\over q}\right)\Gamma\left({1\over q} +{h\over 2}\right)\Gamma\left({1\over 2}+ {1\over q}-{h\over 2}\right)\over \Gamma\left({1\over 2}+{1\over q}\right)\Gamma\left({1\over q}\right) \Gamma\left({3\over 2}-{1\over q} -{h\over 2}\right)\Gamma\left(1-{1\over q}+{h\over 2}\right) }\ ,\label{eq: km function}\\
    k_{\anti +}(h)=&-{\Gamma\left({3\over 2} -{1\over q}\right)\Gamma\left(1-{1\over q}\right)\Gamma\left({1\over q}-{1\over 2}+{h\over 2}\right)\Gamma\left({1\over q}-{h\over 2}\right)\over \Gamma\left({1\over 2}+{1\over q}\right)\Gamma\left({1\over q}\right) \Gamma\left(1-{1\over q} -{h\over 2}\right)\Gamma\left({1\over 2}-{1\over q}+{h\over 2}\right) }\ .\label{eq: kp function}
\end{align}
We also define $k_{\sym -}(h)$ for symmetric representation by
\begin{equation}    
     k_{\sym -}(h)\equiv{1\over q-1}k_-(h)=- {\Gamma\left({3\over 2} -{1\over q}\right)\Gamma\left(1-{1\over q}\right)\Gamma\left({1\over q} +{h\over 2}\right)\Gamma\left({1\over 2}+ {1\over q}-{h\over 2}\right)\over \Gamma\left({1\over 2}+{1\over q}\right)\Gamma\left({1\over q}\right) \Gamma\left({3\over 2}-{1\over q} -{h\over 2}\right)\Gamma\left(1-{1\over q}+{h\over 2}\right)}\label{eq: sym km function}
\end{equation}
such that
\begin{equation}
J^2 \psi_{cl}\star ([\psi_{cl}]^qu^-_{\nu w})  \star \psi_{cl}= k_{\sym -}(h) u^-_{\nu w}\ .
\end{equation}
Note that $k_-(h), k_{\anti +}(h)$ and $k_{\sym -}(h)$ are important in the four point functions as you will see in the next section since they determine the spectrum of our model. In particular, $k_-(h)$ is identical to that of the original SYK model. On the other hand, $k_{\anti -}(h)$ does not appear in the (real) SYK model. But, this function has been found in the complex SYK model which has $U(1)$ symmetry~\cite{Bulycheva:2017uqj} and the generalized SYK model for $q=4$ case~\cite{Peng:2017kro}. Furthermore, $k_{\sym -}(h)$ appears in the Pillow channel of the rank~$(q-1)$ tensor model~\cite{Narayan:2017qtw}.

Using \eqref{def: utilde}, \eqref{eq: utilde m sol}, \eqref{eq: utilde p sol} and \eqref{eq: sym km function}, the first terms in each line of~\eqref{eq: quadratic action} are also reduced to the inner product of the two eigenvectors with ${1\over k}$ factor. \ie
\begin{align}
    \tr (\psi_{cl}^{-1}\star u^-_{\nu w} \psi_{cl}^{-1}\star u^-_{\nu' w'} )=&{q-1\over k_-(h)}J^2\tr ([\psi_{cl}^{q-2}u^-_{\nu w}]\star u^-_{\nu' w'} )\cr
    =& -{(q-1)\coeff^{q-2}\over 2^{2-{4\over q}}k_-(h)}  J^2\widetilde{N}^-_{\nu} \delta_{w, -w'}\delta_{\nu, \nu'}\label{eq: 1st term of quadratic action m}\ ,\\
    \tr (\psi_{cl}^{-1}\star u^{ +}_{\nu w} \psi_{cl}^{-1}\star u^{ +}_{\nu' w'} )=&{1\over k_{\anti +}(h)}J^2\tr ([\psi_{cl}^{q-2}u^{ +}_{\nu w}]\star u^{ +}_{\nu' w'} )\cr
    =& {\coeff^{q-2}\over 2^{2-{4\over q}}k_{\anti +}(h)} J^2 \widetilde{N}^+_{\nu} \delta_{w, -w'}\delta_{\nu, \nu'}
\end{align}
where $h\equiv \nu+{1\over 2}$. Note that we have used anti-symmetry of $u^-_{\nu w}$ in~\eqref{eq: 1st term of quadratic action m}.

Now, we expand the singlet and anti-symmetric/symmetric irrep fluctuations in terms of 
\begin{align}
    \fluca(t,z)=&\int dw \sum_{\substack{ \nu=2n+{3\over 2}\\n=0 }}^\infty \fluca_{\nu w} u^-_{\nu w}(t,z) + \int dw \sum_{\substack{\nu=ir \\ r\geqq 0 }} \fluca_{\nu w} u^-_{\nu w}(t,z)\ ,\\
    \flucb^a_{\anti}(t,z)=&\int dw\; \sum_{\substack{ \nu=2n+{1\over 2}\\ n=0}}^\infty  \flucb^a_{\anti, \nu w} u^{+}_{\nu w} (t,z)+ \int dw \sum_{\substack{\nu=ir \\ r\geqq 0 }} \flucb^a_{\anti , \nu w} u^{+}_{\nu w}(t,z)\ ,\\
        \flucb^a_{\sym}(t,z)=&\int dw\; \sum_{\substack{ \nu=2n+{1\over 2}\\ n=0}}^\infty  \flucb^a_{\sym, \nu w} u^{-}_{\nu w} (t,z)+ \int dw \sum_{\substack{\nu=ir \\ r\geqq 0 }} \flucb^a_{\sym, \nu w} u^{-}_{\nu w}(t,z)\ .
\end{align}
The reality condition of the fundamental fermion $\chi^{ia}$ leads to 
\begin{align}
    \overline{\Psi^{\alpha_1\alpha_2}(\tau_1,\tau_2)}=\Psi^{\alpha_2\alpha_1}(\tau_2,\tau_1)=-\Psi^{\alpha_1\alpha_2}(\tau_1,\tau_2)\ .
\end{align}
and, using~\eqref{eq: conjugate of z}, we have
\begin{alignat}{3}
    &\overline{\fluca_{\nu w}}=-\fluca_{\nu, -w}&&\mbox{for}\;\; \nu= 2n+{3\over 2} \hspace{4mm} &&(n=0,1,\cdots)\ ,\\
    &\overline{\flucb^a_{\anti \nu w}}=\flucb^a_{\anti \nu, -w}&&\mbox{for}\;\; \nu= 2n+{1\over 2} &&(n=0,1,\cdots)\ ,\\
    &\overline{\flucb^a_{\sym \nu w}}=-\flucb^a_{\sym \nu, -w}&&\mbox{for}\;\; \nu= 2n+{3\over 2} &&(n=0,1,\cdots)\ ,\\    
    &\overline{\fluca_{\nu w}}=-\xi_\nu \fluca_{\nu, -w}&&\mbox{for}\;\; \nu= ir && (r\geqq 0))\ ,\\
    &\overline{\flucb^a_{\anti \nu w}}=\xi_{-\nu}\flucb^a_{\anti  \nu, -w}\hspace{6mm}&&\mbox{for}\;\; \nu= ir  &&(r\geqq 0)\ ,\\
    &\overline{\flucb^a_{\sym \nu w}}=-\xi_{-\nu}\flucb^a_{\sym \nu, -w}\hspace{6mm}&&\mbox{for}\;\; \nu= ir  &&(r\geqq 0)    \ .
\end{alignat}
Therefore, the quadratic action can be written as
\begin{align}
    S_{col}^{(2)}=&-{(q-1)J^2 \coeff^{q-2}\over 2^{2-{4\over q}}} \int_{w\geqq 0} dw \left.\left(\sum_{\nu=2n+{3\over 2}}+\sum_{\nu=ir} \right) {1-k_-(h)\over k_-(h)}N_\nu |\fluca_{\nu,w}|^2\right|_{h=\nu+{1\over 2}} \cr
    &- { J^2 \coeff^{q-2}\over 2^{2-{4\over q}}} \sum_{a}\int_{w\geqq 0} dw \left.\left(\sum_{\nu=2n+{1\over 2}}+\sum_{\nu=ir} \right) {1-k_{\anti +}(h)\over k_{\anti +}(h)}N_\nu |\flucb^a_{\anti, \nu,w}|^2\right|_{h=\nu+{1\over 2}}\cr
    &- { J^2 \coeff^{q-2}\over 2^{2-{4\over q}}} \sum_{a}\int_{w\geqq 0} dw \left.\left(\sum_{\nu=2n+{3\over 2}}+\sum_{\nu=ir} \right) {1-k_{\sym -}(h)\over k_{\sym -}(h)}N_\nu |\flucb^a_{\sym, \nu,w}|^2\right|_{h=\nu+{1\over 2}}\ .
    \label{eq:Quadratic Action}
\end{align}
It is easy to see that $\nu={3\over 2}$ and $\nu={1\over 2}$ corresponds to zero mode of the singlet and the anti-symmetric fluctuation respectively because 
\begin{equation}
    k_-(2)=1\quad,\quad k_{\anti +}(1)=1\hspace{10mm}(h=\nu+{1\over 2})\ .
\end{equation}
For these zero modes, one has to consider the perturbation of the classical solution $\maPsi$ at finite coupling, and evaluate the correction to the quadratic action~\cite{Jevicki:2016ito}. On the other hand, note that the symmetric irrep fluctuation does not have divergence. For now, ignoring the zero mode contributions, we write the propagators of the bi-local fluctuations, which corresponds to four point functions of the fundamental fermions as we will see in the next section, as follows.
\begin{align}
    &\langle \fluca(t,z)\fluca(t',z')\rangle\cr
    =& {2^{2-{4\over q}}\over (q-1)J^2 \coeff^{q-2}}\int dw \left.\left(\sum_{\substack{\nu=2n+{3\over 2}\\n=1,2,\cdots}}+\sum_{\nu=ir} \right) {k_-(h)\over 1-k_-(h)}{1\over N_\nu }\overline{u^-_{\nu w}(t,z)}u^-_{\nu w}(t',z')\right|_{h=\nu+{1\over 2}}\ ,\\
    &\langle \flucb^a_\anti(t,z)\flucb^b_\anti(t',z')\rangle\cr
    =&-\delta^{ab} {2^{2-{4\over q}}\over J^2 \coeff^{q-2}}\int dw \left.\left(\sum_{\substack{\nu=2n+{1\over 2}\\n=1,2,\cdots}}+\sum_{\nu=ir} \right) {k_{\anti +}(h)\over 1-k_{\anti +}(h)}{1\over N_\nu }\overline{u^+_{\nu w}(t,z)}u^+_{ \nu w}(t',z')\right|_{h=\nu+{1\over 2}}\ ,\\
        &\langle \flucb^a_\sym(t,z)\flucb^b_\sym(t',z')\rangle\cr
    =&\delta^{ab} {2^{2-{4\over q}}\over J^2 \coeff^{q-2}}\int dw \left.\left(\sum_{\substack{\nu=2n+{3\over 2}\\n=1,2,\cdots}}+\sum_{\nu=ir} \right) {k_{\sym -}(h)\over 1-k_{\sym -}(h)}{1\over N_\nu }\overline{u^-_{ \nu w}(t,z)}u^-_{ \nu w}(t',z')\right|_{h=\nu+{1\over 2}}\ .
\end{align}
%
%
One can proceed to evaluate these propagators as in~\cite{Jevicki:2016bwu}. However, in order to analyze the chaotic behavior of the out-of-time-order correlators, it is useful to utilize a conformal eigenfunction in calculating the four point functions~\cite{Maldacena:2016hyu,Murugan:2017eto,Peng:2017spg,Bulycheva:2017uqj}.

\subsection{Kernel and Four Point Function}
\label{sec: kernel}

In the previous section, we have diagonalized the quadratic action by using the eigenfunctions which diagonalize the bi-local time translation $\dP=\partial_{\tau_1}+\partial_{\tau_2}=\partial_t$. Though this basis could imply a close connection to the bulk, it is not easy to utilize the conformal symmetry.

Therefore, in this section, we will diagonalize the kernel generating the ladder diagrams by using a basis\footnote{In this paper, we use a different notation for the conformal eigenfunction from~\cite{Maldacena:2016hyu,Murugan:2017eto,Peng:2017spg,Bulycheva:2017uqj} for consistency in our notation. Hence, readers should keep in mind that $\Phi^-_h(\chi)=\Phi^s_h(\chi)$ and $\Phi^+_h(\chi)=\Phi^a_h(\chi)$.} $\Phi^\mp_h(\chi)$ related conformal block with the dimension $h$ following~\cite{Maldacena:2016hyu,Murugan:2017eto,Peng:2017spg,Bulycheva:2017uqj}. For this, let us first consider the four point function of the fundamental fermions, and express it in terms of the bi-local fields:
\begin{equation}
    {1\over N^2}\sum_{i,j=1}^N\left\langle \chi^{i\alpha_1}(\tau_1)\chi^{i\alpha_2}(\tau_1)\chi^{j\alpha_3}(\tau_1)\chi^{j\alpha_4}(\tau_1) \right\rangle=\left\langle \Psi^{\alpha_1\alpha_2}(\tau_1,\tau_2)\Psi^{\alpha_3\alpha_4}(\tau_3,\tau_4) \right\rangle\ .
\end{equation}
As mentioned in~\eqref{eq: decomposition of bi-fundamental}, one can decompose the bi-local field into the singlet, anti-symmetric and symmetric representations of $SO(M)$. Furthermore, since these representations are decoupled in the quadratic action~\eqref{eq: quadratic action}, one can consider three representations separately up to quadratic level. Hence, we consider the three correlation functions corresponding to the singlet, anti-symmetric and symmetric representations, and expand them around the large $N$ classical solution~\eqref{eq: large n expansion of classical solution}:
\begin{align}
    &{1\over M}\left\langle \tr\left[\maPsi(\tau_1,\tau_2)\right] \tr\left[\maPsi(\tau_3,\tau_4)\right] \right\rangle=M\psi_{cl}(\tau_1,\tau_2)\psi_{cl}(\tau_1,\tau_2)+{1\over N} \mathcal{F}_-(\tau_1,\tau_2,\tau_3,\tau_4)\ ,\\
    &{1\over 2\ind_\anti}\left\langle\tr\left[ \maPsi(\tau_1,\tau_2)\mT^a_\anti\right]\tr \left[\maPsi(\tau_3,\tau_4)\mT^b_\anti\right] \right\rangle={1\over N}\mathcal{F}^{ab}_{\anti +}(\tau_1,\tau_2,\tau_3,\tau_4)\ ,\\
       &{1\over 2\ind_\sym}\left\langle\tr\left[ \maPsi(\tau_1,\tau_2)\mT^a_\sym\right]\tr \left[\maPsi(\tau_3,\tau_4)\mT^b_\sym\right] \right\rangle={1\over N}\mathcal{F}^{ab}_{\sym -}(\tau_1,\tau_2,\tau_3,\tau_4)
\end{align}
where $\mathcal{F}_-(\tau_1,\tau_2,\tau_3,\tau_4)$ and $\mathcal{F}_{\anti/\sym +}(\tau_1,\tau_2,\tau_3,\tau_4)$ denote the two point function of the corresponding fluctuations:
\begin{align}
    \mathcal{F}_-(\tau_1,\tau_2,\tau_3,\tau_4)=&2\left\langle \fluca(\tau_1,\tau_2)\fluca(\tau_3,\tau_4) \right\rangle\label{def: correlation function1}\ ,\\
    \mathcal{F}^{ab}_{\anti +} (\tau_1,\tau_2,\tau_3,\tau_4)=&2\left\langle \flucb^a_{\anti}(\tau_1,\tau_2)\flucb^b_{\anti} (\tau_3,\tau_4) \right\rangle\label{def: correlation function2}\ ,\\
    \mathcal{F}^{ab}_{ \sym -} (\tau_1,\tau_2,\tau_3,\tau_4)=&2\left\langle \flucb^a_{\sym}(\tau_1,\tau_2)\flucb^b_{\sym} (\tau_3,\tau_4) \right\rangle\label{def: correlation function3}\ .
\end{align}
First, the factor $2$ in \eqref{def: correlation function1}$\sim$\eqref{def: correlation function3} comes from the large $N$ expansion of the bi-local field in \eqref{eq: large n expansion of classical solution}. Also, note that the correlation function of the anti-symmetric/symmetric representation does not have the leading disconnected diagram of order $\mathcal{O}(N^0)$. This is because the classical solution belong to the singlet representation of $SO(M)$. Furthermore, from the quadratic action~\eqref{eq: quadratic action}, one can expect that $\mathcal{F}_{\anti/\sym +}^{ab}$ is proportional to $\delta^{ab}$. \ie
\begin{align}
    \mathcal{F}_{\anti +}^{ab}(\tau_1,\tau_2,\tau_3,\tau_4)=\delta^{ab}\mathcal{F}_{\anti +}(\tau_1,\tau_2,\tau_3,\tau_4)\ ,\\
        \mathcal{F}_{\sym -}^{ab}(\tau_1,\tau_2,\tau_3,\tau_4)=\delta^{ab}\mathcal{F}_{\sym -}(\tau_1,\tau_2,\tau_3,\tau_4)\ .
\end{align}
Hence, it is enough to consider $\mathcal{F}_{\anti +}$ and $\mathcal{F}_{\sym -}$ without index $a$ and $b$ up to quadratic level. 

The correlation functions $\mathcal{F}_-$, $\mathcal{F}_{\anti +}$ and $\mathcal{F}_{\sym -}$ correspond to the ladder diagrams. For now, we will derive the Schwinger-Dyson equation without diagrammatics. We also work out diagrammatics in Section~\eqref{sec: diagrammatic}, and obtain the same result.

\paragraph{Schwinger Dyson Equation for four point function: }In order to obtain the Schwinger-Dyson equation for the two point function of the bi-local fluctuations, let us consider the following identity:
\begin{equation}
    \int \mathcal{D}\fluca\mathcal{D}\flucb_\anti\mathcal{D}\flucb_\sym {\delta\over \fluca(\tau_6,\tau_5)}\left( \fluca(\tau_3,\tau_4)e^{-S_{col}^{(2)}[\fluca,\flucb_\anti,\flucb_\sym]} \right) =0\label{eq: functional identity1}
\end{equation}
where we include terms up to of order $\mathcal{O}(N^0)$. Recall that the classical solution~$\Psi_{cl}(\tau_1,\tau_2)$ and the singlet fluctuation~$\fluca(\tau_1,\tau_2)$ are anti-symmetric under $\tau_1\leftrightarrow \tau_2$ (See~\eqref{eq: symmetry of fluctuation1}). Then, the variation of the quadratic action~\eqref{eq: quadratic action} with respect to $\fluca(\tau_6,\tau_5)$ is 
\begin{equation}
    {\delta S_{col}^{(2)} \over \delta \fluca(\tau_6,\tau_5)}=-(\psi_{cl}^{-1} \star \fluca \star \psi_{cl}^{-1})(\tau_5,\tau_6)+J^2(q-1) [\psi_{cl}(\tau_{56})]^{q-2} \fluca(\tau_5,\tau_6)\label{eq: variation of quadratic action}\ .
\end{equation}
Note that the anti-symmetry of the bi-local fluctuation gives
\begin{equation}
    {\delta \fluca(\tau_3,\tau_4)\over \delta \fluca(\tau_6,\tau_5)}={1\over 2}\delta(\tau_{36})\delta(\tau_{45})-{1\over 2}\delta(\tau_{35})\delta(\tau_{46})\label{eq: functional derivative of fluctuation1}\ .
\end{equation}
Substituting \eqref{eq: variation of quadratic action} and \eqref{eq: functional derivative of fluctuation1} into \eqref{eq: functional identity1}, and then multiplying $\psi_{cl}(\tau_1,\tau_5)\psi_{cl}(\tau_6,\tau_2)$, we have
\begin{equation}
    \mathcal{F}_-(\tau_1,\tau_2,\tau_3,\tau_4)-\mathcal{F}_{-,0}(\tau_1,\tau_2,\tau_3,\tau_4)=\int d\tau_5\tau_6 K_-(\tau_1,\tau_2,\tau_5,\tau_6)\mathcal{F}_-(\tau_5,\tau_6,\tau_3,\tau_4)\label{eq: SD for 2pt function1}
\end{equation}
where $\mathcal{F}_{-,0}(\tau_1,\tau_2,\tau_3,\tau_4)$ is found to be
\begin{equation}
    \mathcal{F}_{-,0}(\tau_1,\tau_2,\tau_3,\tau_4)\equiv -\psi_{cl}(\tau_1,\tau_3)\psi_{cl}(\tau_2,\tau_4)+\psi_{cl}(\tau_1,\tau_4)\psi_{cl}(\tau_2,\tau_3)\label{eq: fm0 term}\ .
\end{equation}
Note that $\mathcal{F}_{-,0}$ corresponds to the zero-rung ladder diagram. Also, the kernel $K_-$ is given by
\begin{equation}
    K_-(\tau_1,\tau_2,\tau_3,\tau_4)=-J^2(q-1) \int d\tau_3d\tau_4\; \psi_{cl}(\tau_{13})\psi_{cl}(\tau_{24})[\psi_{cl}(\tau_{34})]^{q-2} \label{def: kernel m}\ .
\end{equation}
This is identical to the kernel from the diagrammatics in Section~\ref{sec: diagrammatic}. In the same way, using
\begin{align}
    {\delta \flucb^a_\anti(\tau_3,\tau_4)\over \delta \flucb^b_\anti(\tau_6,\tau_5)}=&{1\over 2}\delta^{ab}\delta(\tau_{36})\delta(\tau_{45}) + {1\over 2}\delta^{ab}\delta(\tau_{35})\delta(\tau_{46})\label{eq: functional derivative of fluctuation2}\ ,\\
    {\delta \flucb^a_\sym(\tau_3,\tau_4)\over \delta \flucb^b_\sym(\tau_6,\tau_5)}=&{1\over 2}\delta^{ab}\delta(\tau_{36})\delta(\tau_{45}) - {1\over 2}\delta^{ab}\delta(\tau_{35})\delta(\tau_{46})\label{eq: functional derivative of fluctuation3}\ ,
\end{align}
we obtain
\begin{align}
    \mathcal{F}_{\anti +}(\tau_1,\tau_2,\tau_3,\tau_4)-\mathcal{F}_{\anti +,0}(\tau_1,\tau_2,\tau_3,\tau_4)=&\int d\tau_5\tau_6 K_+(\tau_1,\tau_2,\tau_5,\tau_6)\mathcal{F}_{\anti +}(\tau_5,\tau_6,\tau_3,\tau_4)\label{eq: SD for 2pt function2}\ ,\\
    \mathcal{F}_{\sym -}(\tau_1,\tau_2,\tau_3,\tau_4)-\mathcal{F}_{\sym -,0}(\tau_1,\tau_2,\tau_3,\tau_4)=&\int d\tau_5\tau_6 K_+(\tau_1,\tau_2,\tau_5,\tau_6)\mathcal{F}_{\sym -}(\tau_5,\tau_6,\tau_3,\tau_4)\label{eq: SD for 2pt function3}
\end{align}
where the kernel $K_{+}$, the zero-rung ladder diagrams $\mathcal{F}_{\anti +,0}$ and $\mathcal{F}_{\sym -,0}$ of order $\mathcal{O}(N^0)$ are found to be
\begin{align}
    K_{+}(\tau_1,\tau_2,\tau_3,\tau_4)\equiv & - J^2  \psi_{cl}(\tau_{13})\psi_{cl}(\tau_{24})[\psi_{cl}(\tau_{34})]^{q-2} \label{def: kernel p}\ , \\
    \mathcal{F}_{\anti+,0}(\tau_1,\tau_2,\tau_3,\tau_4)\equiv& \psi_{cl}(\tau_1,\tau_3)\psi_{cl}(\tau_2,\tau_4)+\psi_{cl}(\tau_1,\tau_4)\psi_{cl}(\tau_2,\tau_3)\equiv \mathcal{F}_{+,0}\label{eq: fp0 term} \ ,\\
    \mathcal{F}_{\sym - ,0}(\tau_1,\tau_2,\tau_3,\tau_4)\equiv& -\psi_{cl}(\tau_1,\tau_3)\psi_{cl}(\tau_2,\tau_4)+\psi_{cl}(\tau_1,\tau_4)\psi_{cl}(\tau_2,\tau_3)=\mathcal{F}_{-,0}\ .
\end{align}
Here, we dropped the trivial matrix structure of the kernels and the ladder diagrams of non-singlet representation because they are diagonal in $(a,b)$ space up to quadratic level. Note that the anti-symmetric and symmetric fluctuations have the same common ratio $K_+$. Also, note that the zero-rung ladder diagram $\mathcal{F}_{s +,0}$ of the symmetric representation is the same as $\mathcal{F}_{-,0}$ of the singlet representation. Moreover, we will denote by $\mathcal{F}_{+,0}$  the zero-rung diagram $\mathcal{F}_{\anti +,0}$ of the anti-symmetric representation for simplicity.

The Schwinger-Dyson equations~\eqref{eq: SD for 2pt function1} and \eqref{eq: SD for 2pt function2} for the two point function of the bi-local fluctuations generates geometric series, which leads to
\begin{equation}
    \mathcal{F}={1\over 1-K} \mathcal{F}_{0}
\end{equation}
for each $\mathcal{F}_-$, $\mathcal{F}_{\anti, +}$ and $\mathcal{F}_{\sym, -}$. Using $SL(2,\mathbb{R})$ symmetry, one can express the correlation functions in terms of the cross ratio. Hence, we define
\begin{equation}
    \mathcal{F}_-(\chi)\equiv{\mathcal{F}_-(\tau_1,\tau_2,\tau_3,\tau_4)\over \psi_{cl}(\tau_{12})\psi_{cl}(\tau_{34})}\quad,\quad \mathcal{F}_{\anti +}(\chi)\equiv{\mathcal{F}_{\anti +}(\tau_1,\tau_2,\tau_3,\tau_4)\over \psi_{cl}(\tau_{12})\psi_{cl}(\tau_{34})}\ ,\label{eq: correlation cross ratio}
\end{equation}
and similar for $\mathcal{F}_{\sym -}(\chi)$ where the cross ratio is given by
\begin{equation}
    \chi\equiv {\tau_{12}\tau_{34}\over \tau_{13}\tau_{24}}\ .
\end{equation}

Following \cite{Maldacena:2016hyu}, we will expand the correlation functions $\mathcal{F}_-(\chi)$, $\mathcal{F}_{\anti +}(\chi)$ and $\mathcal{F}_{\sym -}(\chi)$ in term of the hypergeometric basis $\Phi^\mp_h(\chi)$ called as conformal eigenfunctions~\cite{Maldacena:2016hyu,Murugan:2017eto,Peng:2017spg,Bulycheva:2017uqj}. 
\begin{align}
    \mathcal{F}_{- }(\chi)=&\sum_h \Phi^-_h(\chi) {1\over 1-k_-(h)} {\langle \Phi^-_h,\mathcal{F}_{-,0}\rangle\over \langle \Phi_h^\mp,\Phi_h^\mp\rangle }\ ,\\
    \mathcal{F}_{\anti + }(\chi)=&\sum_h \Phi^+_h(\chi) {1\over 1-k_{\anti +}(h)} {\langle \Phi^+_h,\mathcal{F}_{ +,0}\rangle\over \langle \Phi_h^\mp,\Phi_h^\mp\rangle }\ ,\\
    \mathcal{F}_{\sym - }(\chi)=&\sum_h \Phi^-_h(\chi) {1\over 1-k_{\sym -}(h)} {\langle \Phi^-_h,\mathcal{F}_{-,0}\rangle\over \langle \Phi_h^-,\Phi_h^-\rangle }
\end{align}
where the inner product is defined~\cite{Maldacena:2016hyu} by
\begin{equation}
    \langle g,f\rangle \equiv \int_0^2 {d\chi\over \chi^2} \overline{g(\chi)} f(\chi)\ ,
\end{equation}
and we used $\mathcal{F}_{\sym, 0}=\mathcal{F}_{-,0}$. Here, $k_-(h)$, $k_{\anti +}(h)$ and $k_{\sym -}(h)$ are given by
\begin{align}
    k_-(h)=&-(q-1){\Gamma\left({3\over 2} -{1\over q}\right)\Gamma\left(1-{1\over q}\right)\Gamma\left({1\over q} +{h\over 2}\right)\Gamma\left({1\over 2}+ {1\over q}-{h\over 2}\right)\over \Gamma\left({1\over 2}+{1\over q}\right)\Gamma\left({1\over q}\right) \Gamma\left({3\over 2}-{1\over q} -{h\over 2}\right)\Gamma\left(1-{1\over q}+{h\over 2}\right) }\ ,\\
    k_{\anti +}(h)=&-{\Gamma\left({3\over 2} -{1\over q}\right)\Gamma\left(1-{1\over q}\right)\Gamma\left({1\over q}-{1\over 2}+{h\over 2}\right)\Gamma\left({1\over q}-{h\over 2}\right)\over \Gamma\left({1\over 2}+{1\over q}\right)\Gamma\left({1\over q}\right) \Gamma\left(1-{1\over q} -{h\over 2}\right)\Gamma\left({1\over 2}-{1\over q}+{h\over 2}\right) }\ ,\\
     k_{\sym -}(h)=&-{\Gamma\left({3\over 2} -{1\over q}\right)\Gamma\left(1-{1\over q}\right)\Gamma\left({1\over q} +{h\over 2}\right)\Gamma\left({1\over 2}+ {1\over q}-{h\over 2}\right)\over \Gamma\left({1\over 2}+{1\over q}\right)\Gamma\left({1\over q}\right) \Gamma\left({3\over 2}-{1\over q} -{h\over 2}\right)\Gamma\left(1-{1\over q}+{h\over 2}\right)}={1\over q-1}k_-(h)\ ,
\end{align}
which we have obtained in diagonalization of the quadratic action in Section~\ref{sec: quadratic action}. 
%
%
Also, $\mathcal{F}_{\mp,0}(\chi)$ is obtained from \eqref{eq: fm0 term} and \eqref{eq: fp0 term} in the same way as in~\eqref{eq: correlation cross ratio}:
\begin{equation}
    \mathcal{F}_{\mp,0}(\chi)=\begin{cases}
    \;\;\mp \chi^{2\over q} +\left(\chi\over 1-\chi\right)^{2\over q}&\quad (0<\chi<1)\\
    \;\;\mp \chi^{2\over q} - \left(\chi\over \chi-1\right)^{2\over q}&\quad (1<\chi)\\
    \end{cases}\ .
\end{equation}
The conformal eigenfunction $\Phi^\mp_h(\chi)$ was introduced by \cite{Maldacena:2016hyu,Murugan:2017eto,Peng:2017spg,Bulycheva:2017uqj}, and their integral representations are given by
%
%
\begin{align}
    \Phi_h^-=&{1\over 2}\int_{-\infty}^\infty {|\chi|^h \over |y|^h |y-1|^{1-h} |y-\chi|^h}dy\ ,\\
    \Phi_h^+=&{\sgn(\chi)\over 2}\int_{-\infty}^\infty {|\chi|^h \sgn(y)\sgn(y-1) \sgn(y-\chi)\over |y|^h |y-1|^{1-h} |y-\chi|^h}dy \ ,
\end{align}
and, their inner products were evaluated in~\cite{Maldacena:2016hyu,Murugan:2017eto,Peng:2017spg,Bulycheva:2017uqj}:
%
%
\begin{align}
    \langle \Phi^\pm_h,\Phi^\pm_{h'}\rangle=&{\pi \tan(\pi h)\over 4h-2} 2\pi \delta(h-h')\hspace{10mm} (h={1\over 2}+ir)\ ,\\
    \langle \Phi^\pm_h,\Phi^\pm_{h'}\rangle=&{\pi^2\delta_{h h'} \over 4h-2}\hspace{10mm} (h=2n+2\;\;\mbox{for}\; \Phi^-_h\;,\;\;h=2n+1\;\;\mbox{for}\; \Phi^+_h)\ .
\end{align}
In addition, the inner product $\langle \Phi^-_h, \mathcal{F}_{-,0}\rangle$ was shown~\cite{Maldacena:2016hyu} to be
%
%
\begin{equation}
    \langle \Phi^-_h, \mathcal{F}_{-,0}\rangle ={\alpha_{-,0}\over 2} k_-(h)     
\end{equation}
where $\alpha_{-,0}$ is defined by
\begin{equation}
    \alpha_{-,0}\equiv {2\pi q\over (q-1)(q-2) \tan{\pi \over q}}={1\over (q-1) J^2 \coeff^q}\ .
\end{equation}
In a similar way as in~\cite{Maldacena:2016hyu,Murugan:2017eto,Peng:2017spg,Bulycheva:2017uqj}, we evaluate the inner product of $\Phi^+_h$ and $\mathcal{F}_{+,0}$. For this, one can use the anti-symmetry of the integral representation $\Phi^+_h$ under $\chi\rightarrow {\chi\over \chi-1}$ and symmetry/anti-symmetry properties of the zero-rung $\mathcal{F}_{+,0}(\chi)$ in~\eqref{eq: fp0 term} to extend the integration region $0<\chi<2$ in the inner product to the entire line $\chi\in \mathbb{R}$. Like $\Phi_h^-$ case in~\cite{Maldacena:2016hyu}, this extension picks up $\sgn(\chi)$ factor, and the inner product can be expressed as
\begin{equation}
    \langle \Phi_h^+,\mathcal{F}_{+,0}\rangle
    ={1\over 2}\int_{-\infty}^\infty  d\chi dy  { \sgn(y)\sgn(y-1) \sgn(y-\chi)\over |\chi|^{2-h-{2\over q} } |y|^h |y-1|^{1-h} |y-\chi|^h}\ .
\end{equation}
In order to evaluate this integral, we derive the following integral formula by using (2.23) in~\cite{Fu:2016vas}:
\begin{equation}
    \int_{-\infty}^\infty dx {\sgn(x-b)\over |x-a|^{2\alpha}|x-b|^{2\beta}}
    ={\pi \sin(\pi(2\alpha+2\beta-1))\Gamma(2\alpha+2\beta-1) \over  \cos(\pi \alpha)\sin (\pi \beta) \Gamma(2\alpha)\Gamma(2\beta)}  {\sgn(a-b)\over |a-b|^{2\alpha+2\beta-1}}\ .
\end{equation}
Using this integral formula, we have
\begin{equation}
    \langle \Phi_h^+,\mathcal{F}_{+,0}\rangle
    ={\alpha_{+,0}\over 2}k_+(h)
\end{equation}
where $\alpha_{+,0}$ is defined by
\begin{equation}
    \alpha_{+,0}\equiv {2\pi q \over (q-2)\tan{\pi \over q}}={1\over J^2\coeff^q}=(q-1)\alpha_{-,0}\label{def: alpha p0}\ .
\end{equation}
%
%
%
%
Finally, the correlation functions $\mathcal{F}_-(\chi)$, $\mathcal{F}_{\anti +}(\chi)$ and $\mathcal{F}_{\sym -}(\chi)$  can be expressed as 
\begin{align}
    \mathcal{F}_-(\chi)
    =&\left.\alpha_{-,0} \int_0^\infty {ds\over 2\pi }{2h-1\over \pi \tan (\pi h)}{k_-(h)\over 1-k_-(h)}\Phi^-_h(\chi) \right|_{h={1\over 2}+is}\cr
    &\hspace{10mm}+\alpha_{-,0} \sum_{n=1}^\infty \left[ {2h-1\over \pi^2} {k_-(h)\over 1-k_- (h)}\Phi^-_h(\chi)\right]_{h=2n}\label{eq: correlation function first result1}\ ,\\
        \mathcal{F}_{\anti +}(\chi)
    =&\left.\alpha_{+,0} \int_0^\infty {ds\over 2\pi }{2h-1\over \pi \tan (\pi h)}{k_{\anti +}(h)\over 1-k_{\anti +}(h)}\Phi^+_h(\chi) \right|_{h={1\over 2}+is}\cr
    &\hspace{10mm}+\alpha_{+,0} \sum_{n=1}^\infty \left[ {2h-1\over \pi^2} {k_{\anti +}(h)\over 1-k_{\anti +} (h)}\Phi^+_h(\chi)\right]_{h=2n-1 }\label{eq: correlation function first result2}\ ,\\
        \mathcal{F}_{\sym -}(\chi)
    =&\left.\alpha_{-,0} \int_0^\infty {ds\over 2\pi }{2h-1\over \pi \tan (\pi h)}{k_{\sym -}(h)\over 1-k_{\sym -}(h)}\Phi^-_h(\chi) \right|_{h={1\over 2}+is}\cr
    &\hspace{10mm}+\alpha_{-,0} \sum_{n=1}^\infty \left[ {2h-1\over \pi^2} {k_{\sym -}(h)\over 1-k_{\sym -} (h)}\Phi^-_h(\chi)\right]_{ h=2n }\label{eq: correlation function first result3}\ .
\end{align}
As mentioned before, the singlet channel correlation function $\mathcal{F}_-(\chi)$ is the same as that of the original SYK model. Also, the symmetric channel $\mathcal{F}_{\sym -}(\chi)$ is also almost the same as the original SYK model except that there is no divergence because of $k_{\sym -}={1\over q-1}k_-$. The anti-symmetric channel $\mathcal{F}_{\anti +}(\chi)$ are also of the same form, but since the properties of $k_{\anti +}(h)$ and $\Phi^+_h(\chi)$ are different, the results are slightly different. Nevertheless, the calculations are exactly parallel to those in Section~3.2.5~of~\cite{Maldacena:2016hyu}. Hence, we will not repeat the details here, but we point out the difference and present the results. We also note here that a similar analysis was also done in~\cite{Murugan:2017eto,Peng:2017spg,Bulycheva:2017uqj}.

The basic idea is to change the continuous and the discrete sum into a contour integral, and express the correlation functions as a sum of the residues. For this, in addition to ${\scriptscriptstyle{2\over \tan \pi h }={1\over \tan {\pi h\over 2} } -{1\over \tan {\pi(1-h)\over 2}}}$ used for $\mathcal{F}_-$, we will also utilize
\begin{equation}
    {2\over \tan \pi h }=\tan {\pi(1-h)\over 2}- \tan{\pi h\over 2}
\end{equation}
to manipulate $\mathcal{F}_{\anti +}$ in~\eqref{eq: correlation function first result2} into 
\begin{align}
    {\mathcal{F}_{\anti +,h\ne1}(\chi)\over \alpha_{+,0}}=&-\int_{-\infty}^\infty {ds\over 2\pi }{h-{1\over 2}\over \pi  }\tan {\pi h\over 2}{k_{\anti +}(h)\over 1-k_{\anti +}(h)}\Phi_h^+(\chi)\cr
    &- \sum_{n=2}^\infty \mbox{Res}\left[ {h-{1\over 2}\over \pi}\tan{\pi h\over 2} {k_{\anti +}(h)\over 1-k_{\anti +}(h)}\Phi_h^+(\chi)\right]_{h=2n-1}\label{eq: correlation intermediate}
\end{align}
where we neglect $h=1$ zero mode which make $\mathcal{F}_{\anti,+}$ divergent. Note that the simple pole from $\tan{\pi h\over 2}$ enables us to write the discrete series as the residue sum. Also, note that the poles at $h=2n$ of $\Phi_h^+$ are cancelled with the zeros of $\tan{\pi h\over 2}$. 

In the same way as in the singlet channel, the symmetric channel can also be written as
\begin{align}
    &{\mathcal{F}_{\sym -}(\chi)\over \alpha_{-,0}}\cr
    =&\int_{-\infty}^\infty {ds\over 2\pi }{h-{1\over 2}\over \pi  \tan {\pi h\over 2}}{k_{\sym -}(h)\over 1-k_{\sym -}(h)}\Phi_h^-(\chi) + \sum_{n=1}^\infty \mbox{Res}\left[ {h-{1\over 2}\over \pi \tan{\pi h\over 2}} {k_{\sym -}(h)\over 1-k_{\sym -}(h)}\Phi_h^-(\chi)\right]_{h=2n}
\end{align}

Now, by moving the contour, we will pick up the residues which were not included in \eqref{eq: correlation intermediate}. These poles are located at
\begin{alignat}{3}
    &h_{-,m}\quad&&:\quad k_-(h_{-,m})=1 \hspace{10mm} &&(m=0,1,2,\cdots)\label{eq: spectrum of the model1}\ ,\\
    &h_{\anti +,m}\quad&&:\quad k_{\anti +}(h_{\anti +,m})=1 \hspace{10mm} &&(m=0,1,2,\cdots)\label{eq: spectrum of the model2}\ ,\\
    &h_{\sym -,m}\quad&&:\quad k_{\sym -} (h_{\sym -,m})=1 \hspace{10mm} &&(m=0,1,2,\cdots)\label{eq: spectrum of the model3}\ .
\end{alignat}
Note that the integrands have double pole at $h=h_{-,0}=2$ and $h=h_{\anti +,0}=1$ which correspond to the zero modes related to reparametrization and $SO(M)$ symmetry.  The lowest dimension in the symmetric irrep channel is $h_{\sym -,0}=1.24340\cdots$. Using the property of the conformal eigenfunction $\Phi^\mp$~\cite{Murugan:2017eto,Peng:2017spg,Bulycheva:2017uqj}
\begin{equation}
    \Phi^{\mp}_{1-h}(\chi)=\pm\Phi^\mp_{h}(\chi)\ ,
\end{equation}
%
%
%
%
%
%
we finally have (together with $\mathcal{F}_{-,h\ne 2}(\chi)$ found in~\cite{Maldacena:2016hyu})
%
%
\begin{align}
    {\mathcal{F}_{-,h\ne 2}(\chi)\over \alpha_{-,0}}=& -\sum_{m=0}^\infty \mbox{Res}\left[ {h-{1\over 2}\over \pi \tan{\pi h \over 2}}{k_- (h)\over 1-k_- (h)}{\Gamma(h)^2\over \Gamma(2h) }\chi^h{}_2F_1(h,h,2h,\chi)\right]_{h=h_{-,m}}\!\!\!\!\!\!\!(\chi<1)\label{eq: four point function small1}\\
    {\mathcal{F}_{\anti +,h\ne 1}(\chi)\over \alpha_{+,0}}=& \sum_{m=0}^\infty \mbox{Res}\left[ {h-{1\over 2}\over \pi}\tan{\pi h \over 2} {k_\mp (h)\over 1-k_\mp (h)}{\Gamma(h)^2\over \Gamma(2h) }\chi^h{}_2F_1(h,h,2h,\chi)\right]_{h=h_{\anti +,m}}\hspace{-10mm}(\chi<1)\label{eq: four point function small2}\\
    {\mathcal{F}_{\sym -}(\chi)\over \alpha_{-,0}}=& -\sum_{m=0}^\infty \mbox{Res}\left[ {h-{1\over 2}\over \pi \tan{\pi h \over 2}} {k_{\sym -} (h)\over 1-k_{\sym -} (h)}{\Gamma(h)^2\over \Gamma(2h) }\chi^h{}_2F_1(h,h,2h,\chi)\right]_{h=h_{\sym -,m}}\hspace{-7mm}(\chi<1)\label{eq: four point function small3}
\end{align}
and
\begin{align}
     {\mathcal{F}_{-,h\ne 2}(\chi)\over\alpha_{-,0}}=& -\sum_{m=0}^\infty \mbox{Res}\left[{h-{1\over 2}\over \pi \tan {\pi h\over 2}}  {k_- (h)\over 1-k_-(h)}\Phi_h^- (\chi)\right]_{h=h_{-,m}}\hspace{7mm} (\chi>1)\label{eq: four point function large1}\\
          {\mathcal{F}_{\anti +,h\ne 1}(\chi)\over\alpha_{+,0}}=& \sum_{m=0}^\infty \mbox{Res}\left[{h-{1\over 2}\over \pi }\tan {\pi h\over 2} {k_{\anti +} (h)\over 1-k_{\anti +}(h)}\Phi_h^+ (\chi)\right]_{h=h_{\anti +,m}}\hspace{0mm} (\chi>1)\label{eq: four point function large2}\\
               {\mathcal{F}_{\sym -}(\chi)\over\alpha_{-,0}}=& -\sum_{m=0}^\infty \mbox{Res}\left[{h-{1\over 2}\over \pi \tan {\pi h\over 2} } {k_{\sym -} (h)\over 1-k_{\sym -}(h)}\Phi_h^- (\chi)\right]_{h=h_{\sym -,m}}\hspace{5mm} (\chi>1)\label{eq: four point function large3}
\end{align}
%
%
%
where $\Phi_h^\mp(\chi)$ ($\chi>1$) are given by~\cite{Maldacena:2016hyu,Peng:2017spg} 
\begin{align}
    \Phi_h^-(\chi)=&{\Gamma\left({1\over 2}-{h\over2}\right)\Gamma\left({h\over 2}\right) \over \sqrt{\pi} } {}_2F_1\left({h\over 2},{1-h\over 2},{1\over 2},{(\chi-2)^2\over \chi^2}\right)\label{eq: phi m greater}\ ,\\
    \Phi_h^+(\chi)=&-{2\Gamma\left(1-{h\over2}\right)\Gamma\left({h\over 2}+{1\over 2}\right) \over \sqrt{\pi} }{\chi-2\over \chi} {}_2F_1\left({2-h\over 2},{h+1\over 2},{3\over 2},{(\chi-2)^2\over \chi^2}\right)\label{eq: phi p greater}\ .
\end{align}

Now, we will take the OPE limit ($\chi\rightarrow 0$) of \eqref{eq: four point function small1}, \eqref{eq: four point function small2} and \eqref{eq: four point function small3} in which the two point function of the bi-locals (or, the four point function of the fermions) is decomposed into conformal blocks as follow.
\begin{align}
     \mathcal{F}_{-,h\ne 2}(\tau_1,\tau_2,\tau_3,\tau_4)=&\sum_{m=1}^\infty c_{-,m}^2 \left[\chi^{h_{-,m} } {}_2 F_1(h_{-,m},h_{-,m},2h_{-,m},\chi)\right]\label{eq: OPE expansion1}\ ,\\
     \mathcal{F}^{ab}_{\anti +,h\ne 1}(\tau_1,\tau_2,\tau_3,\tau_4)=&\sum_{m=1}^\infty c_{\anti +,m}^2\! \left[\chi^{h_{\anti +,m} } {}_2 F_1(h_{\anti +,m},h_{\anti +,m},2h_{\anti +,m},\chi)\right]\label{eq: OPE expansion2}\ ,\\
     \mathcal{F}^{ab}_{\sym -}(\tau_1,\tau_2,\tau_3,\tau_4)=&\sum_{m=0}^\infty c_{\sym -,m}^2\! \left[\chi^{h_{\sym -,m} } {}_2 F_1(h_{\sym -,m},h_{\sym -,m},2h_{\sym -,m},\chi)\right]\label{eq: OPE expansion3}
\end{align}
where $h_{-,m}$, $h_{\anti +,m}$ and $h_{\sym -, m}$ are defined in \eqref{eq: spectrum of the model1}$\sim$\eqref{eq: spectrum of the model3}, and they can be understood as the conformal dimensions of the operators in the OPE channel. One can write the conformal dimensions as
\begin{align}
    h_{-,m}=&{2\over q} +1 +2m +\epsilon^-_m\ , \\
    h_{\anti +,m}=&{2\over q}  +2m +\epsilon^{\anti +}_m \ ,\\
    h_{\sym -,m}=&{2\over q} +1 +2m +\epsilon^{\sym -}_m\ ,
\end{align}
and, the asymptotic behavior for large $m$ are given by
\begin{align}
    \epsilon^-_m=&{2\Gamma\left(3-{2\over q}\right)\sin {2\pi \over q}\over \pi \Gamma\left(1+{2\over q}\right)  \; (2m)^{2-{4\over q}}}\ ,\\
    \epsilon^{\anti +}_m=&{2\Gamma\left(2-{2\over q}\right)\sin {2\pi \over q}\over \pi \Gamma\left({2\over q}\right)  \; (2m)^{2-{4\over q}}}\ ,\\
    \epsilon^{\sym -}_m=&{2\Gamma\left(2-{2\over q}\right)\sin {2\pi \over q}\over \pi \Gamma\left({2\over q}\right)  \; (2m)^{2-{4\over q}}}  \ . 
\end{align}
The OPE coefficients $c_{-,m}$, $c_{\anti +,m}$ and $c_{\sym -,m}$ in \eqref{eq: OPE expansion1}, \eqref{eq: OPE expansion2} and \eqref{eq: OPE expansion3} are found to be
\begin{align}
    c_{-,m}^2=&- {\alpha_{- ,0}\over N} {h_{-,m}-{1\over 2}\over \pi\tan{\pi h_{-,m} \over 2}}{\Gamma(h_{-,m} )^2\over \Gamma(2h_{-,m})}{1\over - k'_\mp(h_{-,m})}\ ,\\
        c_{\anti +,m}^2=& {\alpha_{+ ,0}\over N} {h_{\anti +,m}-{1\over 2}\over \pi}\tan{\pi h_{\anti +,m} \over 2}{\Gamma(h_{\anti +,m} )^2\over \Gamma(2h_{\anti +,m})}{1\over - k'_\mp(h_{\anti +,m})}\ ,\\
            c_{\sym +,m}^2=& -{\alpha_{- ,0}\over N} {h_{\sym +,m}-{1\over 2}\over \pi \tan{\pi h_{\sym +,m} \over 2} }{\Gamma(h_{\sym +,m} )^2\over \Gamma(2h_{\sym +, m})}{1\over - k'_\mp(h_{\sym + ,m})}\ ,
\end{align}
which are positive.


\subsection{Diagrammatic Derivation of Two Point Function and Kernels}
\label{sec: diagrammatic}


In this subsection\footnote{I thank Prithvi Narayan for extensive discussion and collaboration on diagrammatics in this subsection.}, we give an alternate derivation for the two point function and for the kernel (relevant for the four point function calculation) given in~\eqref{def: kernel m} and \eqref{def: kernel p}. The diagrammatics is done at large $N$. 

\subsection*{Two Point Function}
\label{app: two point function diagrammatics}

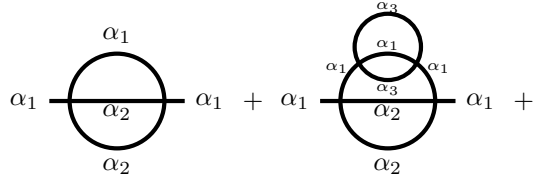
\begin{figure}[t!]
\centering

\begin{tikzpicture}[scale=0.9]

\def\c{0}  \def\s{1} \def\r{0.7}

\def\c{3} 
\draw[ultra thick,color=black] (-\s+\c,0) --  (\s+\c,0);
\draw[ultra thick,color=black] (\c,0) circle (\r);
\draw[thick, color=black]{ 
(-\s+\c,0) node [left] {\small{$\alpha_1$}}   
(\s+\c,0) node [right] {\small{$\alpha_1$}}   
(\c,\r) node [above] {\small{$\alpha_1$}}   
(\c,-\r) node [below] {\small{$\alpha_2$}}   
(\c,0+0.1) node [below] {\small{$\alpha_2$}}   
};

\draw[thick, color=black]
{
(\s + \c+ 0.7,0) node [right] {\small{$ + $}}
};

\def\c{7} \def\v{0.8}
\draw[ultra thick,color=black] (-\s+\c,0) --  (\s+\c,0);
\draw[ultra thick,color=black] (\c,0) circle (\r); 
\draw[ultra thick,color=black] (\c,\v) circle (0.7*\r);

\draw[thick, color=black]
{
(\c-\r-0.05,0.25) node [above] {\tiny{$\alpha_1$}} 
(\c+\r+0.05,0.25) node [above] {\tiny{$\alpha_1$}} 
(-\s+\c,0) node [left] {\small{$\alpha_1$}}   
(\s+\c,0) node [right] {\small{$\alpha_1$}}   
(\c,\r-0.1) node [above] {\tiny{$\alpha_1$}}   
(\c,-\r) node [below] {\small{$\alpha_2$}}   
(\c,0+0.1) node [below] {\small{$\alpha_2$}}  
(\s + \c+ 0.7,0) node [right] {\small{$ + $}}
( \c, \v + \r-0.35) node [above]  {\tiny{$\alpha_3$}}
( \c, \v - \r+0.3) node [below]  {\tiny{$\alpha_3$}}
};

\end{tikzpicture} 
\caption{Contributions to two point function self energy}
\label{fig:Self Energy}
\end{figure}

We begin our analysis with the two point function. We will find the Schwinger-Dyson equation satisfied by the disorder averaged two point function  
\begin{equation}
    \Psi_{cl}^{\alpha_1\alpha_2}(\tau_1,\tau_2)  = {1 \over N }\sum_{i=1}^N \langle  \chi^{i\alpha_1}(\tau_1) \chi^{i\alpha_2}(\tau_2) \rangle\label{eq:TwoPointdef}
\end{equation}
%
where we used the notation $\Psi_{cl}^{\alpha_1\alpha_2}$ because the two point function of the fermions corresponds to the large $N$ classical solution (or, equivalently one point function of the bi-local field). Recall that the two point function of the free theory is given by $\Psi^{\alpha_1\alpha_2}_{cl,\text{free}}(\tau_1,\tau_2) = {1\over2}\delta^{\alpha_1\alpha_2} \sgn{\tau_{12}}$. The $SO(M)$ invariance guarantees that $\Psi^{\alpha_1\alpha_2}_{cl}(\tau_1,\tau_2) \propto \delta^{\alpha_1\alpha_2}$ to all orders in $J$.  Some diagrams which contribute to the self energy are given in Figure~\ref{fig:Self Energy}.

The only difference from the original SYK model, in fact, is that the fermion propagators carry an extra $SO(M)$ index. From Figure~\ref{fig:Self Energy}, it is obvious that this results in extra factors of $M$ due to the sum over internal indices (\eg sum over $\alpha_2$ and $\alpha_3$ in Figure~\ref{fig:Self Energy}). Let us see how this works for the simplest case of the second diagram in Figure~\ref{fig:Self Energy}:  There are $q-1$ internal legs, of which one carries the same $SO(M)$ index as the external legs. Hence the sum over internal indices gives an enhanced factors of $M^{q/2-1}$ compared to the original SYK model. However, this is offset by the fact that the disorder average is suppressed by a factor of $M^{q/2 -1}$ as in \eqref{eq: gaussian distribution 2}. From the $SO(M)$ invariant ansatz $\Psi_{cl}^{\alpha_1\alpha_2} \equiv \delta^{\alpha_1 \alpha_2} \psi_{cl}(\tau_1,\tau_2)$, we finally get the same Schwinger-Dyson equation as in the original SYK model, \ie
\begin{equation}
\partial_{\tau_1} \psi_{cl}(\tau_1,\tau_2) -   J^2 \int d\tau_3 [\psi_{cl}(\tau_1,\tau_3)]^{q-1} \psi_{cl}(\tau_3,\tau_2) = \delta(\tau_1-\tau_2) \ .
\end{equation}
This agrees with what we already derived in Section~\ref{sec: classical solution}, and the rest of the analysis is the same as given there. 
\subsection*{Four Point Function}

\begin{figure}
\centering 
\begin{tikzpicture}[scale=0.85]
\def\r{2} \def\v{1.7}  
\draw[ultra thick,color=black] (-\r,0) -- (\r,0);
\draw[ultra thick,color=black] (-\r,-\v) --  (\r,-\v);
\draw[thick, color=black]
{
(-\r,0) node [left] { \small{$\alpha_1,\tau_1$} } 
(-\r,-\v) node [left] { \small{$\alpha_2,\tau_2$} } 
(\r,0) node [right] { \small{$\alpha_3,\tau_3$} } 
(\r,-\v) node [right] { \small{$\alpha_4,\tau_4$} } 
(\r+2,-\v/2) node [right] { $+$ }
};

\def\c{8}
\draw[ultra thick,color=black] (\c-\r,0) -- (\c+\r,0);
\draw[ultra thick,color=black] (\c-\r,-\v) --  (\c+\r,-\v);
\draw[ultra thick,color=black] (\c,0) arc (30:-30:1.7);
\draw[ultra thick,color=black] (\c,0) arc (180-30:180+30:1.7);
\draw[dashed,color=black] (\c,0) -- (\c,-\v);
\draw[thick, color=black]
{ 
(\c-\r,0) node [left] { \small{$\alpha_1,\tau_1$} } 
(\c-\r,-\v) node [left] { \small{$\alpha_2,\tau_2$} } 
(\c+\r,0) node [right] { \small{$\alpha_3,\tau_3$} } 
(\c+\r,-\v) node [right] { \small{$\alpha_4,\tau_4$} } 
};
\end{tikzpicture}
\caption{Ladder diagrams. There are also diagrams with $(\alpha_3,\tau_3) \leftrightarrow (\alpha_4,\tau_4)$ with a relative minus sign.}
\label{fig:Ladder Diagram}
\end{figure}
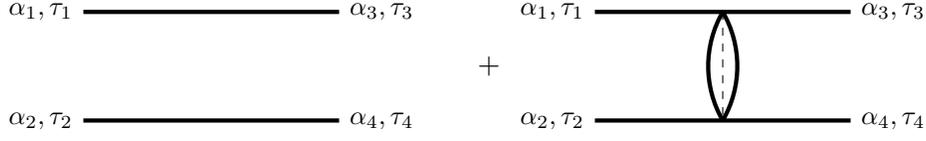

Let us start with the four point function which has the ${1 \over N}$ expansion of the form
\begin{align}
    &{1 \over N^2} \sum_{ij} \langle  \chi^{i\alpha_1}(\tau_1) \chi^{i\alpha_2}(\tau_2) \chi^{j\alpha_3}(\tau_3) \chi^{j\alpha_4}(\tau_4) \rangle \cr
    =& \Psi_{cl}^{\alpha_1\alpha_2}(\tau_1,\tau_2)  \Psi_{cl}^{\alpha_3\alpha_4}(\tau_3,\tau_4) + {1\over N} \mathcal{F}^{\alpha_1 \alpha_2 \alpha_3 \alpha_4}(\tau_1,\dots \tau_4)
\end{align}
where the first term is the leading disconnected diagram made out of two propagators. We will evaluate the second term via diagrams. As in the SYK model the diagrams contributing to $\mathcal{F}$ are the ladder diagrams built out of the propagators $\Psi_{cl}$. The first two diagrams for $\mathcal{F}$ is shown in Figure~\ref{fig:Ladder Diagram}.
Let us denote the ladder with $n$ rungs by $\mathcal{F}_n$. The first diagram in Figure~\ref{fig:Ladder Diagram} gives
\begin{equation}
    {{\cal F}_0}^{\alpha_1 \alpha_2 \alpha_3 \alpha_4}(\tau_1,\dots \tau_4) = - \delta^{\alpha_1 \alpha_3} \delta^{\alpha_2 \alpha_4} \psi_{cl}(\tau_{13}) \psi_{cl}(\tau_{24}) + \delta^{\alpha_1 \alpha_4} \delta^{\alpha_2 \alpha_3} \psi_{cl}(\tau_{14}) \psi_{cl}(\tau_{23}) 
\end{equation}
whereas the second diagram in Figure \ref{fig:Ladder Diagram} gives 
\begin{align}
       &{\mathcal{F}_1}^{\alpha_1 \alpha_2 \alpha_3 \alpha_4}(\tau_1,\dots \tau_4) \cr
       =& J^2  \left[  \delta^{\alpha_1\alpha_3} \delta^{\alpha_2\alpha_4} + {q-2 \over M} \delta^{\alpha_1\alpha_2} \delta^{\alpha_3\alpha_4} \right]\int d\tau d\tilde \tau  \psi_{cl}(\tau_1,\tau) \psi_{cl}(\tau_2,\tilde  \tau)  \psi_{cl}(\tau,\tau_3) \psi_{cl}(\tilde\tau,\tau_4)  \psi_{cl}(\tau,\tilde \tau)^{q-2}\cr
       &-\left[ (\alpha_3,\tau_3) \leftrightarrow (\alpha_4,\tau_4)\right]\label{eq:F_1 equation}\ .
\end{align}
It is useful to define a kernel by 
\begin{align}
    &K^{\alpha_1 \alpha_2 \alpha_3 \alpha_4 }(\tau_1,\tau_2,\tau_3,\tau_4) \cr
    \equiv&  -  J^2  \left[ \delta^{\alpha_1\alpha_4} \delta^{\alpha_2\alpha_3} + {q-2 \over M} \delta^{\alpha_1\alpha_2} \delta^{\alpha_3\alpha_4} \right]  \psi_{cl}(\tau_1,\tau_3) \psi_{cl}(\tau_2, \tau_4)  \psi_{cl}(\tau_3,\tau_4)^{q-2}  \label{eq: kernel from diagrammatics}\ .
\end{align}
It is also convenient to consider $K, F$ as matrices with a collective index given by the time $\tau$ and the $SO(M)$ index. Correspondingly one can define matrix multiplication as an integral over $\tau$ and a sum over $SO(M)$ index. \eg
\begin{equation}
(K\cdot \mathcal{F})(X_1,X_2,X_3,X_4)=\sum_{X_5,X_6}K(X_1,X_2,X_5,X_6)\mathcal{F}(X_6,X_5,X_3,X_4)\label{eq: matrix multiplication syk model}
\end{equation}
where $X_1,\cdots, X_6$ are the collective indices of the time $\tau$ and the $SO(M)$ index. With this compact notation, the sum of all ladder diagrams becomes just a geometric series which can be summed to obtain 
\begin{equation}
    \mathcal{F} = \sum_n {\cal F}_n = {1 \over 1 - K}   {\cal F}_0 \ .
\end{equation}
Let us note the following properties of the kernel $K$ and the first term of the geometric series $\mathcal{F}_0$ (or, the zero-rung ladder diagram).
\begin{itemize}

\item \textbf{Singlet Sector of ${\cal F}_0$ and $K$}
\begin{align}
{{\cal F}_{-,0}}(\tau_1,\tau_2,\tau_3,\tau_4)\equiv&{1\over M}\sum_{\alpha_1,\alpha_2,\alpha_3,\alpha_4} \delta^{\alpha_2\alpha_1}\delta^{ \alpha_4\alpha_3} {\mathcal{F}_{0}}^{\alpha_1\alpha_2 \alpha_3\alpha_4}(\tau_1,\tau_2,\tau_3,\tau_4)  \ ,  \\
K_-(\tau_1,\tau_2,\tau_3,\tau_4)\equiv&{1\over M} \sum_{\alpha_1,\alpha_2,\alpha_3,\alpha_4} \delta^{\alpha_2\alpha_1}\delta^{ \alpha_4\alpha_3} K^{\alpha_1\alpha_2 \alpha_3\alpha_4}(\tau_1,\tau_2,\tau_3,\tau_4) 
\end{align}
where
\begin{align}
{\cal F}_{-,0}(\tau_1,\tau_2,\tau_3,\tau_4) = &   -\psi_{cl}(\tau_{13})\psi_{cl}(\tau_{24})+   \psi_{cl}(\tau_{14})\psi_{cl}(\tau_{23}) \ ,\\
{ K}_{-}(\tau_1,\tau_2,\tau_3,\tau_4) = &  -   J^2 (q-1)     \psi_{cl}(\tau_1,\tau_3) \psi_{cl}(\tau_2, \tau_4)  \psi_{cl}(\tau_3, \tau_4)^{q-2} \ .
\end{align}
\item \textbf{Anti-symmetric Sector of ${\cal F}_0,K$}

\begin{align}
\mathcal{F}^{ab}_{\anti +,0}(\tau_1,\tau_2,\tau_3,\tau_4) \equiv& {1\over 2\ind_\anti }\sum_{\alpha_1,\alpha_2,\alpha_3,\alpha_4} (\mT^a_\anti)_{ \alpha_2\alpha_1} (\mT^b_\anti)_{ \alpha_4\alpha_3} \mathcal{F}_0^{\alpha_1\alpha_2 \alpha_3\alpha_4}(\tau_1,\tau_2,\tau_3,\tau_4)  \ ,  \\
K^{ab}_+(\tau_1,\tau_2,\tau_3,\tau_4)\equiv&{1\over 2\ind_\anti }\sum_{\alpha_3 \alpha_4}(\mT^a_\anti)_{ \alpha_2\alpha_1} (\mT^b_\anti)_{ \alpha_4\alpha_3} K^{\alpha_1\alpha_2 \alpha_3\alpha_4}  (\tau_1,\tau_2,\tau_3,\tau_4) 
\end{align}
where 
\begin{align}
      \mathcal{F}^{ab}_{\anti +,0}(\tau_1,\tau_2,\tau_3,\tau_4) \equiv &\delta^{ab} \mathcal{F}_{\anti +,0}= \delta^{ab}\left( \psi_{cl}(\tau_{13})\psi_{cl}(\tau_{24})+ \psi_{cl}(\tau_{14})\psi_{cl}(\tau_{23})\right) \ ,  \\
    K^{ab}_+(\tau_1,\tau_2,\tau_3,\tau_4) \equiv  & \delta^{ab} K_+ =  -  \delta^{ab} J^2      \psi_{cl}(\tau_1,\tau_3) \psi_{cl}(\tau_2, \tau_4)  \psi_{cl}(\tau_3, \tau_4)^{q-2}  \label{eq: diagram common ratio anti}
\end{align}
where $a,b =1,\cdots, {1\over 2}M(M-1)$ and $\ind_\anti$ is the Dynkin index of the anti-symmetric representation of $G=SO(M)$.

\item \textbf{Symmetric Sector of ${\cal F}_0,K$}

\begin{align}
\mathcal{F}^{ab}_{\sym -,0}(\tau_1,\tau_2,\tau_3,\tau_4) =& {1\over 2\ind_\sym}\sum_{\alpha_1,\alpha_2,\alpha_3,\alpha_4} (\mT^a_\sym)_{ \alpha_2\alpha_1} (\mT^b_\sym)_{ \alpha_4\alpha_3} \mathcal{F}_0^{\alpha_1\alpha_2 \alpha_3\alpha_4}(\tau_1,\tau_2,\tau_3,\tau_4)  \ ,  \\
K ^{ab}_+(\tau_1,\tau_2,\tau_3,\tau_4)=&{1\over 2\ind_\sym }\sum_{\alpha_3 \alpha_4}(\mT^a_\sym)_{ \alpha_2\alpha_1} (\mT^b_\sym)_{ \alpha_4\alpha_3} K^{\alpha_1\alpha_2 \alpha_3\alpha_4}  (\tau_1,\tau_2,\tau_3,\tau_4) 
\end{align}
where 
\begin{align}
      \mathcal{F}^{ab}_{\sym -,0}(\tau_1,\tau_2,\tau_3,\tau_4) \equiv  &\delta^{ab} \mathcal{F}_{\sym -,0}= \delta^{ab}\left( - \psi_{cl}(\tau_{13})\psi_{cl}(\tau_{24})+ \psi_{cl}(\tau_{14})\psi_{cl}(\tau_{23})\right)=\mathcal{F}_-^{ab} \ ,  \\
    K^{ab}_+(\tau_1,\tau_2,\tau_3,\tau_4) \equiv  & \delta^{ab}K_+=  -  \delta^{ab}J^2      \psi_{cl}(\tau_1,\tau_3) \psi_{cl}(\tau_2, \tau_4)  \psi_{cl}(\tau_3, \tau_4)^{q-2}   \label{eq: diagram common ratio sym}
\end{align}
where $a,b=1,\cdots, {1\over 2}M(M+1)-1$ and $\ind_\sym$ is the Dynkin index of the symmetric representation of $G=SO(M)$.

\end{itemize}
With these ingredients, one can easily see that 
\begin{equation}
\mathcal{F}_- =  \sum_n   K_-^n \mathcal{F}_{-,0}   = {1 \over 1 - K_- } \mathcal{F}_{-,0}\ .
\end{equation}
Similarly, we get
\begin{align}
\mathcal{F}^{ab}_{\anti +} = & \delta^{ab}\sum_n K_+^n \mathcal{F}_{\anti +,0}   =  \delta^{ab} {1 \over 1 - K_+ } \mathcal{F}_{\anti +,0}\ ,\\
\mathcal{F}^{ab}_{\sym -} = & \delta^{ab} \sum_n K_+^n \mathcal{F}_{\sym -,0}   =  \delta^{ab} {1 \over 1 - K_+ } \mathcal{F}_{\sym -,0}\ .
\end{align}
Note that all the results are consistent with the results found in section~\ref{sec: kernel}. 

%

\section{Chaos}
\label{sec: chaos and effective action}

\subsection{Lypunov Exponent from Effective Action}
\label{sec: effective action}

In this section, we will deduce the low energy effective action for Pseudo-Nambu-Goldstone boson of the broken reparametrization and the local $SO(M)$ symmetry by using $\epsilon$-expansion technique introduced in~\cite{Jevicki:2016bwu,Jevicki:2016ito}.

At strong coupling limit, the collective action~\eqref{eq: collective action general q} (or, the critical collective action~\eqref{eq: critical collective action general q}) has an emergent reparametrization and a local $\widehat{SO}(M)$ symmetry. They are spontaneously broken by the classical solution~\eqref{eq: classical solution}. In addition, these local symmetries are explicitly broken by the kinetic term of the collective action, which leads to an effective action for the Pseudo-Nambu-Goldstone bosons. In order to obtain the effective action, we transform the classical solution~\eqref{eq: classical solution} by the reparametrization and the local $\widehat{SO}(M)$ transformation with parameters $f(\tau)$ and $\mg(\tau)$, respectively. Then, substituting the transformed classical solution to the kinetic term which break the symmetries explicitly, we obtain the effective action of $f(\tau)$ and $\mg(\tau)$. In order to evaluate this in a controlled manner, we will use the following $\epsilon$-expansion of $q\;(\geqq 2)$ suggested in~\cite{Jevicki:2016ito}:
\begin{equation}
    q\equiv {2\over 1-\epsilon}\label{eq: epsilon and q}
\end{equation}
%
%
where the range of the corresponding $\epsilon$ is $[0,1]$. Let us consider the large $N$ classical solution:
 \begin{align}
    \maPsi_{cl}(\tau_1,\tau_2)=\coeff{\sgn(\tau_1-\tau_2)\over |\tau_1-\tau_2|^{2\over q}}\idm\ .
\end{align}
The coefficient $\coeff$ can be expressed as
\begin{equation}
    \coeff
    ={1\over J^{1-\epsilon}\alpha_{+,0}^{{1\over 2}(1-\epsilon)}}
\end{equation}
where $\alpha_{+,0}$ is defined in~\eqref{def: alpha p0} whose $\epsilon$-expansion is given by
\begin{equation}
    {1\over \alpha_{+,0}^{{1\over 2}(1-\epsilon)}}={1\over \pi  }\left(1 +\epsilon\log \pi +  {\epsilon^2\over 24}(12(\log \pi)^2 -\pi^2) +\cdots \right)\ .
\end{equation}
Under the (finite) transformation of $\diffeo \ltimes \widehat{so}(M)$ in~\eqref{eq: local transformation}, the $\epsilon$-expansion of the classical solution is
 \begin{align}
     &\maPsi_{cl,[f,\mg]}(\tau_1,\tau_2)=\coeff  {|f'(\tau_1)f'(\tau_2)|^{1\over q}\over |f(\tau_1)-f(\tau_2)|^{2\over q}} \sgn(\tau_{12}) \mg(\tau_1)\mg^{-1}(\tau_2)\cr
     =&{1\over J\alpha_{+,0}^{{1\over 2}(1-\epsilon)}} \mg(\tau_1)\mg^{-1}(\tau_2)\sgn(\tau_{12}) \left({\sqrt{f'(\tau_1)f'(\tau_2)}\over |f(\tau_1)-f(\tau_2)|}\right) \cr
     & \times \left[\underbrace{\hspace{4mm}1\hspace{4mm}}_{\text{1st}}-\underbrace{\epsilon \left(\log{\sqrt{f'(\tau_1)f'(\tau_2)}\over J|f(\tau_1)-f(\tau_2)|}\right)}_{\text{2nd}}+\underbrace{{1\over 2} \epsilon^2 \left(\log {\sqrt{f'(\tau_1)f'(\tau_2)}\over J|f(\tau_1)-f(\tau_2)|}\right)^2}_{\text{3rd}}+\cdots  \right]\label{eq: epsilon expansion of classical solution}
 \end{align}
where we did not expand $\alpha_{+,0}^{{1\over 2}(1-\epsilon)}$ in terms of $\epsilon$. We substitute the $\epsilon$-expansion of the transformed classical solution~\eqref{eq: epsilon expansion of classical solution} into the kinetic term of the collective action:
\begin{equation}
    S_{col,\text{kin}}={N\over 2}\Tr[-D\mstar \maPsi]= {N\over 2} \int d\tau_2 \; \tr\left[-\partial_{\tau_1} \maPsi(\tau_1,\tau_2) \right]_{\tau_1\rightarrow \tau_2}\ .
\end{equation}
Then, we will evaluate the effective action order by order. For this, it is convenient to define $\widehat{so}(M)$ current: 
\begin{equation}
    \mcurrent(\tau)\equiv - k \partial_\tau\mg(\tau) \cdot \mg^{-1}(\tau) 
\end{equation}
where $k$ is a constant which is not important for now, and does not appear in the final result. In components, we have
 \begin{equation}
     \mcurrent(\tau)\equiv {1\over \sqrt{2\ind_\anti}} \sum_a \current^a(\tau) \mT^a_\anti\hspace{5mm},\hspace{5mm} \current^a(\tau)\equiv {1\over \sqrt{2\ind_\anti} } \tr(\mcurrent(\tau) \mT^a_\anti)
 \end{equation}
where $\ind_\anti$ is the Dynkin index of the anti-symmetric representation, which is the same as the dual Coxeter number of $SO(M)$. Using the expansion of $\mg(\tau_1)\mg^{-1}(\tau_2)$
\begin{align}
    \mg(\tau_1)\mg^{-1}(\tau_2)=&\idm+\partial_{\tau_2}\mg(\tau_2)\cdot\mg^{-1}(\tau_2) (\tau_1-\tau_2)+{1\over 2}\partial^2_{\tau_2}\mg(\tau_2)\cdot\mg^{-1}(\tau_2) (\tau_1-\tau_2)^2+\cdots\cr
    =&\idm-{1\over k}\mcurrent(\tau_2) (\tau_1-\tau_2)+ {1\over 2k^2} \left(\mcurrent \cdot \mcurrent -k \partial_\tau \mcurrent \right)(\tau_1-\tau_2)^2+\cdots\ ,
\end{align}
%
%
%
%
%
%
the substitution of the first term in~\eqref{eq: epsilon expansion of classical solution} gives  
\begin{equation}
     (\text{1st})=-{1\over J\alpha_{+,0}^{{1\over 2}(1-\epsilon)}}\int d\tau \;\left[  {M\over 12}\left( {f'''(\tau)\over f'(\tau)}-{3\over 2}\left({f''(\tau)\over f'(\tau)}\right)^2 \right)+{1\over 2 k^2} \sum_a \current^a(\tau)\current^a(\tau)\right]\ .
\end{equation} 
Also, for the second and the third terms, we have
\begin{align}
     (\text{2nd})=&-{1\over J\alpha_{+,0}^{{1\over 2}(1-\epsilon)}}\int d\tau \;\left[ -{1\over 2 k^2} \epsilon \sum_a \current^a(\tau)\current^a(\tau)\right]\ ,\\
     (\text{3rd})=&-{1\over J\alpha_{+,0}^{{1\over 2}(1-\epsilon)}}\int d\tau \;\left[ {M\over 12}(-\epsilon^2)\left( {f'''(\tau)\over f'(\tau)}-{3\over 2}\left({f''(\tau)\over f'(\tau)}\right)^2 \right)\right]\ .
\end{align}  
As argued in~\cite{Jevicki:2016ito}, the higher order terms do not give any contribution to the effective action because $(-\log|f(\tau_1)-f(\tau_2)|)^n $ makes the contribution vanish as $\tau_1\rightarrow \tau_2$. Therefore, the effective action for $f(\tau)$ and $\mg(\tau)$ becomes 
\begin{equation}
     S_{\text{eff}}=-{ M N \alpha_{\tdiffeo}\over J}\int d\tau \;\text{Sch}(f,\tau)-{ N \alpha_{\tlie} \over J}\int d\tau\;  {1\over 2 k^2}\sum_a \current^a(\tau)\current^a(\tau)
\end{equation}  
where $\text{Sch}$ denotes the Schwarzian derivative:
\begin{equation}
    \text{Sch}(f,\tau)\equiv {f'''(\tau)\over f'(\tau)}-{3\over 2}\left({f''(\tau)\over f'(\tau)}\right)^2
\end{equation} 
Moreover, the coefficients $\alpha_{\tdiffeo}$ and $\alpha_{\tlie}$ are defined by
\begin{alignat}{2} 
    &\alpha_{\tdiffeo}\equiv {1\over 12 \alpha_{+,0}^{{1\over2}(1-\epsilon)}}(1-\epsilon^2)={(q-1)^{1-{1\over q} }\over 3q^2 \alpha_{-,0}^{1\over q}}\ ,\hspace{5mm}
    \\
    &\alpha_{\tlie}\equiv {1\over \alpha_{+,0}^{{1\over2}(1-\epsilon)}}(1+\epsilon)= {2(q-1)\over q \alpha_{+,0}^{1\over q}}
\end{alignat} 
where we expressed $\epsilon$ in terms of $q$ by using \eqref{eq: epsilon and q}, and also used $\alpha_{+,0}=(q-1)\alpha_{-,0}$ in~\eqref{def: alpha p0}. Following~\cite{Maldacena:2016hyu}, we obtain the effective action at finite temperature. For this, we perform the successive reparametrization $\tau\rightarrow \tan{\pi \tau\over \beta} \rightarrow  \tan{\pi f(\tau)\over \beta}$ using \eqref{eq: local transformation}:
\begin{align}
     &S_{\text{eff}}=-{ M N \alpha_{\tdiffeo}\over J}\int_0^\beta d\tau \text{Sch}\left(\tan{\pi f(\tau)\over \beta},\tau\right)-{ N \alpha_{\tlie} \over J}\int_0^\beta d\tau {1\over 2 k^2}\sum_a \current^a\left(f(\tau)\right)\current^a(f(\tau))\cr
     =&{ M N \alpha_{\tdiffeo}\over 2 J}\int_0^\beta d\tau \left[\left({f''\over f'}\right)^2-\left({2\pi \over \beta}\right)^2\left(f'\right)^2\right]-{ N \alpha_{\tlie} \over J}\int_0^\beta d\tau  {1\over 2 k^2}\sum_a \current^a\left(f(\tau)\right)\current^a(f(\tau))\ .
\end{align}

\paragraph{Contribution of Effective Action} Now, we will consider the contribution of the zero modes to the correlation functions of the bi-local fluctuations. For this, it is convenient to use the dimensionless angular variable~$\theta$ instead of the Euclidean time $\tau$:
\begin{equation}
    \theta\equiv{2\pi \over \beta}\tau\hspace{10mm} 0\leqq \theta\leqq 2\pi\ .
\end{equation}
In this coordinate, the large $N$ classical solution~\eqref{eq: finite temperature classical solution} at finite temperature becomes
\begin{equation}
    \maPsi(\theta)=\Lambda \left(1\over 2\sin{\theta\over 2}\right)^{2\over q}\sgn(\theta)\idm \ .\label{eq: classical solution finite temperature2}
\end{equation}
Let us consider an infinitesimal transformations of the diffeomorphism and the local $\widehat{so}(M)$:
\begin{equation}
        f(\theta) = \theta + \epsilon(\theta) \hspace{5mm},\hspace{5mm} \mg(\theta) = \idm + i\mzero(\theta) \ .
\end{equation}
and, the quadratic effective action\footnote{Note that we do not have the cross term of the reparametrization and $\hat{so}(M)$ zero modes at quadratic level~\eqref{eq: effective action infnitesimal} because the fluctuation $\mzero$ is traceless. When we consider $U(1)$ symmetry with non-zero ``spectral asymmetry''~\cite{Davison:2016ngz}, we expect to have such a cross term.} for these zero modes is given by
\begin{equation}
    S_{\text{eff}}={M N \alpha_{\tdiffeo}\over \beta J}\sum_{n=2}^\infty n^2(n^2-1) \epsilon_n \epsilon_{-n}+{N \alpha_{\tlie}\over \beta J}\sum_{n=1}^\infty n^2\zero^a_n \zero^a_{-n}\label{eq: effective action infnitesimal}
\end{equation}
where we expanded the fluctuations as follows.
\begin{equation}
    \epsilon(\theta) = {1\over 2\pi} \sum_n \epsilon_n e^{-in\theta}\quad,\quad \zero^a(\theta)={1\over 2\pi} \sum_n \zero^a_n e^{-in\theta}\ .
\end{equation}
One can easily read off the propagators of the zero modes: 
\begin{alignat}{2}
    &\langle \epsilon_n \epsilon_{-n}\rangle ={\beta J\over M N\alpha_{\tdiffeo} } {1\over n^2(n^2-1)}\hspace{10mm}&& (n=2,3,4,\cdots)\ ,\label{eq: correlation function of zero mode1}\\
    &\langle \zero^a_n \zero^b_{-n}\rangle =\delta^{ab}{\beta J\over N\alpha_{\tlie}} {1\over n^2}\hspace{10mm} &&(n=1,2,\cdots)\ .\label{eq: correlation function of zero mode2}
\end{alignat}
The corresponding zero mode eigenfunctions can be obtained from the transformation~\eqref{eq: local transformation} of the large $N$ classical solution~\eqref{eq: classical solution finite temperature2} at finite temperature by the corresponding infinitesimal transformation. The eigenfunctions corresponding to the reparametrization were already discussed in~\cite{Maldacena:2016hyu,Jevicki:2016ito,Davison:2016ngz}, which are given by
\begin{equation}
    \eigenz_n^2(\theta_1,\theta_2)\equiv {1\over \sqrt{M}}\tr\left(\delta_{\epsilon_n}\maPsi_{cl}\right)={i \sqrt{M} \over \pi q}\psi_{cl}(\theta_{12}) e^{-i{n\over 2}(\theta_1+\theta_2)}\left[{\sin{n\theta_{12}\over 2}\over \tan{\theta_{12}\over 2}}-n\cos {n\theta_{12}\over 2} \right]
\end{equation}
where $|n|\geqq 2$. Note that $\eigenz^2_0=0$ and $\eigenz^2_{\pm1}=0$ because the classical solution is invariant under $SL(2,\mathbb{R})$. 

In the same way, we will obtain the zero mode eigenfunctions related to the local $\widehat{so}(M)$ symmetry. Under the $\widehat{so}(M)$ transformation with $\mg(\theta)$, the classical solution~\eqref{eq: classical solution finite temperature2} is transformed as
\begin{equation}
    \maPsi_{cl}=\coeff\left[{1 \over 2 \sin{ \theta_{12}\over 2 }}\right]^{2\over q}\sgn(\theta_{12})\longrightarrow \; \coeff\left[{1 \over 2 \sin{ \theta_{12}\over 2 }}\right]^{2\over q}\sgn(\theta_{12})\mg(\theta_1) \mg^{-1}(\theta_2)\ .
\end{equation}
In particular, we consider the following infinitesimal transformation:
\begin{equation}
    \mg(\theta)=\idm + i\mzero(\theta) \ .
\end{equation}
Using the mode expansion
\begin{equation}
    \zero^a(\theta)={1\over \sqrt{2\ind_\anti}} \tr \left(\mzero(\theta)\mT^a_\anti\right)={1\over 2\pi} \sum_{n} \zero^a_n e^{-in\theta}\ ,
\end{equation}
one can get the zero mode eigenfunction for the $\widehat{so}(M)$:
\begin{equation}
    \eigenz_n^{1,a}(\theta_1,\theta_2)\equiv {1\over \sqrt{2\ind_\anti}}\tr \left(\delta_{\zero_n} \maPsi_{cl}\; \mT^a_\anti\right)={1\over \pi}\psi_{cl}(\theta_{12}) e^{-i{n\over 2}(\theta_1+\theta_2)}\sin {n\theta_{12}\over 2}\ .\label{eq: zero mode eigenfunction som}
\end{equation}
We explore the further properties of zero mode eigenfunctions in Appendix~\ref{app: zero mode}.

\paragraph{Singlet Channel: }First, let us consider the contribution of the effective action to the out-of-time-ordered correlation function of the singlet channel. For this, we first consider the correlation function of the bi-local fields without contraction of the $SO(M)$ index:
\begin{equation}
    \left\langle \Psi^{\alpha_1\alpha_2}(\tau_1,\tau_2)\Psi^{\alpha_3\alpha_4}(\tau_3,\tau_4)\right\rangle=\Psi_{cl}^{\alpha_1\alpha_2}(\tau_1,\tau_2)\Psi_{cl}^{\alpha_3\alpha_4}(\tau_3,\tau_4)+\mathcal{O}\left({1\over N}\right)\ .
\end{equation}
Then, we take the singlet channel of the infinitesimal transformations of the leading contribution, which capture the contribution of the zero modes.
%
%
The variation with respect to the zero mode of reparametrization gives
\begin{equation}
    {1\over M} \left\langle  \epsilon_n\epsilon_{-n}\right\rangle \tr\left(\delta_{\epsilon_n} \maPsi_{cl}(\theta_1,\theta_2) \right) \tr \left(\delta_{\epsilon_{-n}} \maPsi_{cl}(\theta_3,\theta_4)\right)\label{eq: big contribution to singlet}\ .
\end{equation}
This is exactly the same as that of the original SYK model, and was evaluated in~\cite{Maldacena:2016hyu}. For the out-of-time-ordered correlator, we take the specific ordering of $\theta$'s following~\cite{Maldacena:2016hyu}:
\begin{equation}
    \theta_1\;<\; \theta_3=0 \;<\;\theta_2\;<\; \theta_4=\pi\ ,\label{eq: ordering of theta}
\end{equation}
and evaluate \eqref{eq: big contribution to singlet}. Then, by taking analytic continuation 
\begin{equation}
    \theta_1=-{\pi\over 2} -{2\pi i \over \beta}t\hspace{5mm},\hspace{5mm} \theta_2={\pi\over 2} -{2\pi i \over \beta}t\ ,\label{eq: analytic continuation to chaos}
\end{equation}
one can repeat the same calculation as in~\cite{Maldacena:2016hyu}:
\begin{equation}
    {\mathcal{F}_{-}^\epsilon (t)\over \psi_{cl}(\pi)\psi_{cl}(\pi)}={\beta J\over \pi^2 q^2\alpha_{\tdiffeo}}\left(1-{\pi \over 2} \cosh {2\pi t\over \beta}\right)\hspace{2mm}\xrightarrow{\hspace{1mm}t\rightarrow \infty \hspace{1mm}}\hspace{2mm} {\mathcal{F}^\epsilon_{-}(t)\over \psi_{cl}(\pi)\psi_{cl}(\pi)}\approx -{\beta J\over 4\pi q^2\alpha_{\tdiffeo}} e^{{2\pi \over \beta}t}
\end{equation}
As expected, it saturates the chaos bound $\lambda_L^{\text{\tiny singlet}}={2\pi \over\beta}$. On the contrary, one can easily see that there is no contribution of the $\hat{so}(M)$ effective action to the singlet channel because the variation with respect to $\zero_n$ is traceless. \ie
\begin{equation}
    {1\over M} \left\langle  \zero_n\zero_{-n}\right\rangle \tr\left(\delta_{\zero_n} \maPsi_{cl}(\theta_1,\theta_2) \right) \tr \left(\delta_{\zero_{-n}} \maPsi_{cl}(\theta_3,\theta_4)\right)=0\ ,
\end{equation}
and therefore, we have
\begin{equation}
    \mathcal{F}_{-}^\zero (t)=0\ .
\end{equation}

\paragraph{Anti-symmetric Channel: }In the same way, we analyze the contribution of the zero modes to the anti-symmetric irrep channel:
\begin{equation}
    {1\over 2\ind}\left\langle\tr\left[ \maPsi(\theta_1,\theta_2)\mT^a_\anti\right]\tr \left[\maPsi(\theta_3,\theta_4)\mT^b_\anti\right] \right\rangle\ .
\end{equation}
The zero mode of the reparametrization does not give any contribution to the anti-symmetric irrep channel because the variation with respect to $\epsilon$ is proportional to the identity matrix. \ie
\begin{equation}
    {1\over 2\ind} \left\langle  \epsilon_n\epsilon_{-n}\right\rangle \tr (\delta_{\epsilon_n} \maPsi_{cl}(\theta_1,\theta_2) \mT^a_\anti  ) \tr  (\delta_{\epsilon_{-n}} \maPsi_{cl}(\theta_3,\theta_4)\mT^b_\anti )=0\ .
\end{equation}
Hence, we have 
\begin{equation}
    \mathcal{F}_{\anti +}^\epsilon (t)=0\ .
\end{equation}

The non-trivial contribution to the anti-symmetric irrep(adjoint) channel comes from the $\hat{so}(M)$ effective action:
\begin{equation}
    {1\over 2\ind_\anti } \left\langle  \zero_n\zero_{-n}\right\rangle \tr (\delta_{\zero_n} \maPsi_{cl}(\theta_1,\theta_2) \mT^a_\anti  ) \tr  (\zero_{\epsilon_{-n}} \maPsi_{cl}(\theta_3,\theta_4)\mT^b_\anti )\ .
\end{equation}
Hence, using \eqref{eq: zero mode eigenfunction som} and \eqref{eq: correlation function of zero mode2}, we obtain
\begin{equation}
     {\mathcal{F}_{\anti +}^{\zero,ab}(\theta_1,\theta_2,\theta_3,\theta_4)\over \psi_{cl}(\theta_{12})\psi_{cl}(\theta_{34})}=-\delta^{ab}{\beta J\over \pi^2\alpha_\tlie}\sum_{|n|\geqq 1} { e^{{in\over 2}(\theta_3+\theta_4-\theta_1-\theta_2)} \over n^2} \sin {n\theta_{12}\over 2}\sin {n\theta_{34}\over 2} \ .
\end{equation}
With the particular ordering of $\theta$'s in \eqref{eq: ordering of theta}, we evaluate the summation, and then take the analytic continuation~\eqref{eq: analytic continuation to chaos}:
\begin{align}
 &{\mathcal{F}_{\anti +}^{\zero,ab}(t)\over \psi_{cl}(\pi)\psi_{cl}(\pi)}\cr
 =&\delta^{ab}{2i\beta J\over \pi^2\alpha_\tlie}\left[e^{{2\pi\over \beta}t}{}_3F_2\left({1\over 2},{1\over 2},1;{3\over 2},{3\over 2}, -e^{{4\pi \over \beta}t}\right)-e^{-{2\pi\over \beta}t}{}_3F_2\left({1\over 2},{1\over 2},1;{3\over 2},{3\over 2}, -e^{-{4\pi \over \beta}t}\right)\right]\ .
\end{align}
As $t\longrightarrow \infty$, one can see that there is no exponential growth. 
\begin{equation}
     {\mathcal{F}_{\anti +}^{\zero,ab}(t)\over \psi_{cl}(\pi)\psi_{cl}(\pi)}\approx \delta^{ab}{2 i \over \alpha_\tlie} Jt\ .
\end{equation}

\paragraph{Symmetric Channel: } It is easy to see that there is no contribution of the low energy effective action to the symmetric irrep channel since the variation of the classical solution is either the singlet or the anti-symmetric irrep:
\begin{equation}
\tr  (\delta_{\epsilon_n}  \maPsi_{cl}(\theta_1,\theta_2) \mT^a_\sym  )=\tr  (\delta_{\zero_n} \maPsi_{cl}(\theta_1,\theta_2) \mT^a_\sym  )=0\ ,
\end{equation}
and therefore, we have
\begin{equation}
\mathcal{F}_{\sym -}^{\epsilon,ab}=\mathcal{F}_{\sym -}^{\zero,ab}=0 \ .
\end{equation}

\paragraph{Summary} We summarize the zero mode contribution to the long time behavior of the out-of-time-ordered correlators for the singlet, anti-symmetric and symmetric irrep channels:
\begin{alignat}{2}
    &{\mathcal{F}^\epsilon_{-}(t)\over \psi_{cl}(\pi)\psi_{cl}(\pi)}\approx -{\beta J\over 4\pi q^2\alpha_{\tdiffeo}} e^{{2\pi \over \beta}t}\hspace{5mm}&&,\qquad \mathcal{F}_{-}^\zero (t)=0\ , \\
    &\mathcal{F}_{\anti +}^{\epsilon,ab} (t)=0  \hspace{5mm} &&,\qquad {\mathcal{F}_{\anti +}^{\zero,ab}(t)\over \psi_{cl}(\pi)\psi_{cl}(\pi)}\approx \delta^{ab}{2 i \over \alpha_\tlie} Jt \ ,\\
    &\mathcal{F}_{\sym -}^{\epsilon,ab} (t)=0  \hspace{5mm} &&,\qquad \mathcal{F}_{\sym -}^{\zero,ab}(t)=0 \ .
\end{alignat}
Here, we found that the $\widehat{so}(M)$ zero mode does not give an exponentially growing contribution to the anti-symmetric irrep channel. In the next section, we will see that the anti-symmetric/symmetric channels do not have any exponential growth even if we take all contributions into account up to quadratic level. Hence, we conclude that
\begin{equation}
    \lambda_L^{\text{\tiny singlet}}={2\pi \over\beta}\hspace{5mm},\hspace{5mm}\lambda_L^{\anti}=0\hspace{5mm},\hspace{5mm}\lambda_L^{\sym}=0\ .
\end{equation}
\subsection{Out-of-time-ordered Correlators by Analytic Continuation}
\label{sec: lyapunov exponent}

In this section, we will analyze the long time behavior of the out-of-time-order correlators from the non-zero modes which we evaluated in Section~\ref{sec: kernel}. The singlet fluctuation~$\fluca$ is the same as that of the original SYK model, and the out-of-time-order correlators of the non-zero modes were already studied in~\cite{Maldacena:2016hyu} by analytic continuation of the Euclidean correlator. Following~\cite{Maldacena:2016hyu}, we find the long time behavior of the out-of-time correlator of the anti-symmetric/symmetric channels. 

\paragraph{Singlet Channel: }Following \cite{Maldacena:2016hyu}, we will consider the following type of the out-of-time-ordered correlator for the singlet channel:
\begin{equation}
     {1\over M}\sum_{\alpha_1,\cdots,\alpha_4}\sum_{i,j}\delta^{\alpha_2\alpha_1}\delta^{\alpha_4\alpha_3}\TR \left[ y \psi^{i \alpha_1}(t)y \psi^{j \alpha_3}(0)y \psi^{i \alpha_2}(t)y \psi^{j \alpha_4}(0)\right]
\end{equation}
where $y\equiv\rho(\beta)^{1\over4}$ and $\rho(\beta)$ is the thermal density matrix. This correlators can be evaluated by analytic continuation of the Euclidean correlator. It was shown in \cite{Maldacena:2016hyu} that this corresponds to the Euclidean correlators with the cross ratio
\begin{equation}
    \chi={2\over 1- i \sinh{2\pi t\over \beta}}\label{eq: analytic continuation of chi}\ .
\end{equation}
Furthermore, since $\chi=2$ for $t=0$ and $\chi\sim e^{-{2\pi t\over\beta}}$ for large $t$, one has to perform the analytic continuation of the Euclidean correlators for $\chi>1$ to $\chi<1$ for large $t$. The out-of-time-ordered correlator of the singlet fluctuation is the same as that of the original SYK model~\cite{Maldacena:2016hyu}:
\begin{equation}
    \mathcal{F}_{-,h\ne2}\sim te^{{2\pi \over \beta }t}\ ,
\end{equation}
and, this corresponds to the ${1\over \beta J}$ correction to the maximal Lyapunov exponent ${2\pi \over \beta}$ of the singlet fluctuation which we evaluated from the effective action of $h=2$ mode in Section~\ref{sec: effective action}.

\paragraph{Anti-symmetric Channel: }For the out-of-time-ordered correlator of the anti-symmetric channels, we consider
\begin{align}
     {1\over 2\ind_\anti}\sum_{\alpha_1,\cdots,\alpha_4}\sum_{i,j}(\mT^a_\anti)^{\alpha_2\alpha_1}(\mT^b_\anti)^{\alpha_4\alpha_3}\TR \left[ y \psi^{i \alpha_1}(t)y \psi^{j \alpha_3}(0)y \psi^{i \alpha_2}(t)y \psi^{j \alpha_4}(0)\right]\ .\label{def: adjoint otoc 1}
\end{align}
The relevant expression of the Euclidean correlator for this analysis is given in~\eqref{eq: correlation intermediate}
\begin{align}
    {\mathcal{F}_{\anti  +,h\ne1}(\chi)\over \alpha_{+,0}}=&\left.-\int_{-\infty}^\infty {ds\over 2\pi }{h-{1\over 2}\over \pi  }\tan {\pi h\over 2}{k_{\anti +}(h)\over 1-k_{\anti +}(h)}\Phi_h^+(\chi)\right|_{h={1\over 2}+is}\cr
    & - \sum_{n=2}^\infty \mbox{Res}\left[ {h-{1\over 2}\over \pi}\tan{\pi h\over 2} {k_{\anti +}(h)\over 1-k_{\anti +}(h)}\Phi_h^+(\chi)\right]_{h=2n-1}\ .\label{eq: otoc integration expression}
\end{align}
For the analytic continuation, we first simplify the residue sum by introducing a function $k_{+,R}$ defined by
\begin{equation}
    {k_{\anti +,R}(1-h)\over k_{\anti +}(h)}\equiv - {\sin \pi\left({1\over q}-{h\over 2}\right)\over \sin \pi\left({1\over q}+{h\over 2}\right) }\ ,
\end{equation}
or, equivalently, it is given by
\begin{equation}
    k_{\anti +,R}(1-h)\equiv{\Gamma\left(2-{2\over q}\right)\Gamma\left(h-1+{2\over q} \right)\over \Gamma\left({2\over q}\right)\Gamma\left(h+1-{2\over q}\right)}\ .\label{eq: retarded kernel eigenvalue anti}
\end{equation}
Note that $k_{\anti +}(h)$ is identical to $k_{+,R}(1-h)$ for $h=2n+1$ $(n=1,2,\cdots)$:
\begin{equation}
    \left.{k_{\anti +,R}(1-h)\over k_{\anti +}(h)}\right|_{h=2n+1}=1\ .
\end{equation}
Therefore, one can replace $k_{\anti +}(h)$ inside of the residue sum with $k_{\anti +,R}(1-h)$. Then, we change the residues into the contour integrals around the $h=2n-1$ ($n=2,3,\cdots$), and push those contours back to the the contour along $h={1\over 2}+is$ ($s\in\mathbb{R}$). This pick up one pole at $h=1$ which did not included in the residue sum of~\eqref{eq: otoc integration expression}. Therefore, we have
\begin{align}
    {\mathcal{F}_{\anti +,h\ne1}(\chi)\over \alpha_{+,0}}=&-\int_{-\infty}^\infty {ds\over 2\pi }\left.{h-{1\over 2}\over \pi  }\tan {\pi h\over 2}\left[{k_{\anti +}(h)\over 1-k_{\anti +}(h)} -{k_{\anti +,R}(1-h)\over 1-k_{+,R}(1-h)}\right]\Phi_h^+(\chi)\right|_{h={1\over 2}+is}\hspace{-5mm}\cr
    &+ \mbox{Res}\left[ {h-{1\over 2}\over \pi}\tan{\pi h\over 2} {k_{\anti +,R}(1-h)\over 1-k_{\anti +,R}(1-h)}\Phi_h^+(\chi)\right]_{h=1}\ .
\end{align}
Using the expression of $\Phi^+_h(\chi)$ for $\chi>1$ in~\eqref{eq: phi p greater}, we consider the analytic continuation to $\chi<1$. As $\chi\rightarrow 0$, $\Phi_h^+(\chi)$ behaves like 
%
%
%
%
\begin{equation}
    \Phi_h^+(\chi)\simeq   - i  {\Gamma\left(1-{h\over 2}\right) \Gamma\left(h-{1\over 2}\right)\over 2^{1-h}\Gamma\left({1\over 2}+{h\over 2}\right)} (-i\chi)^{h}+\left(h\longrightarrow 1-h\right)\ .
\end{equation}
%
%
Noting that 
\begin{equation}
    \chi={2\over 1-i \sinh{2\pi t\over \beta}}\sim e^{-{2\pi t\over \beta}}\hspace{10mm} (t\rightarrow \infty)\ ,
\end{equation}
one can easily see that the contour integral along $h={1\over 2}+is$ does not have an exponential growth as $\chi\rightarrow 0$. In addition, the residue at (double) pole $h=1$ also does not grow exponentially. At most, it grows $-\log(-i\chi)\sim t$ because of the double pole. \ie
\begin{equation}
    \mathcal{F}_{\anti +,h\ne1}(\chi)\sim  {t\over \beta}\ .
\end{equation}
The effective action analysis showed that the $h=1$ mode also give the linear growth of the out-of-time ordered correlator in Section~\ref{sec: effective action}. In this section, we also found that the contribution of the non-zero modes grows linearly. 

\paragraph{Symmetric Channel: }For the out-of-time-ordered correlator of the symmetric channel
\begin{align}
     {1\over 2\ind_\sym}\sum_{\alpha_1,\cdots,\alpha_4}\sum_{i,j}(\mT^a_\sym)^{\alpha_2\alpha_1}(\mT^b_\sym)^{\alpha_4\alpha_3}\TR \left[ y \psi^{i \alpha_1}(t)y \psi^{j \alpha_3}(0)y \psi^{i \alpha_2}(t)y \psi^{j \alpha_4}(0)\right]\label{def: adjoint otoc 2}\ ,
\end{align}
we will consider the following Euclidean correlator
\begin{align}
    {\mathcal{F}_{\sym  - }(\chi)\over \alpha_{-,0}}=&\left.-\int_{-\infty}^\infty {ds\over 2\pi }{h-{1\over 2}\over \pi  \tan {\pi h\over 2}}{k_{\sym -}(h)\over 1-k_{\sym -}(h)}\Phi^-_h(\chi)\right|_{h={1\over 2}+is}\cr
    & - \sum_{n=1}^\infty \mbox{Res}\left[ {h-{1\over 2}\over \pi \tan{\pi h\over 2}} {k_{\sym -}(h)\over 1-k_{\sym -}(h)}\Phi_h^-(\chi)\right]_{h=2n}\ ,\label{eq: otoc integration expression2}
\end{align}
and we will perform the analytic continuation in \eqref{eq: analytic continuation of chi}. Note that \eqref{eq: otoc integration expression2} is the same as that of the singlet channel (or, the original SYK model) except that the contribution of $h=2$ eigenfunction is included because it does not diverge. Therefore, in a similar way, we define $k_{\sym -,R}(1-h)$ such that
\begin{equation}
    {k_{\sym -,R}(1-h)\over k_{\sym -}(h)}\equiv {\cos \pi\left({1\over q}-{h\over 2}\right)\over \cos \pi\left({1\over q}+{h\over 2}\right) }\ .
\end{equation}
Explicitly, one can see that $k_{\sym -,R}(1-h)$ is the same as $k_{\anti +,R}(1-h)$:
\begin{equation}
    k_{\sym -,R}(1-h)\equiv{\Gamma\left(2-{2\over q}\right)\Gamma\left(h-1+{2\over q} \right)\over \Gamma\left({2\over q}\right)\Gamma\left(h+1-{2\over q}\right)}=k_{\anti +,R}(1-h)\ .\label{eq: retarded kernel eigenvalue sym}
\end{equation}
As before, we replace $k_{\sym -}(h)$ with $k_{\sym -,R}(1-h)$ inside of the residue, and we push the contour around the poles to the contour $h={1\over 2}+is$ ($s\in\mathbb{R}$). Then, we have
\begin{align}
    {\mathcal{F}_{\sym -}(\chi)\over \alpha_{0,-}}=&-\int_{-\infty}^\infty {ds\over 2\pi }\left.{h-{1\over 2}\over \pi  \tan {\pi h\over 2}}\left[{k_{\sym -}(h)\over 1-k_{\sym -}(h)} -{k_{\sym -,R}(1-h)\over 1-k_{\sym -,R}(1-h)}\right]\Phi_h^-(\chi)\right|_{h={1\over 2}+is}\hspace{-5mm}\cr
    &+ \mbox{Res}\left[ {h-{1\over 2}\over \pi \tan{\pi h\over 2}} {k_{\sym -,R}(1-h)\over 1-k_{\sym -,R}(1-h)}\Phi_h^-(\chi)\right]_{h=1}\ .
\end{align}
Note that the function inside of the residue has a simple pole at $h=1$ unlike the anti-symmetric channel where the analogous function has a double pole at $h=1$. Therefore, using the asymptotic behavior ($\chi\rightarrow 0$) of $\Phi_h^-(\chi)$ in~\eqref{eq: phi m greater}
\begin{equation}
    \Phi_h^-(\chi)\simeq   - i  {\Gamma\left({1\over 2} -{h\over 2}\right) \Gamma\left(h-{1\over 2}\right)\over 2^{1-h}\Gamma\left({h\over 2}\right)} (-i\chi)^{h}+\left(h\longrightarrow 1-h\right)\ ,
\end{equation}
both contour integral and the residue at the simple pole $h=1$ give at most a constant (or, exponential decay) in the real time $t$. \ie
\begin{equation}
	\mathcal{F}_{\sym -}(t) \sim (\text{constant})\ .
\end{equation}
This is also consistent with the low energy effective action analysis where we have shown that there is no contribution of the low energy effective action to the symmetric irrep channel in Section~\ref{sec: effective action}.

\paragraph{Summary: }We summarize the long time behavior of the out-of-time-ordered correlators from the zero mode contribution in Section~\ref{sec: effective action} and the non-zero mode contributions found in this section:
\begin{alignat}{4}
	&\mbox{Singlet}\quad &&: \hspace{5mm}\mathcal{F}_{-}\sim &&\underbrace{{\beta J}e^{{2\pi \over \beta}t }}_{h=2\; \text{zero mode}}&&+\underbrace{{t\over \beta} e^{{2\pi \over \beta}t}}_{\text{non-zero modes}}\ ,\\
	&\mbox{Anti-symmetric}\quad &&: \hspace{5mm}\mathcal{F}_{\sym +}\sim &&\underbrace{Jt}_{h=1 \; \text{zero mode}}&&+\underbrace{{t\over \beta} }_{\text{non-zero modes}}\ ,\\
	&\mbox{Symmetric}\quad &&: \hspace{5mm}\mathcal{F}_{\anti -}\sim &&\;\;\underbrace{(\mbox{constant})}_{\text{non-zero modes}}
\end{alignat}
where we omitted unnecessary numerical factors.

\subsection{Retarded Kernel}
\label{sec: retarded kernel}

In this section, we will utilize the real time propagators of two point function of the fermions to study the out-of-time-ordered correlators following~\cite{kitaevfirsttalk,Maldacena:2016hyu} instead of analytic continuation of the Euclidean result. We also consider similar out-of-time-ordered correlators to~\eqref{def: adjoint otoc 1} and \eqref{def: adjoint otoc 2} for the anti-symmetric/symmetric fluctuations
\begin{align}
    F_{\anti +}(t_1,t_2)\equiv&{1\over 2\ind_\anti}(T^a_\anti)^{\alpha_2\alpha_1}(T^b_\anti)^{\alpha_4\alpha_3}\TR \left[ y \psi^{i \alpha_1}(t_1)y \psi^{j \alpha_3}(0)y \psi^{i \alpha_2}(t_2)y \psi^{j \alpha_4}(0)\right]\ ,\\
    F_{\sym -}(t_1,t_2)\equiv&{1\over 2\ind_\sym}(T^a_\sym)^{\alpha_2\alpha_1}(T^b_\sym)^{\alpha_4\alpha_3}\TR \left[ y \psi^{i \alpha_1}(t_1)y \psi^{j \alpha_3}(0)y \psi^{i \alpha_2}(t_2)y \psi^{j \alpha_4}(0)\right]
\end{align}
where $y\equiv [\rho(\beta)]^{1\over 4}$ and $\rho(\beta)$ is the thermal density of states, and we omitted summations over $\alpha_1,\cdots, \alpha_4$ and $i, j$. Recall that both anti-symmetric and symmetric irrep channels have the common ratio of their geometric series of the ladder diagrams (See \eqref{eq: diagram common ratio anti},  \eqref{eq: diagram common ratio sym} or \eqref{def: kernel p}). For the configuration of the ordering in the two real-time folds, the relevant retarded kernel for both four point functions is given by
\begin{equation}
    K_{+,R}(t_1,t_2,t_3,t_4)=J^2\Psi_{cl,R}(t_{13})\Psi_{cl,R}(t_{24})\left[\Psi_{cl,lr}(t_{34})\right]^{q-2}
\end{equation}
where $\Psi_{cl,R}(t)$ and $\Psi_{cl,lr}(t)$ are the retarded propagator and the Wightman propagator, respective, which one can obtain by using appropriate analytic continuation~\cite{Maldacena:2016hyu,Peng:2017kro}:
\begin{align}
    \Psi_{cl,R}(t)=&2\coeff\cos \pi\Delta  \left({\pi\over \beta\sinh {\pi t\over \beta} }\right)^{2\Delta}\theta(t)\ ,\\
    \Psi_{cl,lr}(t)=&\coeff\left( {\pi \over \beta \cosh {\pi t\over \beta}}\right)^{2\Delta}\ .
\end{align}
%
%
%
%
%
In order to find the exponential growth, we utilize the following integral equation~\cite{Maldacena:2016hyu}.
\begin{equation}
    \int dt_3dt_4\; K_{+,R}(t_1,t_2,t_3,t_4)F_{\anti/\sym}(t_3,t_4)=F_{\anti/\sym}(t_1,t_2)
\end{equation}
which means that the exponential growing part of the sum of the ladder diagrams does not change by attaching additional rungs. Thus, let us consider the following eigenvalue problem:
\begin{equation}
    K_{+,R}F_{\anti/\sym}= k_{+,R}(h) F_{\anti/\sym}
\end{equation}
where we will find the eigenfunction $F_{\anti/\sym}$ corresponding to the eigenvalue $k_{+,R}(h)=1$. Using the exponentially growing ansatz~\cite{Maldacena:2016hyu}
%
%
\begin{equation}
    F_+(t_1,t_2)={e^{-{h\pi \over \beta}(t_1+t_2)}\over \left[\cosh {\pi  t_{12}\over \beta }\right]^{{2\over q}-h}}\ ,
\end{equation}
we find the eigenvalue to be
\begin{align}
    k_{+,R}(h)
    ={\Gamma\left(2-{2\over q}\right)\Gamma\left({2\over q}-h\right) \over \Gamma\left(2\over q\right) \Gamma\left(2-{2\over q}-h\right)}=k_{\anti +,R}(h)=k_{\sym -,R}(h)={1\over q-1} k_{-,R}(h)\ .\label{eq: eigenvalue of retarded kernel}
\end{align}
Note that $k_{+,R}(h)$ is the same as $k_{\anti +,R}(h)$ and $k_{\sym -,R}(h)$ which were found in the evaluation of the non-zero mode contributions of the anti-symmetric channel~\eqref{eq: retarded kernel eigenvalue anti} and the symmetric channel~\eqref{eq: retarded kernel eigenvalue sym}. In addition, recall that the kernel $K_+$ for the anti-symmetric/symmetric channels is equal to ${1\over q-1}K_-$ where $K_-$ is the kernel of the singlet fluctuation. Hence, the corresponding retarded kernels have the same relation. Since we used the same ansatz as in~\cite{Maldacena:2016hyu}, it is not surprising to have the relation between $k_{+,R}$ and $k_{-,R}$ in \eqref{eq: eigenvalue of retarded kernel}.

It is enough to consider solutions of $k_{+,R}(h)=1$ for $h\leqq 0$ because positive $h$'s do not give exponential growth of the $F_+(t_1,t_2)$. First, one can easily see that $h=0$ is a solution:
\begin{equation}
    k_{+,R}(0)=1\ ,
\end{equation}
and, since the function $k_{+,R}$ is increasing function for $h\leqq 0$, $h=0$ is the only solution for $k_{+,R}(h)=1$ $(h\leqq0)$. Therefore, there is no exponential growth for the out-of-time-ordered correlator of the anti-symmetric/symmetric channels, which is consistent with the results from the analysis in the previous sections. The same result for $U(1)$ case of the complex SYK model was also found in~\cite{Bulycheva:2017uqj}. In order to see the linear growth in the anti-symmetric/symmetric channel from the retarded kernel, one need to consider a different type of eigenfunctions. We leave this for future work.

\newpage

\section{Tensor Model with Global Symmetry}
\label{sec: tensor model}

It has been shown that tensor models also feature the dominance of the ``melonic'' diagrams in large $N$. In~\cite{Witten:2016iux,Gurau:2016lzk}, it was shown that a new class of tensor model (Gurau-Witten (GW) model) reproduces interesting behaviors of the SYK model such as the saturation chaos bound. Also, \cite{Klebanov:2016xxf} suggested the simplest (SYK-like) tensor model (Klebanov-Tarnopolsky (KT) model). The lattice generalizations thereof were investigated in~\cite{Narayan:2017qtw} which developed general techniques to analyze a broad class of tensor models. For example, \cite{Narayan:2017qtw} showed that the GW model can be understood as a KT model on the 4-site lattice after ``uncoloring'' process. However, this analysis is not restricted to the lattice generalizations. One can add any additional structure to the tensor model, and the same analysis can be applied straightforwardly. In particular, we add $SO(M)$ index to the tensors, and we follow~\cite{Narayan:2017qtw} to study tensor model with $SO(M)$ global symmetry.

Let us consider the simplest uncolored $O(N)$ tensor model with $SO(M)$ global symmetry of which Hamiltonian is defined by
\begin{equation}
    H={J N^{-{3\over 2}} \over 4}  \sum_{\alpha_1,\alpha_2=1}^{M} \chi^{\alpha_1}_{i_1 j_1 k_1}\chi^{\alpha_1}_{i_1 j_2 k_2}\chi^{\alpha_2}_{i_2 j_1 k_2}\chi^{\alpha_2}_{i_2 j_2 k_1}\label{def: hamiltonian q4}
\end{equation}
where $\chi^{\alpha}_{i j k}$ denotes $MN^3$ Majorana fermions which transform in the tri-fundamental representation of $O(N)$~($i,j,k=1,2,\cdots, N$) and in the fundamental representation of $SO(M)$~($\alpha=1,2,\cdots, M$). Since the three $O(N)$ indices are distinguishable, we label them by RGB color $\col$. (\eg The first index $i$ corresponds to red color, etc.)

\begin{figure}[t!]
\centering
\begin{tikzpicture}[scale=1]
\draw[thick,color=red] (-1.5,0.5) -- (-0.5,0.5) -- (-0.5,1.5);
\draw[thick,color=blue] (-1.5,-0.5) -- (-0.5,-0.5) -- (-0.5,-1.5);
\draw[thick,color=red] (1.5,-0.5) -- (0.5,-0.5) -- (0.5,-1.5);
\draw[thick,color=blue] (1.5,0.5) -- (0.5,0.5) -- (0.5,1.5);

\draw[thick,color=black!50!green] (-1.5,0.0) -- (1.5,0.0) ;
\draw[thick,color=black!50!green] (0.0,1.5) -- (0.0,0.2) ;
\draw[thick,color=black!50!green] (0.0,-1.5) -- (0.0,-0.2) ;
\draw[color=blue] (0.0,0.2) arc (90:-90:0.2);
\draw[thick, color=black]
{
(0.0,-2) node [below] {\small{$ \Jonsite N^{-{3 \over 2}}=JN^{-{3\over2}}$}}
(-1.5,0) node [left] {\small{$ \alpha  $}}
(2.0,0) node [left] {\small{$ \alpha  $}}
(0.3,1.7) node [left] {\small{$ \alpha  $}}
(0.3,-1.7) node [left] {\small{$ \alpha  $}}
};
\end{tikzpicture}\hspace{10mm}
\begin{tikzpicture}[scale=1]
\draw[thick,color=red] (-1.5,0.5) -- (-0.5,0.5) -- (-0.5,1.5);
\draw[thick,color=blue] (-1.5,-0.5) -- (-0.5,-0.5) -- (-0.5,-1.5);
\draw[thick,color=red] (1.5,-0.5) -- (0.5,-0.5) -- (0.5,-1.5);
\draw[thick,color=blue] (1.5,0.5) -- (0.5,0.5) -- (0.5,1.5);

\draw[thick,color=black!50!green] (-1.5,0.0) -- (1.5,0.0) ;
\draw[thick,color=black!50!green] (0.0,1.5) -- (0.0,0.2) ;
\draw[thick,color=black!50!green] (0.0,-1.5) -- (0.0,-0.2) ;
\draw[color=blue] (0.0,0.2) arc (90:-90:0.2);
\draw[thick, color=black]
{
(0.0,-1.85) node [below] {\small{$ {1\over \sqrt{2}}J_\R N^{-{3 \over 2}}=J N^{-{3 \over 2}}$ ($\alpha_1\ne \alpha_2$)}}
(-1.5,0) node [left] {\small{$ \alpha_1  $}}
(2.35,0) node [left] {\small{$ \alpha_2  $}}
(0.3,1.7) node [left] {\small{$ \alpha_1  $}}
(0.25,-1.7) node [left] {\small{$ \alpha_2  $}}
};
\end{tikzpicture} 
\caption{The Vertices of the on-site and (non-local) hopping interactions.}
\label{fig:vertex}
\end{figure}
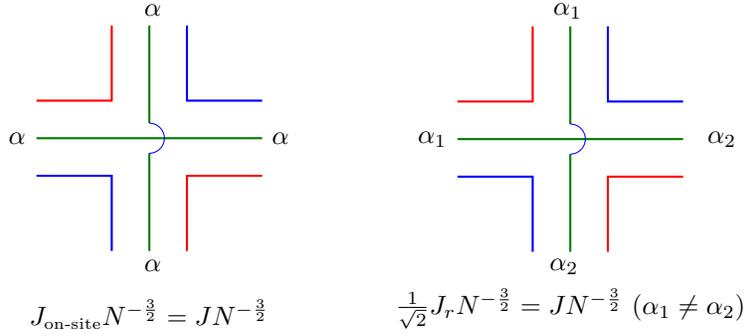

One can see that the form of the Hamiltonian in \eqref{def: hamiltonian q4} is similar to that of the tensor model on the lattice if one think of the $SO(M)$ index as the lattice index. Hence, we can use the analysis carried out in~\cite{Narayan:2017qtw} in this model. To contact with~\cite{Narayan:2017qtw}, we need to write \eqref{def: hamiltonian q4} in the same form as that of~\cite{Narayan:2017qtw}.
\begin{align}
    H=&{J N^{-{3\over 2}}\over  4}  \sum_{\alpha=1}^{M} \chi^{\alpha}_{i_1 j_1 k_1}\chi^{\alpha}_{i_1 j_2 k_2}\chi^{\alpha}_{i_2 j_1 k_2}\chi^{\alpha}_{i_2 j_2 k_1}+{J N^{-{3\over 2}}\over 2 }  \sum_{\alpha_1>\alpha_2} \chi^{\alpha_1}_{i_1 j_1 k_1}\chi^{\alpha_1}_{i_1 j_2 k_2}\psi^{\alpha_2}_{i_2 j_1 k_2}\chi^{\alpha_2}_{i_2 j_2 k_1}\label{eq: hamiltonian lattice}\ .
\end{align}
Here, $SO(M)$ invariance is not manifest, but one can directly use the result of~\cite{Narayan:2017qtw}. Comparing to the Hamiltonian in~\cite{Narayan:2017qtw}, \eqref{eq: hamiltonian lattice} can be interpreted as a tensor model on the lattice with non-local hopping interactions. Moreover, the coupling constants of the on-site and hopping interactions are given by
\begin{equation}
    J_{\text{on-site}}=J\hspace{5mm},\hspace{5mm}J_\R= \sqrt{2}J\hspace{5mm},\hspace{5mm} J_\G=J_\B=0\ .
\end{equation}
Note that $\sqrt{2}$ in front of $J_\R$ appears because we follow the normalization of the hopping coupling constant in~\cite{Narayan:2017qtw}. However, this $\sqrt{2}$ factor is cancelled in the Feynmann diagram since the contribution of these vertices to the Feynman diagram is $J_{\text{on-site}}=J$ and ${1\over \sqrt{2}} J_\R= J$, respectively. See the corresponding interaction vertices in Figure~\ref{fig:vertex}. Moreover, we emphasize that the $SO(M)$ index is connected to each other along the red line. Hence, $J_\R$ is the only non-zero hopping interaction constant.

\begin{figure}[t!]
\centering
\begin{tikzpicture}[scale=1]
\draw[thick,color=red] (-2,0.3) -- (-1,0.3);
\draw[thick,color=red] (-1,0.3) arc (170:10:1.04403);
\draw[thick,color=red] (2,0.3) -- (1,0.3); 

\draw[thick,color=blue] (-0.5,0.3) -- (0.5,0.3);
\draw[thick,color=blue] (-0.5,0.3) arc (170:10:0.5095);

\draw[thick,color=black!50!green] (-2,0) -- (2,0);
\draw [thick,color=black!50!green] (-0.75,0.15) arc (180:0:0.8);
\draw [thick,color=black!50!green] (-0.75,-0.15) arc (-180:0:0.8);
\draw [thick,color=black!50!green] (-0.75,0.15) arc (90:-90:0.15); 
\draw [thick,color=black!50!green] (0.8,0.15) arc (90:-90:0.15);

\draw[thick,color=red] (-0.5,-0.3) -- (0.5,-0.3);
\draw[thick,color=red] (-0.5,-0.3) arc (-170:-10:0.5095);

\draw[thick,color=blue] (-2,-0.3) -- (-1,-0.3);
\draw[thick,color=blue] (-1,-0.3) arc (-170:-10:1.04403);
\draw[thick,color=blue] (2,-0.3) -- (1,-0.3);
\draw[thick, color=black]
{
(-2,0) node [left] {\small{$ \alpha $}}
(2,0) node [right] {\small{$ \alpha $}}
(0,-1.1) node [below] {\small{$ \alpha $}}
(0,1.2) node [above] {\small{$ \alpha $}}
(0,-0.1) node [above] {\small{$ \alpha $}}
(0,-1.5) node [below] { {$J_{\text{on-site}}^2=J^2 $}}
};
\end{tikzpicture}\hspace{10mm}
\begin{tikzpicture}[scale=1]
\draw[thick,color=red] (-2,0.3) -- (-1,0.3);
\draw[thick,color=red] (-1,0.3) arc (170:10:1.04403);
\draw[thick,color=red] (2,0.3) -- (1,0.3); 

\draw[thick,color=blue] (-0.5,0.3) -- (0.5,0.3);
\draw[thick,color=blue] (-0.5,0.3) arc (170:10:0.5095);

\draw[thick,color=black!50!green] (-2,0) -- (2,0);
\draw [thick,color=black!50!green] (-0.75,0.15) arc (180:0:0.8);
\draw [thick,color=black!50!green] (-0.75,-0.15) arc (-180:0:0.8);
\draw [thick,color=black!50!green] (-0.75,0.15) arc (90:-90:0.15); 
\draw [thick,color=black!50!green] (0.8,0.15) arc (90:-90:0.15);

\draw[thick,color=red] (-0.5,-0.3) -- (0.5,-0.3);
\draw[thick,color=red] (-0.5,-0.3) arc (-170:-10:0.5095);

\draw[thick,color=blue] (-2,-0.3) -- (-1,-0.3);
\draw[thick,color=blue] (-1,-0.3) arc (-170:-10:1.04403);
\draw[thick,color=blue] (2,-0.3) -- (1,-0.3);
\draw[thick, color=black]
{
(-2,0) node [left] {\small{$ \alpha $}}
(2,0) node [right] {\small{$ \alpha $}}
(0,-1.1) node [below] {\small{$ \alpha_2 $}}
(0,1.2) node [above] {\small{$ \alpha $}}
(0,-0.1) node [above] {\small{$ \alpha_2 $}}
(0,-1.5) node [below] { {$ {1\over 2} J_\R^2=J^2\hspace{3mm} (\alpha_2\ne \alpha) $}}
};
\end{tikzpicture}
\caption{Contribution to the Schwinger-Dyson equation for the two point function. }
\label{fig:melonic diagram vertex contribution}
\end{figure}
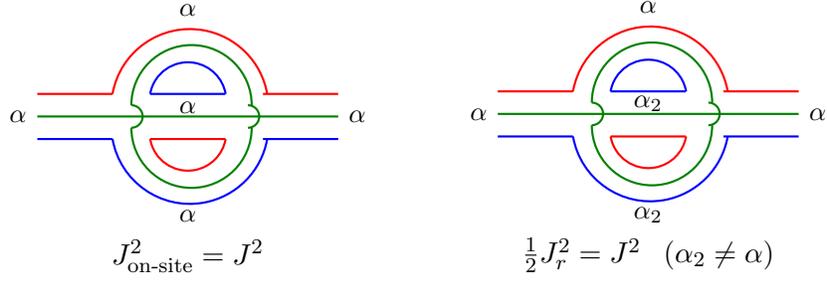

First, let us consider the $O(N)$ invariant two point function:
\begin{equation}
{1 \over N^3}  \langle \chi^{\alpha_1}_{ijk}(\tau_1) \chi^{\alpha_2}_{ijk} (\tau_2) \rangle 
\end{equation}
where the summations over $O(N)$ indices are omitted. It was shown in~\cite{Narayan:2017qtw} that the large $N$ diagrammatics of the two point function leads to a geometric series of the melonic diagrams which corresponds to the same Schwinger-Dyson equation of the two point function as in the SYK model. For the tensor model with $SO(M)$ global symmetry, we have two types of the contribution to the melonic diagrams: one melonic diagram with the on-site interactions and $(M-1)$ melonic diagrams with the hopping interactions of red color. See Figure~\ref{fig:melonic diagram vertex contribution}. Hence, considering all contributions, one can obtain the same Schwinger-Dyson equation as in the $q=4$ SYK model with the coupling constant $J$ of the SYK model replaced by the effective coupling constant~$\effcoupling$ given by
\begin{equation}
    \effcoupling\equiv J_{\text{on-site}}^2+(M-1){1\over 2}J_\R^2=M J^2\label{eq: effective coupling constant}\ .
\end{equation}
At strong coupling limit, the two point function is found to be
\begin{equation}\label{Two Point Function with effective coupling J}
{1 \over N^3}  \langle \chi^{\alpha_1}_{ijk}(\tau_1) \chi^{\alpha_2}_{ijk} (\tau_2) \rangle =\coeff{\sgn(\tau_{12})\over |\tau_{12}|^{1\over 2}}\delta^{\alpha_1\alpha_2}=\psi_{cl}(\tau_1,\tau_2)\delta^{\alpha_1\alpha_2}\ .
\end{equation} 
Here, $\psi_{cl}(\tau_,\tau_2)$ is the same as the classical solution $\psi_{cl}(\tau_1,\tau_2)$ for $q=4$ SYK model with the the effective coupling constant~$\effcoupling$ in \eqref{eq: effective coupling constant}.

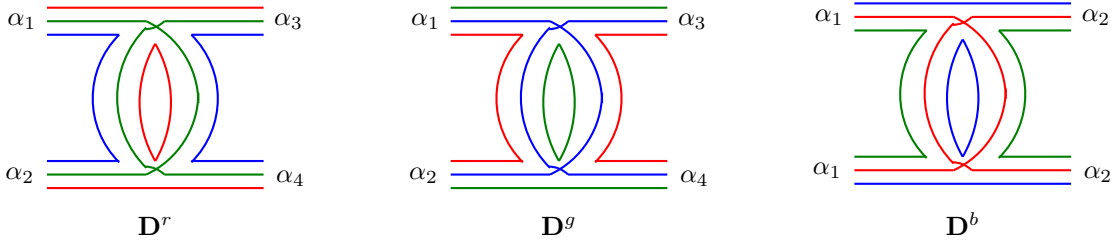
\begin{figure}[t!]
\centering
\begin{tikzpicture}[scale=1.2]
\def\xa{0}

\draw[thick,color=red] (\xa+-1.2,1) -- (\xa+1.2,1);

\draw[thick,color=blue]  (\xa+-1.2,0.7) -- (\xa+-0.4,0.7);
\draw[thick,color=blue]  (\xa+1.2,0.7) --  (\xa+0.4,0.7);
\draw[thick,color=blue]  (\xa+0.4,0.7) arc (45:-45:1);
\draw[thick,color=blue]  (\xa+-0.4,0.7) arc (135:225:1);

\draw[thick,color=black!50!green]  (\xa+-1.2,0.85)  -- (\xa+-0.1,0.85);
\draw[thick,color=black!50!green] (\xa+1.2,0.85)  -- (\xa+0.1,0.85);
\draw[thick,color=black!50!green] (\xa+-1.2,-0.85)  -- (\xa+-0.1,-0.85);
\draw[thick,color=black!50!green] (\xa+1.2,-0.85)  -- (\xa+0.1,-0.85);

\draw[thick,color=black!50!green] (\xa+-0.1,0.85) arc (65:0:1.0) ;
\draw[thick,color=black!50!green] (\xa+0.1,0.85) arc (-45:-45-45:0.3) ;
\draw[thick,color=black!50!green] (\xa+-0.1,0.78) arc (135:180+45:1.1) ;

\draw[thick,color=black!50!green] (\xa+-0.1,-0.85) arc (-65:0:1.0) ;
\draw[thick,color=black!50!green] (\xa+0.1,-0.85) arc (45:45+45:0.3) ;

\draw[thick,color=blue] (\xa+-1.2,-0.7) -- (\xa+-0.4,-0.7);
\draw[thick,color=blue] (\xa+1.2,-0.7) -- (\xa+0.4,-0.7);

\draw[thick,color=red] (\xa+0,0.6) arc (30:-30:1.3) ;
\draw[thick,color=red] (\xa+0,0.6) arc (180-30:180+30:1.3) ;
\draw[thick,color=red] (\xa+-1.2,-1) -- (\xa+1.2,-1);
\draw[thick, color=black]{
(\xa,-1.2) node [below] {\small{$ \dipole^\R$}}
};
\draw[thick, color=black]
{
(-1.2,0.85) node [left] {\small{$ \alpha_1 $}}
(1.2,0.85) node [right] {\small{$ \alpha_3 $}}
(-1.5, -0.65) node [below] {\small{$ \alpha_2 $}}
(1.5, -1.1) node [above] {\small{$ \alpha_4 $}}
};
\end{tikzpicture}\hspace{1cm}
\begin{tikzpicture}[scale=1.2]
\def\xa{0}

\draw[thick,color=black!50!green] (\xa+-1.2,1) -- (\xa+1.2,1);

\draw[thick,color=red]  (\xa+-1.2,0.7) -- (\xa+-0.4,0.7);
\draw[thick,color=red]  (\xa+1.2,0.7) --  (\xa+0.4,0.7);
\draw[thick,color=red]  (\xa+0.4,0.7) arc (45:-45:1);
\draw[thick,color=red]  (\xa+-0.4,0.7) arc (135:225:1);

\draw[thick,color=blue]  (\xa+-1.2,0.85)  -- (\xa+-0.1,0.85);
\draw[thick,color=blue] (\xa+1.2,0.85)  -- (\xa+0.1,0.85);
\draw[thick,color=blue] (\xa+-1.2,-0.85)  -- (\xa+-0.1,-0.85);
\draw[thick,color=blue] (\xa+1.2,-0.85)  -- (\xa+0.1,-0.85);

\draw[thick,color=blue] (\xa+-0.1,0.85) arc (65:0:1.0) ;
\draw[thick,color=blue] (\xa+0.1,0.85) arc (-45:-45-45:0.3) ;
\draw[thick,color=blue] (\xa+-0.1,0.78) arc (135:180+45:1.1) ;

\draw[thick,color=blue] (\xa+-0.1,-0.85) arc (-65:0:1.0) ;
\draw[thick,color=blue] (\xa+0.1,-0.85) arc (45:45+45:0.3) ;

\draw[thick,color=red] (\xa+-1.2,-0.7) -- (\xa+-0.4,-0.7);
\draw[thick,color=red] (\xa+1.2,-0.7) -- (\xa+0.4,-0.7);

\draw[thick,color=black!50!green] (\xa+0,0.6) arc (30:-30:1.3) ;
\draw[thick,color=black!50!green] (\xa+0,0.6) arc (180-30:180+30:1.3) ;
\draw[thick,color=black!50!green] (\xa+-1.2,-1) -- (\xa+1.2,-1);
\draw[thick, color=black]{
(\xa,-1.2) node [below] {\small{$ \dipole^\G$}}
};
\draw[thick, color=black]
{
(-1.2,0.85) node [left] {\small{$ \alpha_1 $}}
(1.2,0.85) node [right] {\small{$ \alpha_3 $}}
(-1.5, -0.65) node [below] {\small{$ \alpha_2 $}}
(1.5, -1.1) node [above] {\small{$ \alpha_4 $}}
};
\end{tikzpicture}\hspace{1cm}
\begin{tikzpicture}[scale=1.2]
\def\xa{0}

\draw[thick,color=blue] (\xa+-1.2,1) -- (\xa+1.2,1);

\draw[thick,color=black!50!green]  (\xa+-1.2,0.7) -- (\xa+-0.4,0.7);
\draw[thick,color=black!50!green]  (\xa+1.2,0.7) --  (\xa+0.4,0.7);
\draw[thick,color=black!50!green]  (\xa+0.4,0.7) arc (45:-45:1);
\draw[thick,color=black!50!green]  (\xa+-0.4,0.7) arc (135:225:1);

\draw[thick,color=red]  (\xa+-1.2,0.85)  -- (\xa+-0.1,0.85);
\draw[thick,color=red] (\xa+1.2,0.85)  -- (\xa+0.1,0.85);
\draw[thick,color=red] (\xa+-1.2,-0.85)  -- (\xa+-0.1,-0.85);
\draw[thick,color=red] (\xa+1.2,-0.85)  -- (\xa+0.1,-0.85);

\draw[thick,color=red] (\xa+-0.1,0.85) arc (65:0:1.0) ;
\draw[thick,color=red] (\xa+0.1,0.85) arc (-45:-45-45:0.3) ;
\draw[thick,color=red] (\xa+-0.1,0.78) arc (135:180+45:1.1) ;

\draw[thick,color=red] (\xa+-0.1,-0.85) arc (-65:0:1.0) ;
\draw[thick,color=red] (\xa+0.1,-0.85) arc (45:45+45:0.3) ;

\draw[thick,color=black!50!green] (\xa+-1.2,-0.7) -- (\xa+-0.4,-0.7);
\draw[thick,color=black!50!green] (\xa+1.2,-0.7) -- (\xa+0.4,-0.7);

\draw[thick,color=blue] (\xa+0,0.6) arc (30:-30:1.3) ;
\draw[thick,color=blue] (\xa+0,0.6) arc (180-30:180+30:1.3) ;
\draw[thick,color=blue] (\xa+-1.2,-1) -- (\xa+1.2,-1);
\draw[thick, color=black]{
(\xa,-1.2) node [below] {\small{$\dipole^\B$}}
};
\draw[thick, color=black]
{
(-1.2,0.85) node [left] {\small{$ \alpha_1 $}}
(1.2,0.85) node [right] {\small{$ \alpha_2 $}}
(-1.5, -0.65) node [below] {\small{$ \alpha_1 $}}
(1.5, -1.1) node [above] {\small{$ \alpha_2 $}}
};
\end{tikzpicture}
\caption{Dipoles $\dipole_{\col}^{\alpha_1\alpha_2\alpha_3\alpha_4}$ of three colors}
\label{fig:dipole}
\end{figure}

Now, we will analyze the ladder diagrams. \cite{Gurau:2010ba,Gurau:2011xq,Gurau:2011xp,Gurau:2011kk,Gurau:2016lzk} introduced a dipole as the basic building block of the ladder diagrams. In addition, \cite{Narayan:2017qtw} generalized the dipole for the tensor model on the lattice (See Figure~\ref{fig:dipole}), which have tensor structure in the lattice space. For our model, the dipole is a tensor in the $SO(M)$ index. 

\paragraph{Dipoles: }There are three dipoles $\dipole_\col$ ($\col=\R, \G, \B$) depending on the color that is transmitted along the ladder. See Figure~\ref{fig:dipole}. The ladder diagrams can be constructed by connecting these dipoles. Moreover, depending on how we connect the dipoles, we will have different types of ladder diagrams.

Hence, let us evaluate the dipoles first. Recall that the red color is distinct from green and blue colors in that the $SO(M)$ indices are contracted along the red line in the interaction vertex. For the dipole of red color, one can consider four configurations of the $SO(M)$ index which can give a contribution to the dipole of red color (See Figure~\ref{fig: red color dipole1} and Figure~\ref{fig: red color dipole2}). By summing up all contribution, we have
\begin{align}
\dipole_\R^{\alpha_1\alpha_2\alpha_3\alpha_4}(\tau_1,\tau_2,\tau_3,\tau_4) =&  -N^{-2} MJ^2\delta^{\alpha_1,\alpha_3}\delta^{\alpha_2,\alpha_4} \psi_{cl}(\tau_{13}) \psi_{cl}(\tau_{24}) [\psi_{cl}(\tau_{34})]^2 \cr
=& - \effcoupling^2\delta^{\alpha_1,\alpha_3}\delta^{\alpha_2,\alpha_4} \psi_{cl}(\tau_{13}) \psi_{cl}(\tau_{24}) [\psi_{cl}(\tau_{34})]^2
\end{align}
where we express the dipole in terms of the effective coupling constant~\eqref{eq: effective coupling constant}. Also, note that $\psi_{cl}(\tau)$ denotes the full propagator obtained in~\eqref{Two Point Function with effective coupling J}.

\begin{figure}[t!]
\centering
\begin{tikzpicture}[scale=1.2]
\def\xa{0}

\draw[thick,color=red] (\xa+-1.2,1) -- (\xa+1.2,1);

\draw[thick,color=blue]  (\xa+-1.2,0.7) -- (\xa+-0.4,0.7);
\draw[thick,color=blue]  (\xa+1.2,0.7) --  (\xa+0.4,0.7);
\draw[thick,color=blue]  (\xa+0.4,0.7) arc (45:-45:1);
\draw[thick,color=blue]  (\xa+-0.4,0.7) arc (135:225:1);

\draw[thick,color=black!50!green]  (\xa+-1.2,0.85)  -- (\xa+-0.1,0.85);
\draw[thick,color=black!50!green] (\xa+1.2,0.85)  -- (\xa+0.1,0.85);
\draw[thick,color=black!50!green] (\xa+-1.2,-0.85)  -- (\xa+-0.1,-0.85);
\draw[thick,color=black!50!green] (\xa+1.2,-0.85)  -- (\xa+0.1,-0.85);

\draw[thick,color=black!50!green] (\xa+-0.1,0.85) arc (65:0:1.0) ;
\draw[thick,color=black!50!green] (\xa+0.1,0.85) arc (-45:-45-45:0.3) ;
\draw[thick,color=black!50!green] (\xa+-0.1,0.78) arc (135:180+45:1.1) ;

\draw[thick,color=black!50!green] (\xa+-0.1,-0.85) arc (-65:0:1.0) ;
\draw[thick,color=black!50!green] (\xa+0.1,-0.85) arc (45:45+45:0.3) ;

\draw[thick,color=blue] (\xa+-1.2,-0.7) -- (\xa+-0.4,-0.7);
\draw[thick,color=blue] (\xa+1.2,-0.7) -- (\xa+0.4,-0.7);

\draw[thick,color=red] (\xa+0,0.6) arc (30:-30:1.3) ;
\draw[thick,color=red] (\xa+0,0.6) arc (180-30:180+30:1.3) ;
\draw[thick,color=red] (\xa+-1.2,-1) -- (\xa+1.2,-1);
\draw[thick, color=black]{
(\xa,-1.5) node [below] {\small{$ \Jonsite^2 N^{-2}=J^2N^{-2}$}}
};
\draw[thick, color=black]
{
(-1.2,0.85) node [left] {\small{$ \alpha_1 $}}
(1.2,0.85) node [right] {\small{$ \alpha_1 $}}
(-1.4, -0.65) node [below] {\small{$ \alpha_1 $}}
(1.4, -1.05) node [above] {\small{$ \alpha_1 $}}
(-1.0,-0.3) node [above] {\small{$ \alpha_1 $}}
(0.95,-0.3) node [above] {\small{$ \alpha_1 $}}
};
\end{tikzpicture}\hspace{2cm}
\begin{tikzpicture}[scale=1.2]
\def\xa{0}

\draw[thick,color=red] (\xa+-1.2,1) -- (\xa+1.2,1);

\draw[thick,color=blue]  (\xa+-1.2,0.7) -- (\xa+-0.4,0.7);
\draw[thick,color=blue]  (\xa+1.2,0.7) --  (\xa+0.4,0.7);
\draw[thick,color=blue]  (\xa+0.4,0.7) arc (45:-45:1);
\draw[thick,color=blue]  (\xa+-0.4,0.7) arc (135:225:1);

\draw[thick,color=black!50!green]  (\xa+-1.2,0.85)  -- (\xa+-0.1,0.85);
\draw[thick,color=black!50!green] (\xa+1.2,0.85)  -- (\xa+0.1,0.85);
\draw[thick,color=black!50!green] (\xa+-1.2,-0.85)  -- (\xa+-0.1,-0.85);
\draw[thick,color=black!50!green] (\xa+1.2,-0.85)  -- (\xa+0.1,-0.85);

\draw[thick,color=black!50!green] (\xa+-0.1,0.85) arc (65:0:1.0) ;
\draw[thick,color=black!50!green] (\xa+0.1,0.85) arc (-45:-45-45:0.3) ;
\draw[thick,color=black!50!green] (\xa+-0.1,0.78) arc (135:180+45:1.1) ;

\draw[thick,color=black!50!green] (\xa+-0.1,-0.85) arc (-65:0:1.0) ;
\draw[thick,color=black!50!green] (\xa+0.1,-0.85) arc (45:45+45:0.3) ;

\draw[thick,color=blue] (\xa+-1.2,-0.7) -- (\xa+-0.4,-0.7);
\draw[thick,color=blue] (\xa+1.2,-0.7) -- (\xa+0.4,-0.7);

\draw[thick,color=red] (\xa+0,0.6) arc (30:-30:1.3) ;
\draw[thick,color=red] (\xa+0,0.6) arc (180-30:180+30:1.3) ;
\draw[thick,color=red] (\xa+-1.2,-1) -- (\xa+1.2,-1);
\draw[thick, color=black]{
(\xa,-1.5) node [below] {\small{${1\over 2} J_\R^2 N^{-2}=J^2 N^{-2}$  $(\alpha_2\ne \alpha_1)$}}
};
\draw[thick, color=black]{
(-1.2,0.85) node [left] {\small{$ \alpha_1 $}}
(1.2,0.85) node [right] {\small{$ \alpha_1 $}}
(-1.4, -0.65) node [below] {\small{$ \alpha_1 $}}
(1.4, -1.05) node [above] {\small{$ \alpha_1 $}}
(-1.15,-0.3) node [above] {\small{$ \alpha_2 $}}
(1.15,-0.3) node [above] {\small{$ \alpha_2 $}}
};
\end{tikzpicture}
\caption{Contribution of interactions to dipoles of red color. For $\alpha_1=\alpha_3\ne\alpha_2=\alpha_4$, there are two configurations that can give a contribution to the dipole of red color}
\label{fig: red color dipole1}
\end{figure}
%

\begin{figure}[t!]
\centering
\begin{tikzpicture}[scale=1.2]
\def\xa{0}

\draw[thick,color=red] (\xa+-1.2,1) -- (\xa+1.2,1);

\draw[thick,color=blue]  (\xa+-1.2,0.7) -- (\xa+-0.4,0.7);
\draw[thick,color=blue]  (\xa+1.2,0.7) --  (\xa+0.4,0.7);
\draw[thick,color=blue]  (\xa+0.4,0.7) arc (45:-45:1);
\draw[thick,color=blue]  (\xa+-0.4,0.7) arc (135:225:1);

\draw[thick,color=black!50!green]  (\xa+-1.2,0.85)  -- (\xa+-0.1,0.85);
\draw[thick,color=black!50!green] (\xa+1.2,0.85)  -- (\xa+0.1,0.85);
\draw[thick,color=black!50!green] (\xa+-1.2,-0.85)  -- (\xa+-0.1,-0.85);
\draw[thick,color=black!50!green] (\xa+1.2,-0.85)  -- (\xa+0.1,-0.85);

\draw[thick,color=black!50!green] (\xa+-0.1,0.85) arc (65:0:1.0) ;
\draw[thick,color=black!50!green] (\xa+0.1,0.85) arc (-45:-45-45:0.3) ;
\draw[thick,color=black!50!green] (\xa+-0.1,0.78) arc (135:180+45:1.1) ;

\draw[thick,color=black!50!green] (\xa+-0.1,-0.85) arc (-65:0:1.0) ;
\draw[thick,color=black!50!green] (\xa+0.1,-0.85) arc (45:45+45:0.3) ;

\draw[thick,color=blue] (\xa+-1.2,-0.7) -- (\xa+-0.4,-0.7);
\draw[thick,color=blue] (\xa+1.2,-0.7) -- (\xa+0.4,-0.7);

\draw[thick,color=red] (\xa+0,0.6) arc (30:-30:1.3) ;
\draw[thick,color=red] (\xa+0,0.6) arc (180-30:180+30:1.3) ;
\draw[thick,color=red] (\xa+-1.2,-1) -- (\xa+1.2,-1);
\draw[thick, color=black]{
(\xa,-1.5) node [below] {\small{${1\over \sqrt{2}} J_\R \Jonsite N^{-2}=J^2 N^{-2}$ }}
(\xa,-2) node [below] {\small{  ($\alpha_1\ne \alpha_2$ and $\alpha_3=\alpha_1$ or $\alpha_2$) }}
};
\draw[thick, color=black]
{
(-1.2,0.85) node [left] {\small{$ \alpha_1 $}}
(1.2,0.85) node [right] {\small{$ \alpha_1 $}}
(-1.4, -0.65) node [below] {\small{$ \alpha_2 $}}
(1.4, -1.05) node [above] {\small{$ \alpha_2 $}}
(-1.0,-0.3) node [above] {\small{$ \alpha_3 $}}
(0.95,-0.3) node [above] {\small{$ \alpha_3 $}}
};
\end{tikzpicture}\hspace{2cm}\begin{tikzpicture}[scale=1.2]
\def\xa{0}

\draw[thick,color=red] (\xa+-1.2,1) -- (\xa+1.2,1);

\draw[thick,color=blue]  (\xa+-1.2,0.7) -- (\xa+-0.4,0.7);
\draw[thick,color=blue]  (\xa+1.2,0.7) --  (\xa+0.4,0.7);
\draw[thick,color=blue]  (\xa+0.4,0.7) arc (45:-45:1);
\draw[thick,color=blue]  (\xa+-0.4,0.7) arc (135:225:1);

\draw[thick,color=black!50!green]  (\xa+-1.2,0.85)  -- (\xa+-0.1,0.85);
\draw[thick,color=black!50!green] (\xa+1.2,0.85)  -- (\xa+0.1,0.85);
\draw[thick,color=black!50!green] (\xa+-1.2,-0.85)  -- (\xa+-0.1,-0.85);
\draw[thick,color=black!50!green] (\xa+1.2,-0.85)  -- (\xa+0.1,-0.85);

\draw[thick,color=black!50!green] (\xa+-0.1,0.85) arc (65:0:1.0) ;
\draw[thick,color=black!50!green] (\xa+0.1,0.85) arc (-45:-45-45:0.3) ;
\draw[thick,color=black!50!green] (\xa+-0.1,0.78) arc (135:180+45:1.1) ;

\draw[thick,color=black!50!green] (\xa+-0.1,-0.85) arc (-65:0:1.0) ;
\draw[thick,color=black!50!green] (\xa+0.1,-0.85) arc (45:45+45:0.3) ;

\draw[thick,color=blue] (\xa+-1.2,-0.7) -- (\xa+-0.4,-0.7);
\draw[thick,color=blue] (\xa+1.2,-0.7) -- (\xa+0.4,-0.7);

\draw[thick,color=red] (\xa+0,0.6) arc (30:-30:1.3) ;
\draw[thick,color=red] (\xa+0,0.6) arc (180-30:180+30:1.3) ;
\draw[thick,color=red] (\xa+-1.2,-1) -- (\xa+1.2,-1);
\draw[thick, color=black]{
(\xa,-1.5) node [below] {\small{${1\over 2} J_\R^2 N^{-2}=J^2 N^{-2}$  }}
(\xa,-2) node [below] {\small{ ($\alpha_1\ne \alpha_2$ and $\alpha_3\ne \alpha_1,\alpha_2$)}}
};
\draw[thick, color=black]
{
(-1.2,0.85) node [left] {\small{$ \alpha_1 $}}
(1.2,0.85) node [right] {\small{$ \alpha_1 $}}
(-1.4, -0.65) node [below] {\small{$ \alpha_2 $}}
(1.4, -1.05) node [above] {\small{$ \alpha_2 $}}
(-1.0,-0.3) node [above] {\small{$ \alpha_3 $}}
(0.95,-0.3) node [above] {\small{$ \alpha_3 $}}
};
\end{tikzpicture}
\caption{Contribution of interactions to dipoles of red color. For $\alpha_1=\alpha_3\ne \alpha_2=\alpha_4$, there are two configurations that can give a contribution to the dipole of red color}
\label{fig: red color dipole2}
\end{figure}
%

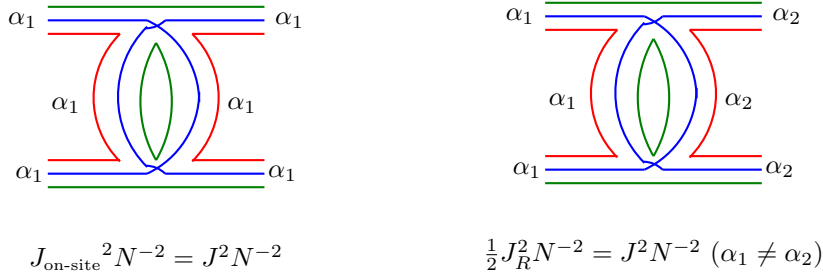
\begin{figure}[t!]
\centering
\begin{tikzpicture}[scale=1.2]
\def\xa{0}

\draw[thick,color=black!50!green] (\xa+-1.2,1) -- (\xa+1.2,1);

\draw[thick,color=red]  (\xa+-1.2,0.7) -- (\xa+-0.4,0.7);
\draw[thick,color=red]  (\xa+1.2,0.7) --  (\xa+0.4,0.7);
\draw[thick,color=red]  (\xa+0.4,0.7) arc (45:-45:1);
\draw[thick,color=red]  (\xa+-0.4,0.7) arc (135:225:1);

\draw[thick,color=blue]  (\xa+-1.2,0.85)  -- (\xa+-0.1,0.85);
\draw[thick,color=blue] (\xa+1.2,0.85)  -- (\xa+0.1,0.85);
\draw[thick,color=blue] (\xa+-1.2,-0.85)  -- (\xa+-0.1,-0.85);
\draw[thick,color=blue] (\xa+1.2,-0.85)  -- (\xa+0.1,-0.85);

\draw[thick,color=blue] (\xa+-0.1,0.85) arc (65:0:1.0) ;
\draw[thick,color=blue] (\xa+0.1,0.85) arc (-45:-45-45:0.3) ;
\draw[thick,color=blue] (\xa+-0.1,0.78) arc (135:180+45:1.1) ;

\draw[thick,color=blue] (\xa+-0.1,-0.85) arc (-65:0:1.0) ;
\draw[thick,color=blue] (\xa+0.1,-0.85) arc (45:45+45:0.3) ;

\draw[thick,color=red] (\xa+-1.2,-0.7) -- (\xa+-0.4,-0.7);
\draw[thick,color=red] (\xa+1.2,-0.7) -- (\xa+0.4,-0.7);

\draw[thick,color=black!50!green] (\xa+0,0.6) arc (30:-30:1.3) ;
\draw[thick,color=black!50!green] (\xa+0,0.6) arc (180-30:180+30:1.3) ;
\draw[thick,color=black!50!green] (\xa+-1.2,-1) -- (\xa+1.2,-1);
\draw[thick, color=black]{
(\xa,-1.5) node [below] {\small{$ \Jonsite^2 N^{-2}=J^2N^{-2}$}}
};
\draw[thick, color=black]
{
(-1.2,0.85) node [left] {\small{$ \alpha_1 $}}
(1.2,0.85) node [right] {\small{$ \alpha_1 $}}
(-1.4, -0.65) node [below] {\small{$ \alpha_1 $}}
(1.4, -1.05) node [above] {\small{$ \alpha_1 $}}
(-1.0,-0.3) node [above] {\small{$ \alpha_1 $}}
(0.95,-0.3) node [above] {\small{$ \alpha_1 $}}
};
\end{tikzpicture}\hspace{2cm}
\begin{tikzpicture}[scale=1.2]
\def\xa{0}

\draw[thick,color=black!50!green] (\xa+-1.2,1) -- (\xa+1.2,1);

\draw[thick,color=red]  (\xa+-1.2,0.7) -- (\xa+-0.4,0.7);
\draw[thick,color=red]  (\xa+1.2,0.7) --  (\xa+0.4,0.7);
\draw[thick,color=red]  (\xa+0.4,0.7) arc (45:-45:1);
\draw[thick,color=red]  (\xa+-0.4,0.7) arc (135:225:1);

\draw[thick,color=blue]  (\xa+-1.2,0.85)  -- (\xa+-0.1,0.85);
\draw[thick,color=blue] (\xa+1.2,0.85)  -- (\xa+0.1,0.85);
\draw[thick,color=blue] (\xa+-1.2,-0.85)  -- (\xa+-0.1,-0.85);
\draw[thick,color=blue] (\xa+1.2,-0.85)  -- (\xa+0.1,-0.85);

\draw[thick,color=blue] (\xa+-0.1,0.85) arc (65:0:1.0) ;
\draw[thick,color=blue] (\xa+0.1,0.85) arc (-45:-45-45:0.3) ;
\draw[thick,color=blue] (\xa+-0.1,0.78) arc (135:180+45:1.1) ;

\draw[thick,color=blue] (\xa+-0.1,-0.85) arc (-65:0:1.0) ;
\draw[thick,color=blue] (\xa+0.1,-0.85) arc (45:45+45:0.3) ;

\draw[thick,color=red] (\xa+-1.2,-0.7) -- (\xa+-0.4,-0.7);
\draw[thick,color=red] (\xa+1.2,-0.7) -- (\xa+0.4,-0.7);

\draw[thick,color=black!50!green] (\xa+0,0.6) arc (30:-30:1.3) ;
\draw[thick,color=black!50!green] (\xa+0,0.6) arc (180-30:180+30:1.3) ;
\draw[thick,color=black!50!green] (\xa+-1.2,-1) -- (\xa+1.2,-1);
\draw[thick, color=black]{
(\xa,-1.5) node [below] {\small{$ {1\over 2}J_R^2 N^{-2}=J^2N^{-2}$ ($\alpha_1\ne \alpha_2$)}}
};
\draw[thick, color=black]
{
(-1.2,0.85) node [left] {\small{$ \alpha_1 $}}
(1.2,0.85) node [right] {\small{$ \alpha_2 $}}
(-1.4, -0.65) node [below] {\small{$ \alpha_1 $}}
(1.4, -1.05) node [above] {\small{$ \alpha_2 $}}
(-1.0,-0.3) node [above] {\small{$ \alpha_1 $}}
(0.95,-0.3) node [above] {\small{$ \alpha_2 $}}
};
\end{tikzpicture}
\caption{Contribution of interactions to dipole of green color.}
\label{fig: green color dipole}
\end{figure}

On the other hand, there are two configurations which give a contribution to the dipoles of green and blue color. \eg See Figure~\ref{fig: green color dipole} for dipole of green color. Therefore, we have
\begin{align}
\dipole_\G^{\alpha_1\alpha_2\alpha_3\alpha_4}(\tau_1,\tau_2,\tau_3,\tau_4) =&  -  J^2\delta^{\alpha_1,\alpha_2}\delta^{\alpha_3,\alpha_4} \psi_{cl}(\tau_{13}) \psi_{cl}(\tau_{24}) [\psi_{cl}(\tau_{34})]^2\cr
=&- {\effcoupling^2\over M}\delta^{\alpha_1,\alpha_2}\delta^{\alpha_3,\alpha_4} \psi_{cl}(\tau_{13}) \psi_{cl}(\tau_{24}) [\psi_{cl}(\tau_{34})]^2\ ,\\
\dipole_\B^{\alpha_1\alpha_2\alpha_3\alpha_4}(\tau_1,\tau_2,\tau_3,\tau_4) =&  -  J^2 \delta^{\alpha_1,\alpha_2}\delta^{\alpha_3,\alpha_4} \psi_{cl}(\tau_{13}) \psi_{cl}(\tau_{24}) [\psi_{cl}(\tau_{34})]^2 \cr
=&- {\effcoupling^2\over M}\delta^{\alpha_1,\alpha_2}\delta^{\alpha_3,\alpha_4} \psi_{cl}(\tau_{13}) \psi_{cl}(\tau_{24}) [\psi_{cl}(\tau_{34})]^2 \ .
\end{align}

%

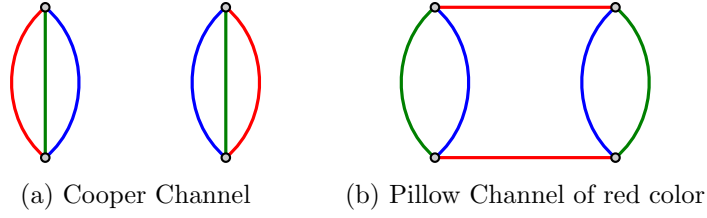
\begin{figure}
\centering
\begin{subfigure}[t]{0.3\linewidth}
\centering
\begin{tikzpicture}[scale=0.8]
\def\xa{0}  \def\ya{0}    
 
    \coordinate (a1) at (\xa,\ya);
    \coordinate (b1) at (\xa,\ya+2.5);
    \coordinate (c1) at (\xa+3,\ya+2.5);
    \coordinate (d1) at (\xa+3,\ya);
    
    \draw[bend left=50,very thick,color=red]  (a1) to   (b1);
    \draw[bend right=50,very thick,color=blue]  (a1) to   (b1);
    \draw[bend right=50,very thick,color=blue]  (c1) to   (d1);  
    \draw[bend left=50,very thick,color=red]  (c1) to   (d1);

      \draw[very thick,color=black!50!green]  (b1) -- (a1) ;
      \draw[very thick,color=black!50!green]   (d1) -- (c1);

	\fill[black!20, draw=black, thick] (a1) circle (2pt) ;
    \fill[black!20, draw=black, thick] (b1) circle (2pt);
    \fill[black!20, draw=black, thick] (c1) circle (2pt)  ;
    \fill[black!20, draw=black, thick] (d1) circle (2pt);      
   
\end{tikzpicture}
\caption{Cooper Channel}
\label{fig:cooper contraction}
\end{subfigure}
\begin{subfigure}[t]{0.35\linewidth}
\centering
\begin{tikzpicture}[scale=0.8]
\def\xa{0}  \def\ya{0}    
 
    \coordinate (a1) at (\xa,\ya);
    \coordinate (b1) at (\xa,\ya+2.5);
    \coordinate (c1) at (\xa+3,\ya+2.5);
    \coordinate (d1) at (\xa+3,\ya);
    
    \draw[bend left=50,very thick,color=black!50!green]  (a1) to   (b1);
    \draw[bend right=50,very thick,color=blue]  (a1) to   (b1);
    \draw[bend right=50,very thick,color=blue]  (c1) to   (d1);  
    \draw[bend left=50,very thick,color=black!50!green]  (c1) to   (d1);

      \draw[very thick,color=red]  (d1) -- (a1) ;
      \draw[very thick,color=red]   (c1) -- (b1);

	\fill[black!20, draw=black, thick] (a1) circle (2pt) ;
    \fill[black!20, draw=black, thick] (b1) circle (2pt);
    \fill[black!20, draw=black, thick] (c1) circle (2pt)  ;
    \fill[black!20, draw=black, thick] (d1) circle (2pt);         
\end{tikzpicture}
\caption{Pillow Channel of red color}
\label{fig:pillow contraction}
\end{subfigure}
%
%
\caption{Schematic representation of $O(N)$ contraction of the Cooper and Pillow four point functions. Each vertex represents a fermion, and the colored edge denotes the contraction of $O(N)$ index of the corresponding color in the two fermions.}
\label{fig:contractions}
\end{figure}

\paragraph{Four Point Functions: }From these basic building blocks, we will construct the ladder diagrams and four point functions. Unlike the SYK model, tensor models have various $O(N)$ invariant operators because of $O(N)$ tensor structure. In particular, there are three types of $O(N)$ invariant four point functions. In this work, we consider the following two types of four point functions:
\begin{align}
F_{C}^{\alpha_1\alpha_2\alpha_3\alpha_4} \equiv & \langle \chi^{\alpha_1}_{i_1 j_1 k_1}(\tau_1) \chi^{\alpha_2}_{i_1 j_1 k_1}(\tau_2) \chi^{\alpha_3}_{i_2 j_2 k_2}(\tau_3) \chi^{\alpha_4}_{i_2 j_2 k_2}(\tau_4) \rangle\ , \\
F_{P,\R}^{\alpha_1\alpha_2\alpha_3\alpha_4} \equiv &  \langle \chi^{a_1}_{i_1 j_1 k_1}(\tau_1) \chi^{\alpha_2}_{i_2 j_1 k_1}(\tau_2) \chi^{\alpha_3}_{i_1 j_2 k_2}(\tau_3) \chi^{\alpha_4}_{i_2 j_2 k_2}(\tau_4) \rangle \ ,\label{eq: pillow channel of red color}\\
F_{P,\G}^{\alpha_1\alpha_2\alpha_3\alpha_4} \equiv &  \langle \chi^{\alpha_1}_{i_1 j_1 k_1}(\tau_1) \chi^{\alpha_2}_{i_1 j_2 k_1}(\tau_2) \chi^{\alpha_3}_{i_2 j_1 k_2}(\tau_3) \chi^{\alpha_4}_{i_2 j_2 k_2}(\tau_4) \rangle\ , \\
F_{P,\B}^{\alpha_1\alpha_2\alpha_3\alpha_4} \equiv &  \langle \chi^{\alpha_1}_{i_1 j_1 k_1}(\tau_1) \chi^{\alpha_2}_{i_1 j_1 k_2}(\tau_2) \chi^{\alpha_3}_{i_2 j_2 k_1}(\tau_3) \chi^{\alpha_4}_{i_2 j_2 k_2}(\tau_4) \rangle 
\end{align}
where we omitted the summation over the $O(N)$ indices. Note that they have different contraction of the $O(N)$ indices, which is schematically shown in Figure~\ref{fig:pillow contraction}. Depending on the structure of the $O(N)$ contraction, we call $F_{C}^{\alpha_1\alpha_2\alpha_3\alpha_4}$ to be Cooper channel, and $F_{P,\col}^{\alpha_1\alpha_2\alpha_3\alpha_4}$ to be Pillow channel of color $\col$ $(\in \{\R,\G,\B\})$. The Cooper channel is the analogous to the four point function of the SYK model. Hence, one can expect that the Cooper channel exhibits the same behavior as that of the SYK model. We emphasize that the Pillow channel is a new $O(N)$ invariant observable which does not appear in the original SYK model. Note that in the Pillow channel there exist two fermions that are contracted by only one $O(N)$ index. The color of this $O(N)$ index characterizes the Pillow channel. Hence, we have three Pillow channels: $F_{P,\R}, F_{P,\G}$ and $F_{P,\B}$ (\eg See the Pillow channel of the red color in Figure~\ref{fig:pillow contraction}.) Furthermore, the ladder diagrams $\mathcal{F}_C$ and $\mathcal{F}_{P,\col}$ are defined by excluding the leading disconnected diagram from each four point function:
\begin{align}
F_C^{\alpha_1\alpha_2\alpha_3\alpha_4} (\tau_1,\tau_2,\tau_3,\tau_4)=& N^6\delta^{\alpha_1\alpha_2}\delta^{\alpha_3\alpha_4} \psi_{cl}(\tau_{12})\psi_{cl}(\tau_{34}) + N^3\mathcal{F}_C^{\alpha_1\alpha_2\alpha_3\alpha_4} (\tau_1,\tau_2,\tau_3,\tau_4)\ , \label{def:ladder diagram cooper channel}\\
F_{P,\col}^{\alpha_1\alpha_2\alpha_3\alpha_4} (\tau_1,\tau_2,\tau_3,\tau_4)=& N^5\delta^{\alpha_1\alpha_2}\delta^{\alpha_3\alpha_4} \psi_{cl}(\tau_{12})\psi_{cl}(\tau_{34}) + N^4 \mathcal{F}_{P,\col}^{\alpha_1\alpha_2\alpha_3\alpha_4}(\tau_1,\tau_2,\tau_3,\tau_4)\label{def:ladder diagram pillow channel}
\end{align}
where the first term of the RHS corresponds to the leading disconnected diagram, and the second term is the ladder diagram which we will analyze.

\paragraph{Cooper Channel} First, we consider the ladder diagram in the Cooper channel. In~\cite{Narayan:2017qtw}, it was shown that the leading ladder diagrams of order $\mathcal{O}(N^3)$ are composed of any combination of dipoles. This leads to the geometric series of the dipoles, and one can write
\begin{equation}
\boldsymbol{\mathcal{F}}_C= \sum_{n=0}^\infty (\dipole_\R+\dipole_\G+\dipole_\B)^n \boldsymbol{\mathcal{F}}_0={1\over 1-\boldsymbol{K}_C}\boldsymbol{\mathcal{F}}_0\ .\label{eq:geometric series for Cooper channel}
\end{equation}
Here, the multiplication of dipoles is the matrix multiplication in which the first two $SO(M)$ indices and $\tau$'s form one matrix index, and the second two $SO(M)$ indices and $\tau$'s become the other matrix index. Also, $\boldsymbol{\mathcal{F}}_0$ denotes the first term of the geometric series, and is given by
\begin{equation}
\boldsymbol{\mathcal{F}}_0= -\delta^{\alpha_1\alpha_3}\delta^{\alpha_2\alpha_4}\psi_{cl}(\tau_{13})\psi_{cl}(\tau_{24})+\delta^{\alpha_1\alpha_4}\delta^{\alpha_2\alpha_3}\psi_{cl}(\tau_{14})\psi_{cl}(\tau_{23})\ .\label{eq: first term in geometric series}
\end{equation}
The common ratio of the geometric series~\eqref{eq:geometric series for Cooper channel} is defined by
\begin{equation}
    \boldsymbol{K}_C\equiv\dipole_\R+\dipole_\G+\dipole_\B=- \effcoupling^2\left(\delta^{\alpha_1\alpha_3}\delta^{\alpha_2\alpha_4} +{2\over M}\delta^{\alpha_1,\alpha_2}\delta^{\alpha_3,\alpha_4}\right)\psi_{cl}(\tau_{13}) \psi_{cl}(\tau_{24}) [\psi_{cl}(\tau_{34})]^2
\end{equation}
This is the same as the kernel~\eqref{eq: kernel from diagrammatics} of the SYK model with $SO(M)$ global symmetry found by diagrammatics in Section~\eqref{sec: diagrammatic} except for the structure of $SO(M)$ indices $\alpha_3$ and $\alpha_4$. Note that the matrix multiplication of the tensor model is slightly different from \eqref{eq: matrix multiplication syk model} of the SYK model with global symmetry:
\begin{equation}
(A\cdot B)(X_1,X_2,X_3,X_4)=\sum_{X_5,X_6}A(X_1,X_2,X_5,X_6)B(X_5,X_6,X_3,X_4)\label{eq: matrix multiplication for tensor model}
\end{equation}
where $X_1,\cdots, X_6$ are the collective indices of the time coordinate and the $SO(M)$ index. We define this multiplication for the tensor model because it is more natural when we connect the dipoles. These two differences in the tensor model lead to the same final results as those of the SYK model with global symmetry after the contraction with $SO(M)$ generators.

\paragraph{Pillow Channel} The leading ladder diagrams of the Pillow channel have a different feature. The specific combination of dipoles becomes the leading ladder diagram~\cite{Narayan:2017qtw}. For example, consider the Pillow channel of red color. The leading ladder diagram of order $\mathcal{O}(N^4)$ comes from the series of dipoles of red color. Hence, one can write the leading ladder diagram as follows.
\begin{equation}
\boldsymbol{\mathcal{F}}_{P,\col}= \sum_{n=0}^\infty (\dipole_\col)^n \boldsymbol{\mathcal{F}}_0={1\over 1-\boldsymbol{K}_{P,\col}}\boldsymbol{\mathcal{F}}_{P,0}\hspace{5mm}(\col=\R,\G,\B)\label{eq:geometric series for Pillow channel}
\end{equation}
The common ratio $\boldsymbol{K}_{P,\col}$ are found to be
\begin{align}
    \boldsymbol{K}_{P,\R}=&\dipole_\R= -\effcoupling^2\delta^{\alpha_1\alpha_3}\delta^{\alpha_2\alpha_4} \psi_{cl}(\tau_{13}) \psi_{cl}(\tau_{24}) [\psi_{cl}(\tau_{34})]^2 \ , \\
    \boldsymbol{K}_{P,\G}=&\dipole_\G= -{\effcoupling^2\over M}\delta^{\alpha_1,\alpha_2}\delta^{\alpha_3,\alpha_4}\psi_{cl}(\tau_{13}) \psi_{cl}(\tau_{24}) [\psi_{cl}(\tau_{34})]^2\ , \\\
    \boldsymbol{K}_{P,\B}=&\dipole_\B=-{\effcoupling^2\over M}\delta^{\alpha_1,\alpha_2}\delta^{\alpha_3,\alpha_4}\psi_{cl}(\tau_{13}) \psi_{cl}(\tau_{24}) [\psi_{cl}(\tau_{34})]^2 \ .
\end{align}
Note that the Pillow channel of red color has different tensor structure from those of green and blue colors. 

The Pillow channel (\eg of red color) can be further decomposed into the symmetric part (the singlet and the symmetric irrep) and the anti-symmetric part (the anti-symmetric irrep) of $O(N)$ index (\eg of red color). For example, one can decompose the Pillow channel of red color $\boldsymbol{\mathcal{F}}_{P,\R}(\tau_1,\tau_2,\tau_3,\tau_4)$ (See~\eqref{eq: pillow channel of red color}) into the symmetric and anti-symmetric parts in $(i_1,i_2)$ which are the contracted external $O(N)$ indices of the red color. This decomposition can simply be taken into account by the same decomposition of the first term of the geometric series:
\begin{equation}
\boldsymbol{\mathcal{F}}_{P^\mp,0}= \mp\delta^{\alpha_1\alpha_3}\delta^{\alpha_2\alpha_4}\psi_{cl}(\tau_{13})\psi_{cl}(\tau_{24})+\delta^{\alpha_1\alpha_4}\delta^{\alpha_2\alpha_3}\psi_{cl}(\tau_{14})\psi_{cl}(\tau_{23})\ .\label{eq: first term in geometric series pillow}
\end{equation}
where $\boldsymbol{\mathcal{F}}_{P^-,0}$ (or, $\boldsymbol{\mathcal{F}}_{P^+,0}$) corresponds to the symmetric part (or, the anti-symmetric part, respectively) under $i_1\leftrightarrow i_2$.

\paragraph{Rank $(q-1)$ Tensor Model with $SO(M)$ Global Symmetry} As in the SYK model, it is easy to generalized to rank $D=(q-1)$ tensor model with global symmetry. In particular, \cite{Narayan:2017qtw} introduced ``uncoloring'' process to obtain the explicit Hamiltonian of the KT tensor model from the GW model. And, it is straightforward to generalize the diagrammatics of rank-3 KT tensor model. For our model, we also repeat the same process, and quickly generalize the above results. For details, we refer readers to~\cite{Narayan:2017qtw}.

In rank $(q-1)$ tensor model, RGB color is generalized into $(q-1)$ colors: $\col_1,\col_2,\cdots, \col_{q-1}$. Without loss of generality, let us choose one color $\col_1$ such that the $SO(M)$ index are connected along the $\col_1$ line in the interaction vertex. (\ie $\col_1$ is analogous to the red color in the rank-3 case.) Then, the dipole of the color $\col_1$ is given by
\begin{equation}
\dipole_{\col_1}^{\alpha_1\alpha_2\alpha_3\alpha_4}(\tau_1,\tau_2,\tau_3,\tau_4) = - \effcoupling^2\delta^{\alpha_1,\alpha_3}\delta^{\alpha_2,\alpha_4} \psi_{cl}(\tau_{13}) \psi_{cl}(\tau_{24}) [\psi_{cl}(\tau_{34})]^{q-2}
\end{equation}
while the dipole of the other colors $\col_m$ ($m=2,3,\cdots, q-1$) is found to be
\begin{equation}
\dipole_{\col_m}^{\alpha_1\alpha_2\alpha_3\alpha_4}(\tau_1,\tau_2,\tau_3,\tau_4) =- {\effcoupling^2\over M}\delta^{\alpha_1,\alpha_2}\delta^{\alpha_3,\alpha_4} \psi_{cl}(\tau_{13}) \psi_{cl}(\tau_{24}) [\psi_{cl}(\tau_{34})]^{q-2}\hspace{5mm} (m\ne 1)\ .
\end{equation}
Note that the full propagator denoted by $\psi_{cl}$ is the same as that of the SYK model for the case of general $q$ with effective coupling constant $\effcoupling\equiv\sqrt{M}J$.

As before, all combinations of the dipoles form the leading ladder diagrams of the Cooper channel, and the corresponding geometric series is given by
\begin{equation}
\boldsymbol{\mathcal{F}}_C= \sum_{n=0}^\infty \left(\sum_{m=1}^{q-1} \dipole_{\col_m}\right)^n \boldsymbol{\mathcal{F}}_0={1\over 1-\boldsymbol{K}_C}\boldsymbol{\mathcal{F}}_0\label{eq:geometric series for Cooper channel rank}
\end{equation}
where the kernel is found to be
\begin{equation}
    \boldsymbol{K}_C=\sum_{m=1}^{q-1} \dipole_{\col_m}= -\effcoupling^2\left(\delta^{\alpha_1\alpha_3}\delta^{\alpha_2\alpha_4} +{q-2\over M}\delta^{\alpha_1,\alpha_2}\delta^{\alpha_3,\alpha_4}\right)\psi_{cl}(\tau_{13}) \psi_{cl}(\tau_{24}) [\psi_{cl}(\tau_{34})]^{q-2}\ .
\end{equation}
As expected, this kernel (together with the zero-rung ladder diagram $\boldsymbol{\mathcal{F}}_0$ in~\eqref{eq: first term in geometric series}) agrees with \eqref{eq: kernel from diagrammatics} in the SYK model with global symmetry for general $q$ case except for the structure of the $SO(M)$ indices $\alpha_3$ and $\alpha_4$ in the kernel. But, they will give the same result as in the SYK model with global symmetry after the contraction with $SO(M)$ generators.

Furthermore, the leading Pillow channel of the given color $c_m$ is consist of combinations of the dipoles of color $c_m$:
\begin{equation}
\boldsymbol{\mathcal{F}}_{P,\col_m} = \sum_{n=0}^\infty  (\dipole_{\col_m})^n \boldsymbol{\mathcal{F}}_0={1\over 1-\boldsymbol{K}_{P,\col_m}}\boldsymbol{\mathcal{F}}_0\ ,\label{eq:geometric series for Pillow channel rank}
\end{equation}
and therefore, the kernels are found to be
\begin{align}
    \boldsymbol{K}_{P,\col_1}=&\dipole_{\col_1}= -\effcoupling^2\delta^{\alpha_1\alpha_3}\delta^{\alpha_2\alpha_4} \psi_{cl}(\tau_{13}) \psi_{cl}(\tau_{24}) [\psi_{cl}(\tau_{34})]^2 \ ,\\
    \boldsymbol{K}_{P,\col_m}=&\dipole_{\col_m}= -{\effcoupling^2\over M}\delta^{\alpha_1,\alpha_2}\delta^{\alpha_3,\alpha_4}\psi_{cl}(\tau_{13}) \psi_{cl}(\tau_{24}) [\psi_{cl}(\tau_{34})]^{q-2} \hspace{4mm} (m=2,3,\cdots, q-1)\ .
\end{align}

\paragraph{Decomposition into Irreducible Representation: }As in the SYK model with global symmetry, one can decomposed into the singlet and anti-symmetric/symmetric channels of the $SO(M)$. We summarize our result:

\begin{itemize}
 
\item \textbf{Cooper Channel of order $\mathcal{O}(N^{q-1})$}: The Cooper channel is decomposed into the singlet, anti-symmetric and symmetric irreducible representation of $SO(M)$:
\begin{align}
\mathcal{F}_{C, -}\equiv& {1\over M}\delta^{\alpha_2\alpha_1}\delta^{\alpha_4\alpha_3} \boldsymbol{\mathcal{F}}_C^{\alpha_1\alpha_2\alpha_3\alpha_4}\ ,\\
\mathcal{F}_{C,\anti +}^{ab}\equiv&  \mathcal{F}_{C,\anti +}\equiv {1\over 2\ind_\anti}(\mT_\anti^a)^{\alpha_2\alpha_1}(\mT_\anti^b)^{\alpha_4\alpha_3} \boldsymbol{\mathcal{F}}_C^{\alpha_1\alpha_2\alpha_3\alpha_4}\ ,\\
\mathcal{F}_{C,\sym -}^{ab}\equiv&  \mathcal{F}_{C,\sym -}\equiv {1\over 2\ind_\sym}(\mT_\sym^a)^{\alpha_2\alpha_1}(\mT_\sym^b)^{\alpha_4\alpha_3} \boldsymbol{\mathcal{F}}_C^{\alpha_1\alpha_2\alpha_3\alpha_4}\ .
\end{align}
Moreover, we decompose the common ratio of the geometric series as follows. 
\begin{align}
    K_{C,-}\equiv&{1\over M}\delta^{\alpha_2\alpha_1}\delta^{\alpha_4\alpha_3} (\boldsymbol{K}_C)^{\alpha_1\alpha_2\alpha_3\alpha_4}\ ,\\
    K_{C,\anti +}^{ab}\equiv&\delta^{ab} K_{C,\anti +}\equiv{1\over 2\ind_\anti}((\mT_\anti^a)^t)^{\alpha_2\alpha_1}(\mT_\anti^b)^{\alpha_4\alpha_3} \boldsymbol(\boldsymbol{K}_C)^{\alpha_1\alpha_2\alpha_3\alpha_4}\ ,\\
    K_{C, \sym -}^{ab}\equiv&\delta^{ab} K_{C, \sym -}\equiv{1\over 2\ind_\sym}((\mT_\sym^a)^t)^{\alpha_2\alpha_1}(\mT_\sym^b)^{\alpha_4\alpha_3} (\boldsymbol{K}_C)^{\alpha_1\alpha_2\alpha_3\alpha_4}\ ,
\end{align}
where we took the transpose of one generator in order to respect the matrix multiplication~\eqref{eq: matrix multiplication for tensor model}. Therefore, we get
\begin{align}
    K_{C,-}=&-(q-1)\effcoupling^2 \psi_{cl}(\tau_{13}) \psi_{cl}(\tau_{24}) [\psi_{cl}(\tau_{34})]^{q-2}=K_-\ ,\\
    K_{C,\anti +}=&-  \effcoupling^2 \psi_{cl}(\tau_{13}) \psi_{cl}(\tau_{24}) [\psi_{cl}(\tau_{34})]^{q-2}=K_+\ ,\\
    K_{C, \sym -}=&-  \effcoupling^2 \psi_{cl}(\tau_{13}) \psi_{cl}(\tau_{24}) [\psi_{cl}(\tau_{34})]^{q-2}=K_+\ .
\end{align}
One can decompose the first term~\eqref{eq: first term in geometric series} of the geometric series of the ladder diagrams into the singlet, anti-symmetric and symmetric irreducible representations of $SO(M)$:
\begin{align}
\mathcal{F}_{-,0}\equiv&{1\over M}\delta^{\alpha_2\alpha_1}\delta^{\alpha_4\alpha_3} (\boldsymbol{\mathcal{F}}_0)^{\alpha_1\alpha_2\alpha_3\alpha_4}\ ,\\
\mathcal{F}_{\anti +,0}^{ab}\equiv&\delta^{ab} \mathcal{F}_{\anti +,0}\equiv{1\over 2\ind_\anti}(\mT^a_\anti)^{\alpha_2\alpha_1}(\mT^b_\anti)^{\alpha_4\alpha_3}(\boldsymbol{\mathcal{F}}_0)^{\alpha_1\alpha_2\alpha_3\alpha_4}\ ,\\
\mathcal{F}_{\sym -, 0}^{ab}\equiv&\delta^{ab}\mathcal{F}_{\sym -, 0}\equiv{1\over 2\ind_\sym}(\mT^a_\sym)^{\alpha_2\alpha_1}(\mT^b_\sym)^{\alpha_4\alpha_3} (\boldsymbol{\mathcal{F}}_0)^{\alpha_1\alpha_2\alpha_3\alpha_4}\ .
\end{align}
where $\ind_R$ is the Dynkin index of representation $R$. Note that the zero-rung ladder diagram (the first term of the geometric series) in the anti-symmetric and symmetric channels are diagonal in $a,b$ space. Hence, we have
\begin{align}
\mathcal{F}_{-,0}=& -\psi_{cl}(\tau_{13})\psi_{cl}(\tau_{24})+\psi_{cl}(\tau_{14})\psi_{cl}(\tau_{23})\ ,\\
\mathcal{F}_{\anti +,0}\equiv& \mathcal{F}_{+,0}= \psi_{cl}(\tau_{13})\psi_{cl}(\tau_{24})+\psi_{cl}(\tau_{14})\psi_{cl}(\tau_{23})\ ,\\
\mathcal{F}_{\sym -, 0}=&-\psi_{cl}(\tau_{13})\psi_{cl}(\tau_{24})+\psi_{cl}(\tau_{14})\psi_{cl}(\tau_{23})=\mathcal{F}_{-,0}\ .
\end{align}
In this section, we will denote by $\mathcal{F}_{+,0}$ the zero-rung diagram of the anti-symmetric channel $\mathcal{F}_{\anti +,0}$. Note that the first terms and the common ratios of the Cooper channel are the same as those in the SYK model with global symmetry after decomposition. Finally, we have
\begin{align}
\mathcal{F}_{C,-}=&{1\over 1-K_{C,-}}\mathcal{F}_{-,0}={1\over 1-K_{-}}\mathcal{F}_{-,0}\ ,\\
\mathcal{F}_{C,\anti +}=&{1\over 1-K_{C,\anti +}}\mathcal{F}_{\anti +,0}={1\over 1-K_+}\mathcal{F}_{ +,0}\ ,\\
\mathcal{F}_{C,\sym -}=&{1\over 1-K_{C,\sym -}}\mathcal{F}_{\sym -,0}={1\over 1-K_+}\mathcal{F}_{ -,0}\ .
\end{align}

\item \textbf{Pillow Channel of order $\mathcal{O}(N^{q})$}: We can also decompose the Pillow channel ladder diagram and the common ratio of the geometric series:
\begin{align}
\mathcal{F}_{P,\col,-}\equiv& {1\over M}\delta^{\alpha_2\alpha_1}\delta^{\alpha_4\alpha_3} \boldsymbol{\mathcal{F}}_{P,\col}^{\alpha_1\alpha_2\alpha_3\alpha_4}\ ,\\
\mathcal{F}_{P,\col, \anti +}^{ab}\equiv&\delta^{ab} \mathcal{F}_{P,\col, \anti +}\equiv {1\over 2\ind_\anti}(\mT^a_\anti)^{\alpha_2\alpha_1}(\mT^b_\anti)^{\alpha_4\alpha_3} \boldsymbol{\mathcal{F}}_{P,\col}^{\alpha_1\alpha_2\alpha_3\alpha_4}\ ,\\
\mathcal{F}_{P,\col,\sym -}^{ab}\equiv&\delta^{ab} \mathcal{F}_{P,\col,\sym -}\equiv {1\over 2\ind_\sym}(\mT^a_\sym)^{\alpha_2\alpha_1}(\mT^b_\sym)^{\alpha_4\alpha_3} \boldsymbol{\mathcal{F}}_{P,\col}^{\alpha_1\alpha_2\alpha_3\alpha_4}\ ,
\end{align}
and
\begin{align}
    K_{P,\col, -}\equiv&{1\over M}\delta^{\alpha_2\alpha_1}\delta^{\alpha_4\alpha_3} (\boldsymbol{K}_{P,\col, -})^{\alpha_1\alpha_2\alpha_3\alpha_4}\ ,\\
    K_{P,\col, \anti +}^{ab}\equiv&\delta^{ab} K_{P,\col, \anti +}\equiv{1\over 2\ind_\anti}((\mT_\anti^a)^t)^{\alpha_2\alpha_1}(\mT_\anti^b)^{\alpha_4\alpha_3} \boldsymbol(\boldsymbol{K}_{P,\col, \anti +})^{\alpha_1\alpha_2\alpha_3\alpha_4}\ ,\\
    K_{P,\col, \sym -}^{ab}\equiv&\delta^{ab}K_{P,\col, \sym -}\equiv{1\over 2\ind_\sym}((\mT_\sym^a)^t)^{\alpha_2\alpha_1}(\mT_\sym^b)^{\alpha_4\alpha_3} \boldsymbol(\boldsymbol{K}_{P,\col, \sym -})^{\alpha_1\alpha_2\alpha_3\alpha_4}\ .
\end{align}
Here, we also took the transpose of the one generator in the decomposition of the common ratio because of the matrix multiplication~\eqref{eq: matrix multiplication for tensor model}. For the Pillow channel of the color $\col_1$, we have
\begin{align}
    K_{P,\col_1, -}=&0\ ,\\
    K_{P,\col_1, \anti +}=&- \effcoupling^2 \psi_{cl}(\tau_{13}) \psi_{cl}(\tau_{24}) [\psi_{cl}(\tau_{34})]^{q-2}=K_+\ ,\\
    K_{P,\col_1, \sym -}=&- \effcoupling^2 \psi_{cl}(\tau_{13}) \psi_{cl}(\tau_{24}) [\psi_{cl}(\tau_{34})]^{q-2}=K_+\ ,
\end{align}
and, for the Pillow channels of other colors $\col_m$ ($m=2,3,\cdots, q-1$), the common ratios are given by
\begin{align}
    K_{P,\col_n, -} =&-\effcoupling^2 \psi_{cl}(\tau_{13}) \psi_{cl}(\tau_{24}) [\psi_{cl}(\tau_{34})]^{q-2}=K_+\ ,\\
    K_{P,\col_n, \anti +}=&0\ ,\\
    K_{P,\col_n, \sym -}=&0\ .
\end{align}
As in the Cooper channel, we decompose the zero-rung ladder diagrams (the first terms of the geometric series) in \eqref{eq: first term in geometric series pillow} into the irreducible representation of $SO(M)$:
\begin{align}
\mathcal{F}_{P^\mp,-,0}=& \mp\psi_{cl}(\tau_{13})\psi_{cl}(\tau_{24})+\psi_{cl}(\tau_{14})\psi_{cl}(\tau_{23})=\mathcal{F}_{\mp,0} \ ,\\
\mathcal{F}_{P^\mp,\anti +,0}=& \pm\psi_{cl}(\tau_{13})\psi_{cl}(\tau_{24})+\psi_{cl}(\tau_{14})\psi_{cl}(\tau_{23})=\mathcal{F}_{\pm,0} \ ,\\
\mathcal{F}_{P^\mp,\sym -, 0}=&\mp\psi_{cl}(\tau_{13})\psi_{cl}(\tau_{24})+\psi_{cl}(\tau_{14})\psi_{cl}(\tau_{23})=\mathcal{F}_{\mp,0} \ .
\end{align}
The decomposition of the zero-rung ladder diagrams with respect to $O(N)$ and $SO(M)$ leads to the various decomposition of the Pillow channels. Note that the common ratio of the Pillow channel is independent of the $O(N)$ decomposition. \eg $\mathcal{F}_{P^\mp,\col,-}$ have the same common ratio, and the difference comes from the zero-rung ladder diagram. Hence, the leading ladder diagrams of the Pillow channels of the color $\col_1$ are given by
\begin{align}
\mathcal{F}_{P^-, \col_1,-}=&0\ ,\\
\mathcal{F}_{P^-, \col_1,\anti +}=&{1\over 1-K_{P,\anti +}}\mathcal{F}_{P^-,\anti +,0}={1\over 1-K_+}\mathcal{F}_{ +,0}\ ,\\
\mathcal{F}_{P^-, \col_1,\sym -}=&{1\over 1-K_{P,\sym -}}\mathcal{F}_{P^-,\sym -,0}={1\over 1-K_+}\mathcal{F}_{ -,0}\ .\\
\mathcal{F}_{P^+, \col_1,-}=&0\ ,\\
\mathcal{F}_{P^+, \col_1,\anti +}=&{1\over 1-K_{P,\anti +}}\mathcal{F}_{P^+,\anti +,0}={1\over 1-K_+}\mathcal{F}_{- ,0}\ ,\\
\mathcal{F}_{P^+, \col_1,\sym -}=&{1\over 1-K_{P,\sym -}}\mathcal{F}_{P^+,\sym -,0}={1\over 1-K_+}\mathcal{F}_{+ ,0}\ .
\end{align}
On the other hand, we obtain those of the other colors $\col_m$ ($m=2,3,\cdots,q-1$):
\begin{align}
\mathcal{F}_{P^-, \col_m , -}=&{1\over 1-K_{P,-}}\mathcal{F}_{P^-,-,0}={1\over 1-K_+}\mathcal{F}_{ -,0}\ ,\\
\mathcal{F}_{P^+, \col_m , -}=&{1\over 1-K_{P,-}}\mathcal{F}_{P^+,-,0}={1\over 1-K_+}\mathcal{F}_{+,0}\ ,\\
\mathcal{F}_{P^\mp, \col_m ,\anti +}=&\mathcal{F}_{P^\mp, \col_m ,\sym -}=0\ .
\end{align}
Note that the ladder diagrams $\mathcal{F}_{P^-, \col_1,\anti +}$, $\mathcal{F}_{P^+, \col_1,\sym -}$ and $\mathcal{F}_{P^+, \col_m, -}$ ($m\ne 1$) are the same as the anti-symmetric channel of the SYK model with $SO(M)$ global symmetry, and the other non-zero ladder diagrams $\mathcal{F}_{P^-, \col_1,\sym -}$, $\mathcal{F}_{P^+, \col_1,\anti +}$ and $\mathcal{F}_{P^-, \col_m,-}$ ($m\ne 1$) are analogous to the symmetric channel of the SYK model with global symmetry.

\end{itemize}

According to the analysis in Section~\ref{sec: lyapunov exponent}, when evaluating the out-of-time-ordered correlator, one can expect that there will be no exponential growth except for the linear growth in various channels which are analogous to anti-symmetric channel in the SYK model with $SO(M)$ global symmetry.

In~\cite{Narayan:2017qtw}, it was shown that although there is no exponential growth in the large $N$ leading ladder diagram of the Pillow channel, the subleading ladder diagram of the Pillow channel would grow exponentially with the growth rate ${2\pi \over \beta}$. Recall that  the leading ladder diagram of the Pillow channel consists of dipoles of the same color as that of the Pillow. The rest of combinations are subleading. \ie Broken ladder diagrams of order $\mathcal{O}(N^{q-2})$. Note that the anti-symmetric part of the external $O(N)$ indices in the Pillow channels will not give a contribution to the broken ladder diagram because of the $O(N)$ tensor structure of the dipoles. Hence, for the the broken ladder diagrams of color $\col_m$, one may consider the symmetric part of the external $O(N)$ indices of color $\col_m$ in the Pillow channels. By summing up those subleading dipoles of order $\mathcal{O}(N^{q-2})$, we have
\begin{align}
\boldsymbol{\mathcal{F}}_{P^-,\col_m}^{\text{\tiny subleading}} =& \sum_{n=0}^\infty \left(\sum_{k=1}^{q-1} \dipole_{\col_k}\right)^n \boldsymbol{\mathcal{F}}_{P^-,0}- \sum_{n=0}^\infty  (\dipole_{\col_m})^n \boldsymbol{\mathcal{F}}_{P^-,0}\cr
=&{1\over 1-\boldsymbol{K}_{C}}\boldsymbol{\mathcal{F}}_{0} -{1\over 1-\boldsymbol{K}_{P,\col_m}}\boldsymbol{\mathcal{F}}_{0}\label{eq: subleading pillow}
\end{align}
where $m=1,2,\cdots, q-1$ and we used $\boldsymbol{\mathcal{F}}_{P^-,0}=\boldsymbol{\mathcal{F}}_{0}$. This seems suggest that the subleading ladder diagram grows exponentially with ${2\pi \over\beta}$ rate. But, one also has to consider other contributions which has not been considered. Namely, in rank-3 tensor model, this subleading ladder diagram is suppressed by the non-melonic diagrams which has not been evaluated yet. However, in rank $(q-1)$ tensor model, \cite{Narayan:2017qtw} showed that the order $N$ of the (melonic) ladder diagrams of both Cooper and Pillow Channel increases with $q$ while the order $N$ of the non-melonic diagrams do not. Schematically, the order $N$ of the four point function of Cooper and Pillow channels are given~\cite{Narayan:2017qtw} by
\begin{alignat}{3}
    &F_{C}\sim&& \underbrace{N^{2(q-1)}\psi_{cl}\psi_{cl}}_{\text{leading disconnected}}\;\; +\;\;\underbrace{N^{q-1}\mathcal{F}_C^{\text{\tiny leading}}}_{\text{ladder}}&&+N^{4}(\text{Non-melonic})\label{eq: cooper order}\ ,\\
    &F_{P,\col}\sim&& \underbrace{N^{2(q-1)-1}\psi_{cl}\psi_{cl}}_{\text{leading disconnected}} \;\;+\;\; \underbrace{N^{q}\mathcal{F}_P^{\text{\tiny leading}}}_{\text{leading ladder}}\;\;+\;\; \underbrace{N^{q-2}\mathcal{F}_P^{\text{\tiny subleading}}}_{\text{subleading ladder}}&&+N^{4}(\text{Non-melonic})\label{eq: pillow order}\ .
\end{alignat}
Therefore, for $q>6$, the non-melonic diagrams are suppressed by the ladder diagrams in~\eqref{eq: subleading pillow} which are indeed the subleading contributions to the ladder diagrams of the Pillow channel and grow exponentially with maximal growth ratio ${ 2 \pi \over \beta }$.

This maximal growth in subleading ladder diagram could be explained by assuming that there exists the Schwarzian effective action. The effective action of the tensor models has not been derived yet because there are too many invariant operators and we do not have collective action for the tensor model. Nevertheless, one can expect Schwarzian action for the tensor model because the pattern of symmetry breaking is reminiscent to that of the SYK model. If exists, the effective action is proportional to $N^{q-1}$. Then, one may estimate the long time behavior of each channel by using method discussed in~Section~\ref{sec: effective action}. Recalling the tensor structure of the leading disconnected diagrams~\eqref{def:ladder diagram cooper channel} and \eqref{def:ladder diagram pillow channel}, one can immediately conclude that the singlet Cooper and singlet Pillow channel will grow exponentially with maximal growth rate ${2\pi \over \beta}$ as in the SYK model. Furthermore, when we evaluate this exponential growth from the leading disconnected diagrams, the order $N$ of them are decreased by $N^{q-1}$ which is the $N$ scaling of the effective action. Hence, from~\eqref{eq: cooper order} and \eqref{eq: pillow order}, one can see that this exponential growth is of order $\mathcal{O}(N^{q-1})$ for the singlet Cooper channel and is of order $\mathcal{O}(N^{q-2})$ for the singlet Pillow channel, which is consistent with \eqref{eq: subleading pillow}.

However, we do not claim that the Pillow channel saturate chaos bound because we have not considered ${1\over \beta \effcoupling}$ correction to the leading ladder diagram. The Cooper channel does saturate chaos bound since the leading ladder diagram has maximal Lyapunov exponent. On contrary, the leading ladder diagram of the Pillow channel does not grow exponentially, and one has to consider the $1/\beta\effcoupling$ corrections which is higher order than the subleading ladder diagram in $N$. Unless one proves that all $1/\beta\effcoupling$ corrections do not grow exponentially, one cannot conclude that the Pillow channel saturate the chaos bound because of the $1/N$ subleading ladder diagram. Our claim is that there exists exponentially growing terms with maximal growth ratio which comes from the effective action if the effective action exists.

\section{3D Gravity : $q=4$}
\label{sec: 3d gravity}

It is highly interesting to investigate the holographic dual of the SYK model. The low energy modes have been captured in the dilaton gravity~\cite{Maldacena:2016upp} and the Liouville theory~\cite{Mandal:2017thl} in the context of 2D gravity. On the other hand, the non-zero modes of the SYK model suggest an infinite tower of particles in the matter sector. In~\cite{Das:2017pif},  for $q=4$ case, this infinite tower of particles was interpreted as Kaluza-Klein modes from 3D gravity. In our model with global symmetry, we have obtained additional infinite towers of spectrums. Hence, following~\cite{Das:2017pif}, we will also interpret them as Kaluza-Klein modes.

In Section~\ref{sec: quadratic action}, we have found the spectrum of our model. In particular, the spectrum of the $q=4$ case is determined by simple trigonometrc functions:
\begin{alignat}{2}
    &p_m^-\hspace{5mm} (m=0,1,2,\cdots)\hspace{5mm} &&:\quad -{3\tan{\pi p_m^- \over 2}\over 2p_m^- }=1\ ,\label{eq: spectrum m q4}\\
    &p_m^{\anti +} \hspace{5mm} (m=0,1,2,\cdots)\hspace{5mm} &&:\quad {\cot{\pi p_m^{\anti +} \over 2}\over 2p_m^{\anti +}}=1\ ,\label{eq: spectrum p q4}\\
    &p_m^{\sym -} \hspace{5mm} (m=0,1,2,\cdots)\hspace{5mm} &&:\quad -{\tan{\pi p_m^{\sym -} \over 2}\over 2p_m^{\sym -}}=1\label{eq: spectrum sym p q4}
\end{alignat}
where we used $p_m^- \equiv h_m^- -{1\over 2}$ and similar for $p_m^{\anti +}$ and $p_m^{\sym -}$. For this case, we will briefly extend the conjecture by~\cite{Das:2017pif} on the 3D bulk dual of the SYK model. \cite{Das:2017pif} interpreted the spectrum of the SYK model as the Kaluza-Klein modes of the three-dimensional scalar field. For this, let us begin by a simple quantum mechanical problem~\cite{Das:2017pif}: two Schr{\"o}dinger equations with delta function potentials
\begin{equation}
    \left[-\partial_y^2 +V_\mp\delta(y)\right]f_\mp(y)=E^\mp f_\mp(y)\hspace{10mm} (i=1,2)
\end{equation}
where the coordinate $y$ is restricted to an interval:
\begin{equation}
    -L\leqq y\leqq L\ .
\end{equation}
At the boundary $y=\pm L$, we demand different boundary conditions:
\begin{alignat}{2}
    &f_-(\pm L)=0\hspace{5mm} &&:\;\; \mbox{Dirichlet}\ ,\\
    &f'_+(\pm L)=0\hspace{5mm} &&:\;\; \mbox{Neumann}\ .
\end{alignat}
%
%
%
For these boundary conditions, general solutions with even parity can be written as
\begin{align}
    f_-(y)=&\begin{cases}
    \; A\sin p^-(y-L) & \quad (0\leqq y\leqq L)\\
    \; -A\sin p^-(y+L) & \quad (-L\leqq y\leqq 0)\\
    \end{cases}\\
    f_+(y)=&\begin{cases}
    \; B\cos p^+(y-L) & \quad (0\leqq y\leqq L)\\
    \; B\cos p^+(y+L) & \quad (-L\leqq y\leqq 0)\\
    \end{cases}
\end{align}
where the normalization is given by
\begin{equation}
    A=\sqrt{2p^-\over 2p^-L-\sin (2p^-L)}\quad,\quad B=\sqrt{2p^+\over 2p^+L+\sin (2p^+L)}\ .
\end{equation}
The continuity condition at $y=0$ leads to quantization condition:
%
%
\begin{align}
    -{2\over V_1}p^-=&\tan p^-L\ ,\\
    {2\over V_2}p^+=&\cot p^+L\ .
\end{align}
By choosing
\begin{equation}
    L={\pi \over 2}\quad,\quad V_-=3\quad,\quad V_{\anti +}=1\quad,\quad V_{\sym -}=1\ ,
\end{equation}
one can get the spectrum $p^-_m$ in \eqref{eq: spectrum m q4}, $p^{\anti +}_m$ in \eqref{eq: spectrum p q4} and $p^{\sym -}_m$ in \eqref{eq: spectrum sym p q4}. To make contact with the SYK model, we generalize the action of the three-dimensional scalar fields in the near-horizon limit of a charged extremal BTZ black hole proposed by~\cite{Das:2017pif}:
\begin{align}
    S_{3D}=&{1\over 2}\int d^3x\sqrt{|g|} \left[-g^{BC}\partial_B \Phi_- \partial_C \Phi_- - m_-^2 \Phi_-^2 -V_-\delta(y) \Phi_-^2\right]\cr
    &+{1\over 2}\int d^3x\sqrt{|g|} \left[-g^{BC}\partial_B \Phi_{\anti +} \partial_C \Phi_{\anti +} - m_{\anti +}^2 \Phi_{\anti +}^2 -V_{\anti +}\delta(y) \Phi_{\anti +}^2\right]\cr
    &+{1\over 2}\int d^3x\sqrt{|g|} \left[-g^{BC}\partial_B \Phi_{\sym -} \partial_C \Phi_{\sym -} - m_{\sym -}^2 \Phi_{\sym -}^2 -V_{\sym -}\delta(y) \Phi_{\sym -}^2\right]\ .\label{eq: 3d gravity action1}
\end{align}
where three-dimensional metric is given by
\begin{equation}
    ds^2= \tilde{g}_{\mu\nu}dx^\mu dx^\nu +dy^2={1\over z^2}\left(-dt^2+dz^2\right) +dy^2\ .\label{eq: 3D metric}
\end{equation}
Note that we choose the constant dilaton $\phi_0=1$ for simplicity, which is enough for our discussion.\footnote{In fact, the ${a\over z}$ correction to the constant dilaton provides a strong evidence for the conjecture. But, we will not discuss in detail in this paper, and refer the readers~\cite{Das:2017pif} for the details.} We also demand the boundary condition on the boundary $y=\pm L$:
\begin{align}
    \Phi_-(t,z;y=\pm L)=0\hspace{2mm},\hspace{5mm} \partial_y \Phi_{\anti +}(t,z;y=\pm L)=0\hspace{2mm},\hspace{5mm}  \Phi_{\sym -}(t,z;y=\pm L)=0\ .
\end{align}
Using the function $f_\mp(y)$ with suitable normalization, one can decompose the three-dimensional scalar field into a tower of the Kaluza-Klein modes~\cite{Das:2017pif}:
\begin{equation}
    \Phi_\mp(t,z;y)=\sum_{m=0}^\infty \phi_{\mp,m}(t,z) f_{\mp,m}(y)\ .
\end{equation}
If we choose the mass to be the BF mass bound $m^2_\mp=-{1\over 4}$~\cite{Das:2017pif}, the action can be written as
\begin{align}
    S_{3D}=&{1\over 2}\sum_{m=0}^\infty  \int d^2x\sqrt{|\tilde{g}|} \left[-\tilde{g}^{\mu\nu}\partial_\mu \phi_{-,m} \partial_\nu \phi_{-,m} -\left( (p_m^-)^2-{1\over 4}  \right)\phi_{-,m}^2 \right]\cr
    &+{1\over 2}\sum_{m=0}^\infty \int d^2x\sqrt{|\tilde{g}|} \left[-\tilde{g}^{\mu\nu}\partial_\mu \phi_{\anti +,m} \partial_\nu \phi_{\anti +,m} - \left((p_m^{\anti +})^2 - {1\over 4} \right)\phi_{\anti +,m}^2 \right]\cr
    &+{1\over 2}\sum_{m=0}^\infty \int d^2x\sqrt{|\tilde{g}|} \left[-\tilde{g}^{\mu\nu}\partial_\mu \phi_{\sym -,m} \partial_\nu \phi_{\sym -,m} - \left((p_m^{\sym -})^2 - {1\over 4} \right)\phi_{\sym -,m}^2 \right]\ .
\end{align}
It was shown in~\cite{Das:2017pif} that the propagators of the three-dimensional scalar field, 
\begin{equation}
\langle\Phi_-(t,z;0)\Phi_-(t',z';0)\rangle
\end{equation}
agrees with the four point function of the SYK model. In this proof, the following identity plays a crucial role in transform the residue of ${k\over 1-k}$ at $\nu=p_m^-$ to the square of eigenfunction at $y=0$:
\begin{equation}
    \underset{\nu=p_m^-}{\mbox{Res}}{k_-\left(\nu+{1\over 2}\right)\over 1 - k_-\left(\nu+{1\over 2}\right)}= {3(p_m^-)^2\over \left[(p_m^-)^2+(3/2)^2\right]\left[\pi p_m^- -\sin(\pi p_m^-)\right]}= {3\over 2 p_m^-}f_{-,m}(0)f_{-,m}(0)\ .
\end{equation}
Likewise, by using the identity
\begin{align}
    \underset{\nu=p_m^{\anti +}}{\mbox{Res}}{k_{\anti +}\left(\nu+{1\over 2}\right)\over 1 - k_{\anti +}\left(\nu+{1\over 2}\right)}=& {(p_m^{\anti +})^2\over \left[(p_m^{\anti +})^2+(1/2)^2\right]\left[\pi p_m^{\anti +} +\sin(\pi p_m^{\anti +})\right]} = {f_{+,m}(0)f_{+,m}(0)\over 2p_m^{\anti +}}\ ,\\
        \underset{\nu=p_m^{\sym -}}{\mbox{Res}}{k_{\sym -}\left(\nu+{1\over 2}\right)\over 1 - k_{\sym -}\left(\nu+{1\over 2}\right)}=& {(p_m^{\sym -})^2\over \left[(p_m^{\sym -})^2+(1/2)^2\right]\left[\pi p_m^{\sym -} - \sin(\pi p_m^{\sym -})\right]} = {f_{-,m}(0)f_{-,m}(0)\over 2p_m^{\sym -}}\ ,
\end{align}
one can easily see that the anti-symmetric and symmetric channel correlation functions agree with 
\begin{equation}
    \langle\Phi_{\anti +}(t,z;0)\Phi_{\anti +}(t',z';0)\rangle\quad, \quad  \langle\Phi_{\sym -}(t,z;0)\Phi_{\sym -}(t',z';0)\rangle\ ,
\end{equation}
respectively. We leave the further analysis related to $1/J$ correction, which was carried out in~\cite{Das:2017pif}, for the future work.

Now, we will consider an action for the three-dimensional scalar fields similar to \eqref{eq: 3d gravity action1}, which does not have delta function potential. This is not a new approach at all, but we follow the same idea as in~\cite{Das:2017pif}, and elaborate it with Robin boundary condition. Let us consider the following action for three scalar fields in $AdS_2\times I$:
\begin{align}
    S'_{3D}=&\int d^2x dy \sqrt{|g|}\left[-{1\over 2} g^{BC} \partial_B\Phi_- \partial_C\Phi_- -\zeta \mathcal{R}\Phi_-^2-{1\over 2} g^{BC} \partial_B\Phi_{\anti +} \partial_C\Phi_{\anti +} -\zeta \mathcal{R}\Phi_{\anti +}^2\right]\cr
    &+\int d^2x dy \sqrt{|g|}\left[-{1\over 2} g^{BC} \partial_B\Phi_{\sym -} \partial_C\Phi_{\sym -} -\zeta \mathcal{R}\Phi_{\sym -}^2\right]
\end{align}
where the metric is the same as before~\eqref{eq: 3D metric}, but we chose a different range of the extra coordinate:
\begin{equation}
    0\leqq y \leqq L\ .
\end{equation}
When $\zeta={1\over 16}$, this action is invariant under the Weyl transformation
\begin{align}
    g_{BC}\quad&\longrightarrow \quad \Omega^2 g_{BC}\ ,\\
    \Phi_-\;,\;\Phi_{\anti +}\;,\; \Phi_{\sym -}  \quad &\longrightarrow \quad \Omega^{-{1\over 2}} \Phi_-\; ,\;  \Omega^{-{1\over 2}} \Phi_{\anti +}\; ,\;  \Omega^{-{1\over 2}}  \Phi_{\sym -} \ .
\end{align}

Now, we impose boundary conditions at $y=\pm L$. Note that the Dirichlet boundary condition is invariant under the Weyl transformation while the Neumann condition is not:
\begin{align}
    \Phi(t,z;y_0)=0\quad&\longrightarrow \quad \Phi(t,z;y_0)=0 \ ,\\
    \left.\partial_y \Phi(t,z;y)\right|_{y=y_0}=0\quad&\longrightarrow \quad  \left.\partial_y \Phi(t,z;y)\right|_{y=y_0}-{1\over 2}\left.\left(\partial_y \log \Omega(y)\right)\right|_{y=y_0}\Phi(t,z;y_0)=0\ .
\end{align}
For the invariant boundary condition, we consider the following Robin boundary condition:
\begin{equation}
    \left.\left(\Xi(y)+\partial_y \right)\Phi(t,z;y)\right|_{y=y_0}=0\ .
\end{equation}
This Robin boundary condition is invariant under the Weyl transformation if $\Xi(y)$ also transforms as~\cite{Kennedy:1979ar,Romeo:2000wt}
\begin{equation}
    \Xi(y)\quad\longrightarrow \quad \Xi(y)+{1\over 2}\left.\left(\partial_y \log \Omega(y)\right)\right|_{y=y_0}\ .
\end{equation}
We demand the following Robin boundary condition assuming that there exists a suitable~$\Xi$:
\begin{alignat}{2}
    &\Phi_-(t,z;0)=0\hspace{5mm},&&\hspace{5mm} \left.\partial_y \Phi_-(t,z;y)\right|_{y=L} - {3\over 2} \Phi_-(t,z;L)=0\ ,\\
    &\left.\partial_y \Phi_{\anti +}(t,z;y)\right|_{y=0}=0\hspace{5mm},&&\hspace{5mm}\left.\partial_y \Phi_{\anti +}(t,z;y)\right|_{y=L} - {1\over 2} \Phi_{\anti +}(t,z;L)=0\ ,\\
    &\Phi_{\sym -}(t,z;0)=0\hspace{5mm},&&\hspace{5mm} \left.\partial_y \Phi_{\sym -}(t,z;y)\right|_{y=L} - {1\over 2} \Phi_{\sym -}(t,z;L)=0\ .
\end{alignat}
One can repeat the same calculations as before, and obtain the same result. For this case, the odd parity solution in the previous KK modes is naturally excluded. Furthermore, the  to Ricci scalar $\mathcal{R}=-2$ of AdS$_2$ leads to $m^2_\pm =-{1\over 4}$. Note that we switch the boundary conditions at $y=0$ and $y=L$ for simplicity. Hence, the correlation functions of the 3D scalar field evaluated at $y=L$ will agree with the correlation functions of our model.
%

\section{Conclusion}
\label{sec: conclusion}

We have explored the SYK model and the SYK-like tensor model with global symmetry. The global symmetry is enhanced to local symmetry at strong coupling limit. Furthermore, the emergent local symmetry together with the reparametrization symmetry is spontaneously broken by the large $N$ classical solution as well as explicitly broken by the kinetic term of the action. This leads to Pseudo-Nambu-Goldstone bosons and the corresponding low energy effective action is found to be a sum of the Schwarzian action and the action of a particle on the group manifold. In both SYK and tensor models with global symmetry, the four point functions have a richer structure, and we have additional infinite towers of spectrum. For the SYK model, the singlet channel four point function saturates the chaos bound\footnote{Not all singlet channel saturate chaos bound. See $U(M)$ case in Appendix~\ref{app: su n}.} due to the zero mode of reparametrization. On the other hand, the anti-symmetric channel does not grow exponentially, but shows a linear growth of which the leading contribution comes from the zero mode of the local $SO(M)$ symmetry. Moreover, there is no contribution of the effective action to the symmetric channel, and it does not exhibit exponential growth.

For the tensor model with global symmetry, we first consider the rank-3 tensor model, and generalize it into rank-$(q-1)$ tensor model by using ``uncoloring'' process~\cite{Narayan:2017qtw}. We showed that the Cooper channel exhibits a similar result as in the SYK model with global symmetry. The Pillow channel does not grow exponentially at the leading order in large $N$. However, for the rank-$(q-1)$ tensor model ($q>6$), the subleading ladder diagram in $N$ shows exponential growth with growth rate ${2\pi \over \beta}$. We also extended the 3D gravity conjecture~\cite{Das:2017pif}, and confirm that the Kaluza-Klein modes could phenomenologically explain the various spectra of our model for $q=4$ case.

Here, we describe future directions and questions that remain to be addressed:

\begin{itemize}

\item One can also consider a large $N$ classical solution which break global $SO(M)$ group. The saddle point equation~\eqref{eq: saddle point equation} implies that such a solution should be proportional to a matrix of which square is the identity matrix. Hence, after diagonalization, the general solution can be written as
\begin{equation}
    \maPsi_{cl}(\tau_1,\tau_2)=\begin{pmatrix}
    \idm_{m} & 0\\
    0 & -\idm_{M-m}\\
    \end{pmatrix}\psi_{cl}(\tau_1,\tau_2)
\end{equation}
where $0\leqq m\leqq M$, and $\psi_{cl}$ is the same as~\eqref{eq: classical solution}. We leave this for future work.

\item It is interesting to study the central extension of $\diffeo(S^1)\ltimes \hat{g}$ and the corresponding coadjoint orbit action explicitly. Furthermore, the generalization to $\mathcal{W}$ symmetry is highly interesting.

\item In a tensor model with global symmetry, the leading ladder diagram of the Pillow channel does not have exponential growth while the subleading ladder diagram in $N$ does grow exponentially with maximal growth rate as mentioned above. In order to prove that the Pillow channel also saturates the chaos bound, one need to show that the ${1\over \beta\effcoupling}$ correction to the leading ladder diagram does not grow exponentially. We also leave this for future work.

\end{itemize}

\acknowledgments

I thank Prithvi Narayan and Victor Ivan Giraldo Rivera for collaboration during the initial stages of this project. I thank Yang Lei, Sang-Jin Shin, Changhyun Ahn, Sungjay Lee, Dario Rosa, Rajesh Gopakumar, Spenta Wadia, Robert de Mello Koch, Antal Jevicki and R. Loganayagam for extensive discussions. I thank Kyung Hee University for the hospitality and the National Research Foundation of Korea for generous support during the completion of this work, within the program ``New ideas on higher spin gravity and holography''. I also thank Korea Institute for Advanced Study~(KIAS) and Hanyang University for the hospitality during the completion of this work. I also thank the organizers of ``Bangalore Area Strings Meeting 2017'' for giving an opportunity to present this work. I gratefully acknowledge support from International Centre for Theoretical Sciences~(ICTS), Tata institute of fundamental research, Bengaluru. I would also like to acknowledge our debt to the people of India for their steady and generous support to research in the basic sciences.

\appendix

\section{Generalized Collective Action: Schur Polynomial}
\label{app: schur collective action}

In this appendix, we will study a general form of the $SO(M)$ invariant actions for $q=2\hq$ case:
\begin{equation}
     S_R = \int d\tau\left[\chi^{i \alpha}  \partial_\tau \chi^{i \alpha} +  i^{q\over 2} \sum_{\rho\in S_\hq} \ch_R(\rho) J_{i_1 \dots i_{q}} \chi^{i_1 \alpha_1}\chi^{i_2 \alpha_{\rho(1)}}  \dots   \chi^{i_{q-1} \alpha_\hq}\chi^{i_{q} \alpha_{\rho(\hq)}}    \right]\label{def: general q action with permutation}
\end{equation}
where we omitted the summation over $SO(N)$ index $i$'s and $SO(M)$ index $\alpha$'s. And, $\ch_R(\rho)$ denotes the $S_\hq$ character of $\rho\in S_\hq$ in the representation $R$. Since we do not assume any symmetry of the random coupling constant $J_{i_1\cdots i_q}$, one might expect $S_q$ permutation of $q$ $SO(N)$ indices\footnote{More precisely, there would be ${q!\over 2^{q\over 2}\left({q\over 2}\right)!}$ independent $SO(M)$ invariant interactions.} rather than $S_\hq=S_{q/2}$. After a disorder average of the random coupling constant $J_{i_1\cdots i_q}$, not all interactions give distinct results, and therefore it is enough to consider $S_\hq=S_{q/2}$ permutations in~\eqref{def: general q action with permutation}. As before, we will treat $J_{i_1\cdots i_q}$ as an additional non-dynamical field and will perform the disorder average over the Gaussian distribution given by
\begin{equation}
    \exp\left[- {N^{q-1}(\hq!)^2\dim_G(R)\over J^2 M\dim_{S_\hq}(R) }\sum_{i_1,\cdots,i_q=1}^N J_{i_1\cdots i_q}J_{i_1\cdots i_q}\right]\ .
\end{equation}
Here, $\dim_G(R)$ denotes the dimension of the representation $R$ of the corresponding group $G$. After the disorder average, we use the following generalized orthogonality of the characters to obtain the collective action:
\begin{equation}
    {1\over |G|} \sum_{g\in G} \ch_R (gh) \ch_S (g^{-1})=\delta_{R,S} {\ch_R(h)\over \ch_R(1)}
\end{equation}
where $|G|$ is the dimension of group $G$, and $R$ and $S$ are irreducible representation.
%
%
%
%
Then, one can show that the bi-local collective action is 
\begin{align}
   &S_{col,R}={N\over 2}\Tr\left[-D\star \maPsi +\log \maPsi \right] - {N M J^2 \over  4\hq \dim_G (R) }\int d\tau_1d\tau_2 \; P_{R}(-\maPsi(\tau_1,\tau_2)\maPsi(\tau_2,\tau_1))\ .
\end{align}
where $P_R(x)$ is the Schur polynomial of $x$ corresponding to the representation $R$, and the bi-local field $\maPsi(\tau_1,\tau_2)$ is defined as before:
\begin{equation}
    [\maPsi(\tau_1,\tau_2)]^{\alpha_1\alpha_2}\equiv {1\over N}\sum_{i=1}^N \chi^{i \alpha_1}(\tau_1)\chi^{i\alpha_2}(\tau_1)\ .
\end{equation}
One may also consider the generic configuration of $SO(M)$ indices. For this case, the interactions are decomposed into irreducible representations of $S_\hq$. This also leads to the decomposition of $J_{i_1\cdots i_q}$ into the irreducible representations which are transpose\footnote{For example, one can treat the original SYK model as if the only possible $\alpha$ is 1. Then, the only possible representation is Young tableau with one row (fully symmetric one). Then, the coupling constant $J_{i_1\cdots \hq}$ which corresponds to Young tableau with one column (fully anti-symmetric one) will survive.} of the representations appearing in the decomposition of the $SO(M)$ indices. Then, the random coupling constant in each representation are averaged separately. Hence, one can obtain a linear combination of the Schur polynomials corresponding to each representation with positive coefficients.

Now, by varying the collective action with respect to the bi-local field, one can obtain the large $N$ saddle-point equation which corresponds to the Schwinger-Dyson equation for the two point function of the fermions. We also take the $SO(M)$ invariant ansatz
\begin{equation}
    \maPsi_{cl}=\idm \psi_{cl}(\tau_1,\tau_2)\ ,
\end{equation}
and the saddle-point equation is reduced to that of the original SYK model:
\begin{align}
    J^2 [\psi]^{q-1}(\tau_1,\tau_3)\star \psi (\tau_3,\tau_2) = -\delta(\tau_{12})\ .
\end{align}
Hence, as before, the solution is given by
\begin{equation}
   \psi_{cl}(\tau_1,\tau_2)= \coeff{\mbox{sgn}(\tau_1-\tau_2) \over |\tau_1 - \tau_2|^{2\Delta}  }
\end{equation}
where the constant $\coeff$ is defined by
\begin{equation}
    J^2 \coeff^q \pi =\left({1\over 2}-{1\over q}\right) \tan {\pi \over q}\ .
\end{equation}

Now, one can repeat the same analysis as in Section~\ref{sec: bi-local collective action}. We expand the bi-local field around the large $N$ classical solution
\begin{equation}
 \maPsi(\tau_1,\tau_2) = \idm \psi_{cl}(\tau_1,\tau_2) +  \sqrt{{2\over N}} \mfluca(\tau_1,\tau_2) +\sqrt{{2\over N}}\mflucb_\anti(\tau_1,\tau_2)+\sqrt{{2\over N}}\mflucb_\sym(\tau_1,\tau_2)
\end{equation}
where the singlet, the anti-symmetric and the symmetric fluctuations are given by 
\begin{equation}
\begin{split}
\mfluca(\tau_1,\tau_2) =& {1\over \sqrt{M} }\fluca(\tau_1,\tau_2) \; \idm\ ,\\  
\mflucb_\anti(\tau_1,\tau_2)=& {1\over\sqrt{2\ind_\anti}}\flucb^{a}_\anti(\tau_1,\tau_2) \mT^a_\anti \hspace{10mm} (a=1,\cdots, {1\over 2}M(M-1)) \ ,\\
\mflucb_\sym(\tau_1,\tau_2)=& {1\over\sqrt{2\ind_\sym}}\flucb_\sym^{a}(\tau_1,\tau_2) \mT^a_\sym \hspace{10mm} (a=1,\cdots, {1\over 2}M(M+1)-1) \ .
\end{split}
\end{equation}
Then, we consider the quadratic action from the large $N$ expansion of the collective action around the classical solution:
\begin{equation}
    S_{col,R}\left[\maPsi_{cl}+  \sqrt{{2/N}}( \mfluca +\mflucb_\anti+\mflucb_\sym )\right]= N S^{(0)}_{col,R}+ S^{(2)}_{col,R}[\mfluca,\mflucb_\anti,\mflucb_\sym]+ {1\over \sqrt{N}} S^{(3)}_{col,R}[\mfluca,\mflucb_\anti,\mflucb_\sym]+\cdots
\end{equation}
Since the classical solution is proportional to the identity matrix, the calculation for the quadratic action is straightforward as before. In particular, using the fact that the Schur polynomial of the identity matrix is equal to the dimension of the representation \ie
\begin{equation}
    P_R(\idm)=\dim_G(R)\ ,
\end{equation}
we obtain exactly the same quadratic action as in Section~\ref{sec: bi-local collective action}:
\begin{align}
    &S_{col,R}^{(2)}\cr
    =&-{1\over 2}\bitr (\psi_{cl}^{-1}\star\fluca \star\psi_{cl}^{-1} \star \fluca )-{q -1\over 2}J^2\int d\tau_1 d\tau_2 [\psi_{cl}(\tau_1,\tau_2)]^{q-2}\fluca(\tau_1,\tau_2)\fluca(\tau_1,\tau_2)\cr
    &-{1\over 2}\bitr (\psi_{cl}^{-1}\star\flucb^a_\anti \star\psi_{cl}^{-1} \star \flucb^a_\anti )+{1\over 2} J^2\int d\tau_1 d\tau_2 \; [\psi_{cl}(\tau_1,\tau_2)]^{q-2}\flucb^a_\anti(\tau_1,\tau_2)\flucb^a_\anti(\tau_1,\tau_2)\cr
    &-{1\over 2}\bitr (\psi_{cl}^{-1}\star\flucb^a_\sym \star\psi_{cl}^{-1} \star \flucb^a_\sym )-{1\over 2} J^2\int d\tau_1 d\tau_2 \; [\psi_{cl}(\tau_1,\tau_2)]^{q-2}\flucb^a_\sym(\tau_1,\tau_2)\flucb^a_\sym(\tau_1,\tau_2)
\end{align}
Hence, up to quadratic level, one can expect the same physics as before (\eg Pseudo-Nambu-Goldstone boson, spectrum, out-of-time-ordered correlators, Lyapunov exponent etc.). The difference might begin to appear in the cubic interactions of the bi-local fields which are related to the 6-point function of the fermions.

\section{SYK Model with $U(M)$ Global Symmetry}
\label{app: su n}

In this appendix, we present the generalization of our model with $U(M)$ global symmetry. The generalization is straightforward, and the most calculations are parallel to the $SO(M)$ case. Hence, we do not repeat the same calculations and briefly show the result.

Let us star with $N M$ complex fermions 
\begin{equation}
    \chi^{i \alpha}(\tau)\quad,\quad \bar{\chi}_{i \alpha} (\tau)\hspace{10mm} (i=1,2\cdots, N\;\;\mbox{and}\;\; \alpha=1,2,\cdots, M)
\end{equation}
where $\chi^{i \alpha}$ (and, $\bar{\chi}^{i \alpha}$) transform in the fundamental(and, anti-fundamental, respectively) representations of $U(N)$ and $U(M)$. 

We consider the simplest $q=2\hq$ complex SYK model with $U(M)$ global symmetry:
\begin{equation}
    S=\int d\tau \left[ \chi^{i \alpha}\partial_\tau \bar{\chi}_{i \alpha } + {J^{j_1 \cdots j_\hq }}_{i_1  \cdots i_\hq}\chi^{i_1 \alpha_1}  \cdots \chi^{i_\hq \alpha_\hq}\bar{\chi}_{j_1 \alpha_1} \cdots \bar{\chi}_{j_\hq \alpha_\hq}\right]
\end{equation}
where the summation over $U(N)$ indices $i$'s and $U(M)$ indices $j$'s are omitted. ${J^{j_1 \cdots j_\hq }}_{i_1  \cdots i_\hq}$ denotes the random coupling constant with Gaussian distribution:
\begin{equation}
    \prod_{\substack{i_1,\cdots, i_\hq\\j_1,\cdots, j_\hq}}\exp \left[ - {  \hq M^{{q\over 2}-1} N^{q-1}\over  2 J^2 } {J^{j_1 \cdots j_\hq}}_{i_1\cdots i_\hq}\overline{{J^{j_1\cdots  j_\hq}}_{i_1\cdots i_\hq }} \right]\ .
\end{equation}
Note that the reality of the action requires ``Hermitian'' condition of the random coupling constant. \ie
\begin{equation}
    \overline{{J^{j_1 \cdots j_\hq }}_{i_1  \cdots i_\hq}}={J^{i_1  \cdots i_\hq}}_{j_1 \cdots j_\hq }\ ,
\end{equation}
and we will not demand any other symmetries of ${J^{j_1 \cdots j_\hq }}_{i_1  \cdots i_\hq}$ as in the $SO(M)$ case. After the disorder average, we have the bi-local collective action for the case of $U(M)$:
\begin{align}
   &S_{col}=N\Tr\left[-D\star \maPsi +\log \maPsi \right] - {N  J^2\over  2\hq M^{\hq-1}}\int d\tau_1d\tau_2 \; \left[\tr\left(-\maPsi(\tau_1,\tau_2)\maPsi(\tau_2,\tau_1)\right)\right]^\hq\ .\label{eq: collective action for su n}
\end{align}
Here, we also defined the bi-local field as follow.
\begin{equation}
    {\Psi^{\alpha_1}}_{\alpha_2}(\tau_1,\tau_2)\equiv {1\over N}\sum_{i=1}^N \chi^{i \alpha_1}(\tau_1)\bar{\chi}_{i \alpha_2}(\tau_2)\ ,
\end{equation}
and, we also introduce the $M\times M$ matrix notation $\maPsi(\tau_1,\tau_2)$ for the bi-local field:
\begin{equation}
    {(\maPsi(\tau_1,\tau_2))^{\alpha_1}}_{\alpha_2}\equiv {\Psi^{\alpha_1}}_{\alpha_2}(\tau_1,\tau_2)\ .
\end{equation}
For the $U(M)$ case, the bi-local field ${\Psi^{\alpha_1}}_{\alpha_2}(\tau_1,\tau_2)$ also have a symmetry analogous to anti-symmetry of the $SO(M)$ case~\eqref{eq:anti-symmetry1}: ${\Psi^{\alpha_1}}_{\alpha_2}(\tau_1,\tau_2)$ is a ``Hermitian'' as a matrix in the $(\tau_1,\alpha_1;\tau_2,\alpha_2)$ space. \ie
\begin{equation}
    \overline{{\Psi^{\alpha_1}}_{\alpha_2}(\tau_1,\tau_2)}= {\Psi^{\alpha_2}}_{\alpha_1}(\tau_2,\tau_1)\ .
\end{equation}
Furthermore, the fermions transforms in the fundamental and anti-fundamental representation, the bi-local field can be decomposed into the singlet and the adjoint representation:
\begin{equation}
     \yng(1)\; \otimes \; \overline{\yng(1)} = (\text{singlet}) \; \oplus \; (\text{adjoint})\ .
\end{equation}
Also, note that $N\tr \log \maPsi$ in~\eqref{eq: collective action for su n} comes from the Jacobian in Hubbard-Stratonovich type transformation to the bi-local field~\cite{deMelloKoch:1996mj,Jevicki:2014mfa,Yoon:2017gut} where we have the overall factor $N$ instead of ${N-1\over 2}$ because of complex vector field.

Now, let us consider the large $N$ classical solution. For this, we vary the collective action~\eqref{eq: collective action for su n} with respect to the bi-local field. We have
\begin{align}
    &-(D\mstar \maPsi_{cl})(\tau_1,\tau_2) + \idm \delta(\tau_1-\tau_2) \cr
    & +  {J^2\over M^{\hq-1}}\int d\tau_3 \; \left[-\tr (  \maPsi_{cl}(\tau_1,\tau_3)\maPsi_{cl}(\tau_3,\tau_1))\right]^{\hq -1} \maPsi_{cl}(\tau_1,\tau_3) \maPsi_{cl}(\tau_3,\tau_2)=0
\end{align}
where we multiplied another bi-local fields after the variation. We take $U(M)$ invariant ansatz:
\begin{equation}
    \maPsi_{cl}(\tau_1,\tau_2)=\idm \psi_{cl}(\tau_1,\tau_2)\ ,
\end{equation}
and the large $N$ saddle point equation becomes the Schwinger-Dyson equaiton for the complex SYK model:
\begin{equation}
    -(D\star \psi_{cl})(\tau_1,\tau_2)+\delta(\tau_1-\tau_2)+J^2 \int d\tau_3 \left[(\psi_{cl}(\tau_1,\tau_3)\psi_{cl}(\tau_3,\tau_1)\right]^{\hq-1}\psi_{cl}(\tau_1,\tau_3)\psi_{cl}(\tau_3,\tau_2) =0
\end{equation}
At strong coupling limit $|Jt|\gg 1$, one can drop the first term which comes from the kinetic term, and a general solution~\cite{Davison:2016ngz} is given by
\begin{equation}
    \psi_{cl}(\tau_1,\tau_2)\sim  {\sgn(\tau_{12})+ \tanh \pi \mathcal{E}  \over |\tau_1-\tau_2|^{2\Delta}}
\end{equation}
where $\mathcal{E}$ is ``spectral asymmetry'' introduced by~\cite{Davison:2016ngz}, which appear because the classical solution need not to be anti-symmetric unlike the real SYK model. It plays an important role in thermodynamics of the complex SYK model. However, in this paper, we consider the anti-symmetric solution $\mathcal{E}=0$ (the particle-hole symmetric case) for simplicity, and leave the thermodynamics of the generic $\mathcal{E}\ne 0$ case for future work. Then, the (anti-symmetric) classical solution is the same as before:
\begin{equation}
    \maPsi_{cl}(\tau_1,\tau_2)=\coeff {\sgn(\tau_{12})\over |\tau_{12}|^{2\over q}}\idm
\end{equation}
where $\Lambda$ is defined by
\begin{equation}
    J^2 \coeff^q \pi =\left({1\over 2}-{1\over q}\right) \tan {\pi \over q}\ .
\end{equation}
We expand the bi-local field around the large $N$ classical solution:
\begin{equation}
 \maPsi(\tau_1,\tau_2) = \idm \psi_{cl}(\tau_1,\tau_2) +  {1\over \sqrt{N}} \mfluca(\tau_1,\tau_2) +{1\over \sqrt{N}}\mflucb(\tau_1,\tau_2)
\end{equation}
where the singlet and the adjoint fluctuation can be written as
\begin{equation}
\mfluca(\tau_1,\tau_2) = {\fluca(\tau_1,\tau_2)\over \sqrt{M} }  \idm \quad,\quad  \mflucb(\tau_1,\tau_2)={\flucb_{a}(\tau_1,\tau_2)\over\sqrt{2\ind}} \mT^a\hspace{5mm} (a=1,\cdots, \dim (su(M))) \ .
\end{equation}
where $\mT^a$ is the $su(M)$ generator of which normalization is given by
\begin{equation}
    \tr(T^a T^b)= 2\ind \delta^{ab}
\end{equation}
where $\ind=M$ is the Dynkin index of the adjoint representation which equals to the dual Coxeter number of the $SU(M)$. Let us consider the complex conjugate of the bi-local fluctuations. One can easily see that the complex conjuate of both the singlet and the adjoint fluctuations becomes
\begin{equation}
    \overline{\fluca(\tau_1,\tau_2)}=\fluca(\tau_2,\tau_1)\quad,\quad \overline{\flucb^a(\tau_1,\tau_2)}=\flucb^a(\tau_2,\tau_1)\ .
\end{equation}
Note the $SU(M)$ case do not have other symmetry property such as \eqref{eq: symmetry of fluctuation1} and \eqref{eq: antisymmetry of fluctuation2}. Using this complex conjugate, one can obtain the quadratic action from the large $N$ expansion of the collective action~\eqref{eq: collective action for su n}:
\begin{align}
    &S_{col}^{(2)}=-{1\over 2}\bitr (\psi_{cl}^{-1}\star\fluca \star\psi_{cl}^{-1} \star \fluca ) + {\hq\over 2}J^2\int d\tau_1 d\tau_2 [\psi_{cl}(\tau_1,\tau_2)]^{q-2}\fluca(\tau_1,\tau_2)\overline{\fluca(\tau_1,\tau_2)}\cr
    &-{\hq -1\over 2}J^2\int d\tau_1 d\tau_2 [\psi_{cl}(\tau_1,\tau_2)]^{q-2}\fluca(\tau_1,\tau_2)\fluca(\tau_1,\tau_2)\cr
    &-{1\over 2}\bitr (\psi_{cl}^{-1}\star\flucb^a \star\psi_{cl}^{-1} \star \flucb^a )+{1\over 2} J^2\int d\tau_1 d\tau_2 \; [\psi_{cl}(\tau_1,\tau_2)]^{q-2}\flucb^a(\tau_1,\tau_2)\overline{\flucb^a(\tau_1,\tau_2)}\ .
\end{align}
To see explicit analogy to the $SO(M)$ case, we further decompose the bi-local fluctuations into symmetric and anti-symmetric components under $\tau_1\leftrightarrow \tau_2$  (or, real and imaginary components).
\begin{align}
    \fluca(\tau_1,\tau_2)= \fluca_+(\tau_1,\tau_2)+\fluca_-(\tau_1,\tau_2)\ ,\\
    \flucb^a(\tau_1,\tau_2)=\flucb_+^a(\tau_1,\tau_2)+\flucb_-^a(\tau_1,\tau_2)
\end{align}
where $\fluca_\pm$ and $\flucb_\pm^a$ are real, and satify
\begin{equation}
    \fluca_\pm(\tau_1,\tau_2)=\pm \fluca_\pm(\tau_2,\tau_1)\quad,\quad\flucb^a_\pm(\tau_1,\tau_2)=\pm \flucb^a_\pm(\tau_2,\tau_1)\ .
\end{equation}
Hence, the quadratic action can be decomposed as follow.
\begin{align}
    &S_{col}^{(2)}=-{1\over 2}\bitr (\psi_{cl}^{-1}\star\fluca_+ \star\psi_{cl}^{-1} \star \fluca_+ ) + {1\over 2}J^2\int d\tau_1 d\tau_2 [\psi_{cl}(\tau_1,\tau_2)]^{q-2}\fluca_+(\tau_1,\tau_2)\fluca_+(\tau_1,\tau_2)\cr
    &-{1\over 2}\bitr (\psi_{cl}^{-1}\star\fluca_- \star\psi_{cl}^{-1} \star \fluca_- ) - {q -1\over 2}J^2\int d\tau_1 d\tau_2 [\psi_{cl}(\tau_1,\tau_2)]^{q-2}\fluca_-(\tau_1,\tau_2)\fluca_-(\tau_1,\tau_2)\cr
    &-{1\over 2}\bitr (\psi_{cl}^{-1}\star\flucb^a_+ \star\psi_{cl}^{-1} \star \flucb^a_+ )+{1\over 2} J^2\int d\tau_1 d\tau_2 \; [\psi_{cl}(\tau_1,\tau_2)]^{q-2}\flucb^a_+(\tau_1,\tau_2)\flucb^a_+(\tau_1,\tau_2)\cr
    &-{1\over 2}\bitr (\psi_{cl}^{-1}\star\flucb^a_- \star\psi_{cl}^{-1} \star \flucb^a_- ) - {1\over 2} J^2\int d\tau_1 d\tau_2 \; [\psi_{cl}(\tau_1,\tau_2)]^{q-2}\flucb^a_-(\tau_1,\tau_2)\flucb^a_-(\tau_1,\tau_2)\ .
\end{align}
Note that $\fluca_-$ and $\flucb_-$ correspond to the singlet and symmetric irrep fluctuations of the $SO(M)$ case, respectively. Moreover, $\fluca_+$ and $\flucb_+^a$ are analogous to the anti-symmetric irrep fluctuation of the $SO(M)$ case. Therefore, one can repeat the same analysis as in Section~\ref{sec: syk model with global symmetry}$\sim $\ref{sec: chaos and effective action} and Section~\ref{sec: 3d gravity} (\ie the spectrum, correlation functions, the low energy effective actions, chaos and 3D gravity, etc.). Also, note that the lowest mode of the $\fluca_+$ is related to the broken $\hat{u}(1)$ symmetry which is a part of $\hat{u}(M)$ symmetry.

It is also straightforward to generalize the diffeomorphism and the Kac-Moody algebra discussed in Section~\ref{sec: algebra} to the $U(M)$ case. At strong coupling limit, we have emergent reparametrization and the local $\hat{U}(M)$ symmetry. These local symmetry is spontaneously broken by the classical solution as well as explicitly broken by the kinetic term of the collective action. Hence, we will have the effective actions related to the reparametrization and the local $\hat{u}(M)$ symmetry. They are of the same form as those of $SO(M)$ case in Section~\ref{sec: effective action}. Furthermore, one can also repeat the analysis of the chaos in Section~\ref{sec: chaos and effective action}, and one can see that the singlet channel will saturate the chaos bound because of the Schwarzian effective action, and the adjoint channel will not. Furthermore, the zero mode from the broken local $\hat{u}(M)$ symmetry will not give any exponential growth.

\section{Eigenfunctions of Quadratic Action}
\label{app: eigenfunctions}

In this appendix, we will study eigenfunctions of the quadratic action $S_{col}^{(2)}$ in~\eqref{eq: quadratic action}. Recall that the classical solution $\maPsi_{cl}$~\eqref{eq: classical solution} is invariant under $SL(2,\mathbb{R})$. Thus, one can easily check that the quadratic action, which is consist of $\psi_{cl}$'s, commutes with the Casimir of $SL(2,\mathbb{R})$ as well as the conformal generators up to total derivative. Therefore, we first diagonalize the Casimir of $SL(2,\mathbb{R})$ to find eingenfunctions of the quadratic action.

For the bi-local field, it is natural to define the bi-local conformal generators $\dP, \dD, \dK$:
\begin{equation}
    \dP\equiv\dP_1 + \dP_2\quad,\quad \dD\equiv\dD_1 + \dD_2\quad,\quad \dK\equiv\dK_1 + \dK_2\label{eq: bi-local generator}
\end{equation}
where each generator is given by
\begin{alignat}{3}
&\dP_1\equiv\partial_{\tau_1}   && , && \dP_2\equiv\partial_{\tau_2}\ ,\\
&\dD_1\equiv\tau_1\partial_{\tau_1}   +{1\over q}  && , && \dD_2\equiv\tau_2\partial_{\tau_2}   +{1\over q}\ ,\\
&\dK_1\equiv\tau_1^2\partial_{\tau_1} + {2\over q} \tau_1 \hspace{10mm} && , \hspace{10mm} && \dK_2\equiv\tau_2^2\partial_{\tau_2} + {2\over q} \tau_2\ .
\end{alignat}
Using the bi-local map
\begin{equation}
    t\equiv {1\over 2}(\tau_1+\tau_2)\quad,\quad z\equiv {1\over 2}(\tau_1-\tau_2)\ ,
\end{equation}
one can write the corresponding $SL(2,\mathbb{R})$ Casimir in terms of $(t,z)$ coordinates:
\begin{equation}
    \mathcal{C}\equiv\dD^2-{1\over 2} (\dP \dK +\dK \dP)={4\over q^2}-{2\over q} +{4\over q} z\partial_z +z^2(-\partial_t^2+\partial_z^2)\ .\label{eq: casimir}
\end{equation}
In order to find eigenfunctions $u(t,z)$ of the Casimir \ie
\begin{equation}
    \mathcal{C} u(t,z)=\lambda u(t,z)\ ,\label{eq: casimir eigenfunction equation}
\end{equation}
we diagoanalize the bi-local generator $\dP$. Then, with ansatz $f(t,z)\sim e^{-iwt} z^{{1\over 2} -{2\over q} } $, the differential equation~\eqref{eq: casimir eigenfunction equation} is reduced to the Bessel's differential equation for $f(z)$. Hence, the eigenfunction of the Casimir is a linear combination of two Bessel functions, $J_\nu(wz)$ and $J_{-\nu}(wz)$ with the corresponding eigenvalue $\lambda$ of the Casimir being $\nu^2-{1\over 4}$. Furthermore, by change of variable $x=\log w z$, this Bessel's differential equation can be understood as a Schrodinger-like equation with potential $V(x)=-e^{2x}$, and the corresponding energy is $-\nu^2$. \cite{Polchinski:2016xgd} argued that there are a continuum spectrum for the positive energy ($\nu=ir$ and $r\geqq0$), and there are also discrete states with negative energy ($\nu$: real.) Therefore, the form of eigenfunction can be written as
\begin{equation}
    u_{\nu w}(t,w)= e^{-i w t }z^{{1\over 2} -{2\over q} }\left(J_\nu(wz)+ c_\nu J_{-\nu}(wz)\right) \hspace{10mm}\begin{cases}
    \nu= ir\quad  (r\geqq 0)\\
    \nu\;:\; \mbox{discrete real number}\\
    \end{cases}
\end{equation}
where $c_\nu$ is a constant. To determine $\nu$ and $c_\nu$, we consider boundary conditions. Because the eigenfunctions are symmetric or anti-symmetric under $\tau_1\;\leftrightarrow \; \tau_2$ (or, equivalently, $z\;\rightarrow \;-z$) (\eg See~\eqref{eq: symmetry of fluctuation1} and~\eqref{eq: antisymmetry of fluctuation2}), one can restrict to $z>0$ space. Then, we demand UV/IR boundary conditions.

First, we demand IR boundary condition in which the eigenfunction behave either $\sin z$ or $\cos z$ as $z\longrightarrow \infty$. For now, we do not have a good argument to justify this boundary condition. However, even if we do not demand this IR boundary condition, one can determine the relative coefficient of the $J_\nu(z)$ and $J_{-\nu}(z)$ by direct calculations of the integration. And, the resulting coefficients agree with what one can obtain from the IR boundary condition. Hence, for the pedagogical purpose, we first demand the IR boundary condition. As $z\longrightarrow \infty$, the eigenfunction asymptote to 
\begin{align}
    J_\nu(z)+c_\nu J_{-\nu}(z)\approx& \sqrt{2\over \pi z} \left(\cos{\pi\over 2}(\nu+1/2) + c_\nu \sin{\pi\over 2}(\nu+1/2) \right)\cos z\cr
    &+\sqrt{2\over \pi z}\left(\sin{\pi\over 2}(\nu+1/2) + c_\nu \cos{\pi\over 2}(\nu+1/2) \right)\sin z\ .
\end{align}
By demanding either $\sin z$ or $\cos z$, one can determine the coefficient $\xi$:
\begin{equation}
    Z^\mp_\nu (z)\equiv J_\nu(z) + \xi_{\pm} J_{-\nu}(z)\quad,\quad \xi_\nu\equiv {\tan{\pi \nu \over 2}+1\over \tan{\pi \nu \over 2}-1 }\ ,
\end{equation}
and, their asymptotic behaviors are 
\begin{equation}
    Z^-_\nu(z)\sim  {\cos z\over \sqrt{z}}\quad,\quad   Z^+_\nu(z)\sim {\sin z \over \sqrt{z}}\ .
\end{equation}

In addition, the eigenfunction should be regular as $z\;\longrightarrow \;0$ (UV boundary condition). Since $J_{-\nu}$ diverges at $z\rightarrow 0$ for the case of the discrete spectrum, this term should be vanishes. Hence, noting the following identities
\begin{equation}
    \xi_{2n+{3\over 2}}=0\quad,\quad \xi_{-\left(2n+{1\over 2}\right)}=0\hspace{10mm} (n=0,1,2,\cdots)\ ,
\end{equation}
one can obtain the quantization condition of the discrete spectrum:
\begin{align}
   Z^-_\nu(z)\quad:\quad \nu=2n+{3\over 2}\hspace{1cm} (n=0,1,2,\cdots)\ ,\\
   Z^+_\nu(z)\quad:\quad \nu=2n+{1\over 2}\hspace{1cm} (n=0,1,2,\cdots)\ ,
\end{align}
and, thus, we have
\begin{equation}
    Z^-_{2n+{3\over 2}}(z)=J_{2n+{3\over 2}}(z)\quad,\quad Z^+_{2n+{1\over 2}}(z)=J_{2n+{1\over 2}}(z)\ .
\end{equation}

Now, using the properties of the singlet and anti-symmetric fluctuations, we will find their eigenfunctions. First, we found that the singlet/anti-symmetric fluctuation is anti-symmetric/symmetric under $\tau_1,\tau_2$ (or, equivalently $z\rightarrow -z$), respectively. Furthermore, the singlet fluctuation have the reparametrization zero mode ($h=2$ or $\nu={3\over 2}$), and the anti-symmetric(adjoint) fluctuation is related to $\hat{SO}(M)$ zero mode ($h=1$ or $\nu={1\over 2}$). Hence, these two properties determine the eigenfunction up to normalization. \ie
\begin{alignat}{2}
    &u^-_{\nu w}(t,z)= {1\over \sqrt{8\pi}} e^{-i w t} \sgn(z) |z|^{{1\over 2}-{2\over q}} Z^-_{\nu}(|wz|)\hspace{2mm} &&,\hspace{3mm} \nu=\begin{cases}
    \; 2n+{3\over 2} \quad &\; (n=0,1,\cdots)\\
    \;ir &\; (r\geqq 0)\\
    \end{cases}\ ,\\
    &u^+_{\nu w}(t,z)= {1\over \sqrt{8\pi}} e^{-i w t} |z|^{{1\over 2}-{2\over q}}  Z^+_{\nu}(|wz|)\hspace{2mm} &&,\hspace{3mm} \nu=\begin{cases}
    \; 2n+{1\over 2} \quad &\; (n=0,1,\cdots)\\
    \;ir &\; (r\geqq 0)\\
    \end{cases}\ .
\end{alignat}
Note that the symmetric irrep fluctuation is analogous to the singlet one. 

Finally, we briefly present the integrals to evaluate the important identity in~\eqref{def: utilde} which relate $\tilde{u}^\mp_{\nu w}$ to $u^\mp_{\nu w}$:
\begin{equation}
    \psi_{cl}\star \tilde{u}^+_{\nu w}  \star \psi_{cl}= k_+(h) u^+_{\nu w}\hspace{10mm} (h\equiv\nu+{1\over 2})\ .\label{eq: utilde in app}
\end{equation}
In particular, we will show $u^+_{\nu w}$ case explicitly because the other has been discussed in the original SYK models. We claim that
\begin{align}
    \tilde{u}^+_{\nu w}(t,z)=&J^2\left[\psi_{cl}(t+z,t-z)\right]^{q-2}u^+_{\nu w}(t,z)\label{eq: utilde p sol app}\ ,
\end{align}
The LHS of \eqref{eq: utilde in app} can be written as (up to trivial factors)
\begin{align}
    &(\mbox{LHS})\sim e^{-iwt_0}2\int dt dz {e^{-iw|z-z_0|t}\sgn\left(1 -t\right)\sgn\left(1+t\right)  |z|^{-{3\over 2}+{2\over q} } Z_\nu^+(|wz|) \over |t^2-1|^{2\over q} |z-z_0|^{{4\over q}-1}  }\cr
    \sim &e^{-iw t_0} 4\int  dz {  |z|^{-{3\over 2}+{2\over q} } Z_\nu^+(|wz|) \over |z-z_0|^{{4\over q}-1}  } \left[-\int_1^\infty dt \; {\cos |w(z-z_0)| t \over |t^2-1|^{2\over q} }+\int_0^1 dt\; {\cos |w(z-z_0)|t \over |t^2-1|^{2\over q} }\right]
\end{align}    
where we used
\begin{alignat}{2}
    &t\equiv{1\over 2}(\tau_3+\tau_4)\quad,\quad&& z\equiv{1\over 2}(\tau_3-\tau_4)\ ,\\
    &t_0\equiv{1\over 2}(\tau_1+\tau_2)\quad,\quad&& z_0\equiv {1\over 2}(\tau_1-\tau_2)\ .
\end{alignat}
Then, using the integral formula
\begin{align}
    &\int_1^\infty dt \; (x^2-1)^{\nu-{1\over 2}}\cos ax= - {\sqrt{\pi}\over 2}\left({2\over a}\right)^\nu \Gamma\left(\nu+{1\over 2}\right)Y_{-\nu} (a)\ ,\\
    &\int_0^1 dx\; (1-x^2)^{\nu-{1\over 2}}\cos ax={\sqrt{\pi}\over 2}\left({2\over a}\right)^\nu \Gamma\left(\nu+{1\over 2}\right)J_\nu (a)\ ,
\end{align}
we have (up to trivial factors)
\begin{equation}
    (\mbox{LHS})\sim \int  dz \;  |wz|^{-{3\over 2}+{2\over q} } Z_\nu^+(|wz|)  |w(z-z_0)|^{{1\over 2}-{2\over q}} \left[J_{{1\over 2}-{2\over q}}(|w(z-z_0)|)+Y_{{2\over q}-{1\over 2}}(|w(z-z_0)|)\right]
\end{equation}
Now, one can consider the Fourier modes of the three Bessel functions by using the following integral formula.
\begin{align*}
    &\int dx\; e^{ipx} |x|^\nu J_{-\nu}(|x|)={2^{1+\nu }\Gamma({1\over 2})\over \Gamma({1\over 2}-\nu)}F({1\over 2},{1\over 2}+\nu ,{1\over 2};p^2)\theta(1-|p|)\ ,\\
    &\int dx\; e^{ipx}|x|^\nu J_{\nu}(|x|)={2^{1+\nu}\Gamma({1\over 2}+\nu)\over \sqrt{\pi} |p^2-1|^{\nu+{1\over 2}}}\left[\theta(1-|p|) -\sin\pi \nu \theta(|p|-1)\right]\ ,\\
    &2\int dx\; |x|^\mu J_\nu (|x|)\cos px ={2^{1+\mu }\Gamma\left({1+\mu +\nu \over 2}\right)\over \Gamma\left({\nu -\mu +1\over 2}\right)}F\left({1+\mu +\nu  \over 2}, {1+\mu -\nu \over 2},{1\over 2},p^2\right)\theta(1-|p|)\cr
    &+{2^{1-\nu }\Gamma(1+\mu +\nu)\cos\left[{\pi\over 2}(1+\mu+\nu)\right]\over \Gamma(\nu+1) |p|^{1+\mu +\nu}}F\left({1+\mu+\nu\over 2},{2+\mu+\nu\over 2},\nu+1;{1\over p^2}\right)\theta(|p|-1)
\end{align*}
%
%
%
%
Also, one can evaluate the Fourier modes of the RHS of~\eqref{eq: utilde in app}. By comparing the Fourier mode on the both sides, one can prove~\eqref{eq: utilde in app} with $k_+(h)$ $(h=\nu+{1\over 2})$ in \eqref{eq: kp function}

\section{Zero Mode Eigenfunctions}
\label{app: zero mode}

In Section~\ref{sec: effective action}, we utilized the zero mode eigenfunctions related to the Kac-Moody algebra and the reparametrization to evaluate the chaotic behavior of the out-of-time-ordered correlators. In this appendix, we will shortly discuss the properties of the zero mode eigenfunctions given by
\begin{equation}
    \Upsilon_n^{1,a}\sim \sin {n\theta_{12}\over 2}\hspace{5mm},\hspace{5mm} \Upsilon_n^{2}\sim {\sin {n\theta_{12}\over 2}\over \tan{n\theta_{12}\over 2}}-n\cos{n\theta_{12}\over 2}
\end{equation}
where we omitted the normalization constant and $e^{{in \over 2}(\theta_1+\theta_2)}$. In~\cite{Maldacena:2016hyu,Davison:2016ngz}, these zero eigenfunctions were normalized by $|\psi_{cl}|^{q-2\over2}$ for the symmetric kernels:
\begin{equation}
    \Upsilon_n^{1,a}\sim {\sin {n\theta_{12}\over 2}\over \sin {\theta\over 2}}\hspace{5mm},\hspace{5mm} \Upsilon_n^{2}\sim {1\over \sin {\theta\over 2}}\left[{\sin {n\theta_{12}\over 2}\over \tan{n\theta_{12}\over 2}}-n\cos{n\theta_{12}\over 2}\right]\ .
\end{equation}
In this appendix, we are interested in the $\theta_{12}$ dependence. Hence, we define
\begin{equation}
    f_{1,n}(\tau,z)\equiv e^{-in \tau}\;{\sin nz\over \sin z}\hspace{5mm},\hspace{5mm} f_{2,n}(t,z) \equiv -{e^{-in \tau} \over \sin z}\left[{\sin nz \over \tan nz}-n\cos nz\right]
\end{equation}
where $\tau \equiv{1\over 2}(\theta_1+\theta_2)$ and $z={1\over 2}(\theta_1-\theta_2)$. By the inner product given by
\begin{equation}
    \langle f,g\rangle ={2\over 2\pi}\int_0^{2\pi} d\tau \int_0^{\pi } dz\; f^\ast g\ ,
\end{equation}
the inner product of two functions are
\begin{align}
    \langle f_{1,m},f_{1,n}\rangle=&2\pi n\delta_{m,n}\ ,\\
    \langle f_{2,m},f_{2,n}\rangle=& {2\pi \over 3}n(n^2-1)\delta_{m,n}\ ,\\
    \langle f_{1,m},f_{2,n}\rangle=&0\ .
\end{align}
Now one can easily see the relation between two zero mode eigenfunctions:
\begin{equation}
    f_{2,n}=\partial_z f_{1,n}\ .
\end{equation}
Note that $\partial_z=\partial_{\theta_1}-\partial_{\theta_2}$ is not part of the $SL(2,\mathcal{R})$.

We construct a function $f_{3,n}$ by acting derivatives on $f_{2,n}$:
\begin{align}
    f_{3,n}\equiv&\partial_z f_{2,n}+{1\over 3} (n^2-1) f_{1,n}=\partial_z f_{2,n}+{1\over 3} (-\partial_\tau^2-1) f_{1,n}\cr
    =&-{e^{-in\tau}\over 3\sin x} \left[\left(1-3{1+\cos^2 x\over \sin^2 x}\right)\sin nx +6n {\cos nx\over \tan x} +2n^2 \sin nx \right]\label{def: f3}
\end{align}
where we included the second term in order to to make $f_{3,n}$ orthogonal to $f_{1,n}$. Furthermore, one can keep constructing such functions. \eg
\begin{align}
    f_{4,n}=&\partial_x f_{3,n} +{4\over 15}(n^2-4)f_{2,n}=\partial_x f_{3,n} +{4\over 15}(-\partial_\tau^2-4)f_{2,n} \\
    =&-{e^{-in\tau}\over 5\sin x} \left[\left(5\cot^3x+25 \cot x \csc^2 x -7\cot x\right)\sin nx \right.\cr
    &\left.+n(7-15 \cot^2x -15\csc^2 x)\cos nx -12n^2\cot x \sin nx+2n^3 \cos nx \right]\ .
\end{align}
And, their inner products are found to be
\begin{align}
    \langle f_{1,n},f_{1,n'}\rangle=&2\pi n\delta_{m,n}\equiv \mathcal{N}^1_n\delta_{n,n'}\label{eq: normalization 1}\ ,\\
    \langle f_{2,n},f_{2,n'}\rangle=& {2\pi \over 3}n(n^2-1)\delta_{n,n'}\equiv \mathcal{N}^2_n\delta_{n,n'}\ ,\\
    \langle f_{3,n},f_{3,n'}\rangle =&{2\pi \over 45} n(n^2-1)(n^2-4)\delta_{n,n'}\equiv \mathcal{N}^3_n\delta_{n,n'}\ ,\\
    \langle f_{4,n},f_{4,n'}\rangle =& {4\pi \over 175}n(n^2-1)(n^2-4)(n^2-9)\delta_{n,n'}\equiv \mathcal{N}^4_n\delta_{n,n'}\label{eq: normalization 4}\ ,\\
    \langle f_{s,n},f_{s',n'}\rangle=&0 \hspace{5mm}(\text{for}\;\; s\ne s' \in \{1,2,3,4\})\ .
\end{align}

Now, we will repeat a calculation performed in Section~\ref{sec: effective action} where we evaluate the long-time behavior of the out-of-time-ordered correlators from the effective action. 

This construction aimed to guess a way to incorporate the larger symmetry than reparametrization into the SYK model (\eg $\mathcal{W}$ symmetry). For now, we assume that there exists an effective action of $f^s_n$ modes with its eigenvalue of the quadratic action being $|n|$. Then, the contribution to the out-of-time-ordered correlator can be obtain from the following expression.
\begin{align}
    \mathcal{F}^s\equiv\sum_{|n|\geqq s} {1\over |n|\mathcal{N}^s_n } f^s_n(\tau,z) f^s_{-n}(\tau',z')
\end{align}
where $\mathcal{N}^s_n$ is the normalization defined in~\eqref{eq: normalization 1}$\sim$\eqref{eq: normalization 4}. As in Section~\ref{sec: effective action}, we will take the particular configuration of $\theta$'s
\begin{align}
    \tau'={1\over 2}(\theta_3+\theta_4)={\pi \over 2}\quad,\quad z'={1\over 2}(\theta_3-\theta_4)=-{\pi\over 2}\ ,
\end{align}
and then perform the analytic continuation:
\begin{align}
    \tau=&{1\over 2}(\theta_1+\theta_2)=-{2\pi i \over \beta }t\ ,\\
    z=&{1\over 2}(\theta_1-\theta_2)=-\pi \ .
\end{align}
For $t\gg 1$, we have
\begin{align}
    \mathcal{F}^1(t)\sim t\ ,\\
    \mathcal{F}^2(t)\sim e^{{2\pi \over \beta}t}\ ,\\
    \mathcal{F}^3(t)\sim e^{{4\pi \over \beta}t}\ ,\\
    \mathcal{F}^4(t)\sim e^{{6\pi \over \beta}t}\ .
\end{align}
Note that $\mathcal{F}^1(t)\sim t$ and $\mathcal{F}^2(t)$ agree what we have obtained in Section~\ref{sec: chaos and effective action}. However, $\mathcal{F}^3(t)\sim t$ and $\mathcal{F}^4(t)$ violate the chaos bound~\cite{Maldacena:2015waa}. This seems to be closely related\footnote{I thank Yang Lei for pointing out this.} to the Lyapunov exponent $2\pi(N-1)/\beta$ of CFTs with higher spin current~\cite{Perlmutter:2016pkf}. Hence, it would be interesting to explore a generalized SYK model with an emergent $\mathcal{W}$ symmetry and its relation to the higher spin gravity. We leave this for the future work.

\bibliographystyle{JHEP}
\bibliography{sykglobal7}

\providecommand{\href}[2]{#2}\begingroup\raggedright\begin{thebibliography}{10}

\bibitem{Sachdev:1992fk}
S.~Sachdev and J.~Ye, {\it {Gapless spin fluid ground state in a random,
  quantum Heisenberg magnet}},  {\em Phys. Rev. Lett.} {\bf 70} (1993) 3339,
  [\href{http://arxiv.org/abs/cond-mat/9212030}{{\tt cond-mat/9212030}}].

\bibitem{KitaevTalks}
A.~Kitaev, {\it {A simple model of quantum holography}},  {\em {}}
  \url{http://online.kitp.ucsb.edu/online/entangled15/kitaev/},
  \url{http://online.kitp.ucsb.edu/online/entangled15/kitaev2/}, Talks at KITP,
  April 7, 2015 and May 27, (2015).

\bibitem{Polchinski:2016xgd}
J.~Polchinski and V.~Rosenhaus, {\it {The Spectrum in the Sachdev-Ye-Kitaev
  Model}},  {\em JHEP} {\bf 04} (2016) 001,
  [\href{http://arxiv.org/abs/1601.06768}{{\tt arXiv:1601.06768}}].

\bibitem{Jevicki:2016bwu}
A.~Jevicki, K.~Suzuki, and J.~Yoon, {\it {Bi-Local Holography in the SYK
  Model}},  {\em JHEP} {\bf 07} (2016) 007,
  [\href{http://arxiv.org/abs/1603.06246}{{\tt arXiv:1603.06246}}].

\bibitem{Maldacena:2016hyu}
J.~Maldacena and D.~Stanford, {\it {Remarks on the Sachdev-Ye-Kitaev model}},
  {\em Phys. Rev.} {\bf D94} (2016), no.~10 106002,
  [\href{http://arxiv.org/abs/1604.07818}{{\tt arXiv:1604.07818}}].

\bibitem{Jevicki:2016ito}
A.~Jevicki and K.~Suzuki, {\it {Bi-Local Holography in the SYK Model:
  Perturbations}},  {\em JHEP} {\bf 11} (2016) 046,
  [\href{http://arxiv.org/abs/1608.07567}{{\tt arXiv:1608.07567}}].

\bibitem{Maldacena:2016upp}
J.~Maldacena, D.~Stanford, and Z.~Yang, {\it {Conformal symmetry and its
  breaking in two dimensional Nearly Anti-de-Sitter space}},  {\em PTEP} {\bf
  2016} (2016), no.~12 12C104, [\href{http://arxiv.org/abs/1606.01857}{{\tt
  arXiv:1606.01857}}].

\bibitem{Mandal:2017thl}
G.~Mandal, P.~Nayak, and S.~R. Wadia, {\it {Virasoro coadjoint orbits of
  SYK/tensor-models and emergent two-dimensional quantum gravity}},
  \href{http://arxiv.org/abs/1702.04266}{{\tt arXiv:1702.04266}}.

\bibitem{Das:2017pif}
S.~R. Das, A.~Jevicki, and K.~Suzuki, {\it {Three Dimensional View of the
  SYK/AdS Duality}},  \href{http://arxiv.org/abs/1704.07208}{{\tt
  arXiv:1704.07208}}.

\bibitem{Gurau:2017xhf}
R.~Gurau, {\it {Quenched equals annealed at leading order in the colored SYK
  model}},  \href{http://arxiv.org/abs/1702.04228}{{\tt arXiv:1702.04228}}.

\bibitem{Gurau:2017qna}
R.~Gurau, {\it {The $\imath \epsilon$ prescription in the SYK model}},
  \href{http://arxiv.org/abs/1705.08581}{{\tt arXiv:1705.08581}}.

\bibitem{Bonzom:2017pqs}
V.~Bonzom, L.~Lionni, and A.~Tanasa, {\it {Diagrammatics of a colored SYK model
  and of an SYK-like tensor model, leading and next-to-leading orders}},
  \href{http://arxiv.org/abs/1702.06944}{{\tt arXiv:1702.06944}}.

\bibitem{Dartois:2017xoe}
S.~Dartois, H.~Erbin, and S.~Mondal, {\it {Conformality of $1/N$ corrections in
  SYK-like models}},  \href{http://arxiv.org/abs/1706.00412}{{\tt
  arXiv:1706.00412}}.

\bibitem{Gu:2016oyy}
Y.~Gu, X.-L. Qi, and D.~Stanford, {\it {Local criticality, diffusion and chaos
  in generalized Sachdev-Ye-Kitaev models}},
  \href{http://arxiv.org/abs/1609.07832}{{\tt arXiv:1609.07832}}.

\bibitem{Berkooz:2016cvq}
M.~Berkooz, P.~Narayan, M.~Rozali, and J.~Sim{\'o}n, {\it {Higher Dimensional
  Generalizations of the SYK Model}},  {\em JHEP} {\bf 01} (2017) 138,
  [\href{http://arxiv.org/abs/1610.02422}{{\tt arXiv:1610.02422}}].

\bibitem{Banerjee:2016ncu}
S.~Banerjee and E.~Altman, {\it {Solvable model for a dynamical quantum phase
  transition from fast to slow scrambling}},
  \href{http://arxiv.org/abs/1610.04619}{{\tt arXiv:1610.04619}}.

\bibitem{Turiaci:2017zwd}
G.~Turiaci and H.~Verlinde, {\it {Towards a 2d QFT Analog of the SYK Model}},
  \href{http://arxiv.org/abs/1701.00528}{{\tt arXiv:1701.00528}}.

\bibitem{Berkooz:2017efq}
M.~Berkooz, P.~Narayan, M.~Rozali, and J.~Sim{\'o}n, {\it {Comments on the
  Random Thirring Model}},  \href{http://arxiv.org/abs/1702.05105}{{\tt
  arXiv:1702.05105}}.

\bibitem{Jian:2017unn}
S.-K. Jian and H.~Yao, {\it {Solvable SYK models in higher dimensions: a new
  type of many-body localization transition}},
  \href{http://arxiv.org/abs/1703.02051}{{\tt arXiv:1703.02051}}.

\bibitem{Murugan:2017eto}
J.~Murugan, D.~Stanford, and E.~Witten, {\it {More On Supersymmetric And 2d
  Analogs of the SYK Model}},  \href{http://arxiv.org/abs/1706.05362}{{\tt
  arXiv:1706.05362}}.

\bibitem{Fu:2016vas}
W.~Fu, D.~Gaiotto, J.~Maldacena, and S.~Sachdev, {\it {Supersymmetric
  Sachdev-Ye-Kitaev models}},  {\em Phys. Rev.} {\bf D95} (2017), no.~2 026009,
  [\href{http://arxiv.org/abs/1610.08917}{{\tt arXiv:1610.08917}}]. [Addendum:
  Phys. Rev.D95,no.6,069904(2017)].

\bibitem{Li:2017hdt}
T.~Li, J.~Liu, Y.~Xin, and Y.~Zhou, {\it {Supersymmetric SYK model and random
  matrix theory}},  \href{http://arxiv.org/abs/1702.01738}{{\tt
  arXiv:1702.01738}}.

\bibitem{Kanazawa:2017dpd}
T.~Kanazawa and T.~Wettig, {\it {Complete random matrix classification of SYK
  models with $\mathcal{N}=0$, $1$ and $2$ supersymmetry}},
  \href{http://arxiv.org/abs/1706.03044}{{\tt arXiv:1706.03044}}.

\bibitem{Yoon:2017gut}
J.~Yoon, {\it {Supersymmetric SYK Model: Bi-local Collective
  Superfield/Supermatrix Formulation}},
  \href{http://arxiv.org/abs/1706.05914}{{\tt arXiv:1706.05914}}.

\bibitem{Peng:2017spg}
C.~Peng, M.~Spradlin, and A.~Volovich, {\it {Correlators in the $\mathcal{N}=2$
  Supersymmetric SYK Model}},  \href{http://arxiv.org/abs/1706.06078}{{\tt
  arXiv:1706.06078}}.

\bibitem{Sachdev:2015efa}
S.~Sachdev, {\it {Bekenstein-Hawking Entropy and Strange Metals}},  {\em Phys.
  Rev.} {\bf X5} (2015), no.~4 041025,
  [\href{http://arxiv.org/abs/1506.05111}{{\tt arXiv:1506.05111}}].

\bibitem{Davison:2016ngz}
R.~A. Davison, W.~Fu, A.~Georges, Y.~Gu, K.~Jensen, and S.~Sachdev, {\it
  {Thermoelectric transport in disordered metals without quasiparticles: The
  Sachdev-Ye-Kitaev models and holography}},  {\em Phys. Rev.} {\bf B95}
  (2017), no.~15 155131, [\href{http://arxiv.org/abs/1612.00849}{{\tt
  arXiv:1612.00849}}].

\bibitem{Gross:2016kjj}
D.~J. Gross and V.~Rosenhaus, {\it {A Generalization of Sachdev-Ye-Kitaev}},
  {\em JHEP} {\bf 02} (2017) 093, [\href{http://arxiv.org/abs/1610.01569}{{\tt
  arXiv:1610.01569}}].

\bibitem{Jevicki:1979mb}
A.~Jevicki and B.~Sakita, {\it {The Quantum Collective Field Method and Its
  Application to the Planar Limit}},  {\em Nucl. Phys.} {\bf B165} (1980) 511.

\bibitem{deMelloKoch:1996mj}
R.~de~Mello~Koch and J.~P. Rodrigues, {\it {Systematic 1/N corrections for
  bosonic and fermionic vector models without auxiliary fields}},  {\em Phys.
  Rev.} {\bf D54} (1996) 7794--7814,
  [\href{http://arxiv.org/abs/hep-th/9605079}{{\tt hep-th/9605079}}].

\bibitem{Das:2003vw}
S.~R. Das and A.~Jevicki, {\it {Large N collective fields and holography}},
  {\em Phys. Rev.} {\bf D68} (2003) 044011,
  [\href{http://arxiv.org/abs/hep-th/0304093}{{\tt hep-th/0304093}}].

\bibitem{Koch:2010cy}
R.~de~Mello~Koch, A.~Jevicki, K.~Jin, and J.~P. Rodrigues, {\it {$AdS_4/CFT_3$
  Construction from Collective Fields}},  {\em Phys. Rev.} {\bf D83} (2011)
  025006, [\href{http://arxiv.org/abs/1008.0633}{{\tt arXiv:1008.0633}}].

\bibitem{Koch:2014aqa}
R.~de~Mello~Koch, A.~Jevicki, J.~P. Rodrigues, and J.~Yoon, {\it {Canonical
  Formulation of $O(N)$ Vector/Higher Spin Correspondence}},  {\em J. Phys.}
  {\bf A48} (2015), no.~10 105403, [\href{http://arxiv.org/abs/1408.4800}{{\tt
  arXiv:1408.4800}}].

\bibitem{Jevicki:2015sla}
A.~Jevicki and J.~Yoon, {\it {Bulk from Bi-locals in Thermo Field CFT}},  {\em
  JHEP} {\bf 02} (2016) 090, [\href{http://arxiv.org/abs/1503.08484}{{\tt
  arXiv:1503.08484}}].

\bibitem{Jevicki:2015pza}
A.~Jevicki and K.~Suzuki, {\it {Thermofield Duality for Higher Spin Rindler
  Gravity}},  {\em JHEP} {\bf 02} (2016) 094,
  [\href{http://arxiv.org/abs/1508.07956}{{\tt arXiv:1508.07956}}].

\bibitem{Gurau:2009tw}
R.~Gurau, {\it {Colored Group Field Theory}},  {\em Commun. Math. Phys.} {\bf
  304} (2011) 69--93, [\href{http://arxiv.org/abs/0907.2582}{{\tt
  arXiv:0907.2582}}].

\bibitem{Gurau:2011aq}
R.~Gurau and V.~Rivasseau, {\it {The 1/N expansion of colored tensor models in
  arbitrary dimension}},  {\em Europhys. Lett.} {\bf 95} (2011) 50004,
  [\href{http://arxiv.org/abs/1101.4182}{{\tt arXiv:1101.4182}}].

\bibitem{Gurau:2011xp}
R.~Gurau and J.~P. Ryan, {\it {Colored Tensor Models - a review}},  {\em SIGMA}
  {\bf 8} (2012) 020, [\href{http://arxiv.org/abs/1109.4812}{{\tt
  arXiv:1109.4812}}].

\bibitem{Bonzom:2012hw}
V.~Bonzom, R.~Gurau, and V.~Rivasseau, {\it {Random tensor models in the large
  N limit: Uncoloring the colored tensor models}},  {\em Phys. Rev.} {\bf D85}
  (2012) 084037, [\href{http://arxiv.org/abs/1202.3637}{{\tt
  arXiv:1202.3637}}].

\bibitem{Gurau:2011xq}
R.~Gurau, {\it {The complete 1/N expansion of colored tensor models in
  arbitrary dimension}},  {\em Annales Henri Poincare} {\bf 13} (2012)
  399--423, [\href{http://arxiv.org/abs/1102.5759}{{\tt arXiv:1102.5759}}].

\bibitem{Carrozza:2015adg}
S.~Carrozza and A.~Tanasa, {\it {$O(N)$ Random Tensor Models}},  {\em Lett.
  Math. Phys.} {\bf 106} (2016), no.~11 1531--1559,
  [\href{http://arxiv.org/abs/1512.06718}{{\tt arXiv:1512.06718}}].

\bibitem{GurauSchaeffer}
R.~Gurau and G.~Schaeffer, {\it {Regular colored graphs of positive degree}},
  {\em Ann. Inst. H. Poincare Comb. Phys. Interact.} {\bf 3} (2016) 257 -- 320,
  [\href{http://arxiv.org/abs/1307.5279}{{\tt arXiv:1307.5279}}].

\bibitem{Witten:2016iux}
E.~Witten, {\it {An SYK-Like Model Without Disorder}},
  \href{http://arxiv.org/abs/1610.09758}{{\tt arXiv:1610.09758}}.

\bibitem{Gurau:2016lzk}
R.~Gurau, {\it {The complete $1/N$ expansion of a SYK--like tensor model}},
  {\em Nucl. Phys.} {\bf B916} (2017) 386--401,
  [\href{http://arxiv.org/abs/1611.04032}{{\tt arXiv:1611.04032}}].

\bibitem{Klebanov:2016xxf}
I.~R. Klebanov and G.~Tarnopolsky, {\it {Uncolored random tensors, melon
  diagrams, and the Sachdev-Ye-Kitaev models}},  {\em Phys. Rev.} {\bf D95}
  (2017), no.~4 046004, [\href{http://arxiv.org/abs/1611.08915}{{\tt
  arXiv:1611.08915}}].

\bibitem{Krishnan:2016bvg}
C.~Krishnan, S.~Sanyal, and P.~N. Bala~Subramanian, {\it {Quantum Chaos and
  Holographic Tensor Models}},  {\em JHEP} {\bf 03} (2017) 056,
  [\href{http://arxiv.org/abs/1612.06330}{{\tt arXiv:1612.06330}}].

\bibitem{Krishnan:2017ztz}
C.~Krishnan, K.~V.~P. Kumar, and S.~Sanyal, {\it {Random Matrices and
  Holographic Tensor Models}},  {\em JHEP} {\bf 06} (2017) 036,
  [\href{http://arxiv.org/abs/1703.08155}{{\tt arXiv:1703.08155}}].

\bibitem{Chaudhuri:2017vrv}
S.~Chaudhuri, V.~I. Giraldo-Rivera, A.~Joseph, R.~Loganayagam, and J.~Yoon,
  {\it {Abelian Tensor Models on the Lattice}},
  \href{http://arxiv.org/abs/1705.01930}{{\tt arXiv:1705.01930}}.

\bibitem{Krishnan:2017txw}
C.~Krishnan and K.~V.~P. Kumar, {\it {Towards a Finite-$N$ Hologram}},
  \href{http://arxiv.org/abs/1706.05364}{{\tt arXiv:1706.05364}}.

\bibitem{Ferrari:2017ryl}
F.~Ferrari, {\it {The Large D Limit of Planar Diagrams}},
  \href{http://arxiv.org/abs/1701.01171}{{\tt arXiv:1701.01171}}.

\bibitem{Itoyama:2017emp}
H.~Itoyama, A.~Mironov, and A.~Morozov, {\it {Rainbow tensor model with
  enhanced symmetry and extreme melonic dominance}},
  \href{http://arxiv.org/abs/1703.04983}{{\tt arXiv:1703.04983}}.

\bibitem{Itoyama:2017xid}
H.~Itoyama, A.~Mironov, and A.~Morozov, {\it {Ward identities and combinatorics
  of rainbow tensor models}},  \href{http://arxiv.org/abs/1704.08648}{{\tt
  arXiv:1704.08648}}.

\bibitem{Narayan:2017qtw}
P.~Narayan and J.~Yoon, {\it {SYK-like Tensor Models on the Lattice}},
  \href{http://arxiv.org/abs/1705.01554}{{\tt arXiv:1705.01554}}.

\bibitem{Klebanov:2017nlk}
I.~R. Klebanov and G.~Tarnopolsky, {\it {On Large $N$ Limit of Symmetric
  Traceless Tensor Models}},  \href{http://arxiv.org/abs/1706.00839}{{\tt
  arXiv:1706.00839}}.

\bibitem{Mironov:2017aqv}
A.~Mironov and A.~Morozov, {\it {Correlators in tensor models from character
  calculus}},  \href{http://arxiv.org/abs/1706.03667}{{\tt arXiv:1706.03667}}.

\bibitem{Gurau:2017qya}
R.~Gurau, {\it {The $1/N$ expansion of tensor models with two symmetric
  tensors}},  \href{http://arxiv.org/abs/1706.05328}{{\tt arXiv:1706.05328}}.

\bibitem{Diaz:2017kub}
P.~Diaz and S.-J. Rey, {\it {Orthogonal Bases of Invariants in Tensor Models}},
   \href{http://arxiv.org/abs/1706.02667}{{\tt arXiv:1706.02667}}.

\bibitem{robert:2017}
R.~de~Mello~Koch, D.~Gossman, and L.~Tribelhorn, {\it Gauge invariants,
  correlators and holography in bosonic and fermionic tensor models},
  \href{http://arxiv.org/abs/1707.01455}{{\tt arXiv:1707.01455}}.

\bibitem{Bulycheva:2017uqj}
K.~Bulycheva, {\it {A note on the SYK model with complex fermions}},
  \href{http://arxiv.org/abs/1706.07411}{{\tt arXiv:1706.07411}}.

\bibitem{Maldacena:2017axo}
J.~Maldacena, D.~Stanford, and Z.~Yang, {\it {Diving into traversable
  wormholes}},  {\em Fortsch. Phys.} {\bf 65} (2017), no.~5 1700034,
  [\href{http://arxiv.org/abs/1704.05333}{{\tt arXiv:1704.05333}}].

\bibitem{kitaevfirsttalk}
A.~Kitaev, {\it {Hidden correlations in the Hawking radiation and thermal
  noise}},  {\em {}} \url{http://online.kitp.ucsb.edu/online/joint98/kitaev/},
  KITP seminar, Feb. 12, (2015).

\bibitem{Maldacena:2015waa}
J.~Maldacena, S.~H. Shenker, and D.~Stanford, {\it {A bound on chaos}},  {\em
  JHEP} {\bf 08} (2016) 106, [\href{http://arxiv.org/abs/1503.01409}{{\tt
  arXiv:1503.01409}}].

\bibitem{Jevicki:2014mfa}
A.~Jevicki, K.~Jin, and J.~Yoon, {\it {1/N and loop corrections in higher spin
  AdS$_4$/CFT$_3$ duality}},  {\em Phys. Rev.} {\bf D89} (2014), no.~8 085039,
  [\href{http://arxiv.org/abs/1401.3318}{{\tt arXiv:1401.3318}}].

\bibitem{Koch:2014mxa}
R.~de~Mello~Koch, A.~Jevicki, J.~P. Rodrigues, and J.~Yoon, {\it {Holography as
  a Gauge Phenomenon in Higher Spin Duality}},  {\em JHEP} {\bf 01} (2015) 055,
  [\href{http://arxiv.org/abs/1408.1255}{{\tt arXiv:1408.1255}}].

\bibitem{Peng:2017kro}
C.~Peng, {\it {Vector models and generalized SYK models}},  {\em JHEP} {\bf 05}
  (2017) 129, [\href{http://arxiv.org/abs/1704.04223}{{\tt arXiv:1704.04223}}].

\bibitem{Gurau:2010ba}
R.~Gurau, {\it {The 1/N expansion of colored tensor models}},  {\em Annales
  Henri Poincare} {\bf 12} (2011) 829--847,
  [\href{http://arxiv.org/abs/1011.2726}{{\tt arXiv:1011.2726}}].

\bibitem{Gurau:2011kk}
R.~Gurau, {\it {Universality for Random Tensors}},  {\em Ann. Inst. H. Poincare
  Probab. Statist.} {\bf 50} (2014), no.~4 1474--1525,
  [\href{http://arxiv.org/abs/1111.0519}{{\tt arXiv:1111.0519}}].

\bibitem{Kennedy:1979ar}
G.~Kennedy, R.~Critchley, and J.~S. Dowker, {\it {Finite Temperature Field
  Theory with Boundaries: Stress Tensor and Surface Action Renormalization}},
  {\em Annals Phys.} {\bf 125} (1980) 346.

\bibitem{Romeo:2000wt}
A.~Romeo and A.~A. Saharian, {\it {Casimir effect for scalar fields under Robin
  boundary conditions on plates}},  {\em J. Phys.} {\bf A35} (2002) 1297--1320,
  [\href{http://arxiv.org/abs/hep-th/0007242}{{\tt hep-th/0007242}}].

\bibitem{Perlmutter:2016pkf}
E.~Perlmutter, {\it {Bounding the Space of Holographic CFTs with Chaos}},  {\em
  JHEP} {\bf 10} (2016) 069, [\href{http://arxiv.org/abs/1602.08272}{{\tt
  arXiv:1602.08272}}].

\end{thebibliography}\endgroup

\end{document}